\documentclass{elsart}
\journal{Theoretical Computer Science}

\newcommand{\predel}{\mbox{\bf CL3}} 
\newcommand{\copredel}{\mbox{\bf CL3}'} 

\newcommand{\Last}{\mbox{\em Last}}
\newcommand{\elz}[1]{\mbox{$\parallel\hspace{-3pt} #1 \hspace{-3pt}\parallel$}} 
\newcommand{\elzi}[1]{\mbox{\scriptsize $\parallel\hspace{-3pt} #1 \hspace{-3pt}\parallel$}}
\newcommand{\gj}[1]{{\bf #1}} 
\newcommand{\chess}{\mbox{\em Chess}}
\newcommand{\rneg}{\neg}               
\newcommand{\pneg}{\neg}               
\newcommand{\emptyrun}{\langle\rangle} 
\newcommand{\oo}{\bot}            
\newcommand{\pp}{\top}            
\newcommand{\xx}{\wp}               
\newcommand{\legal}[2]{\mbox{\bf Lr}^{#1}_{#2}} 
\newcommand{\win}[2]{\mbox{\bf Wn}^{#1}_{#2}} 
\newcommand{\constants}{\{\mbox{\em constants}\}} 
\newcommand{\variables}{\{\mbox{\em variables}\}}    
\newcommand{\players}{\{\mbox{\em players}\}} 
\newcommand{\valuations}{\{\mbox{\em valuations}\}} 
\newcommand{\terms}{\{\mbox{\em terms}\}}
\newcommand{\seq}[1]{\langle #1 \rangle}           
\newcommand{\tuple}[2]{\mbox{$ {#1}_{1},\ldots,{#1}_{#2}$}} 

\newcommand{\gneg}{\neg}                  
\newcommand{\mli}{\rightarrow}                     
\newcommand{\mleq}{\hspace{2pt}\leftrightarrow\hspace{2pt}}   
\newcommand{\cla}{\mbox{\large $\forall$}}      
\newcommand{\cle}{\mbox{\large $\exists$}}        
\newcommand{\mld}{\vee}    
\newcommand{\mlc}{\wedge}  
\newcommand{\ade}{\mbox{\Large $\sqcup$}}      
\newcommand{\ada}{\mbox{\Large $\sqcap$}}      
\newcommand{\add}{\sqcup}                      
\newcommand{\adc}{\sqcap}                      
\newcommand{\clai}{\forall}     
\newcommand{\clei}{\exists}        
\newcommand{\adei}{\mbox{$\sqcup$}}      
\newcommand{\adai}{\mbox{$\sqcap$}}      
\newcommand{\tlg}{\bot}               
\newcommand{\twg}{\top}               
\newcommand{\intimpl}{\mbox{\hspace{2pt}$\circ$\hspace{-0.14cm} \raisebox{-0.058cm}{\Large --}\hspace{2pt}}}
\newcommand{\st}{\mbox{\raisebox{-0.05cm}{$\circ$}\hspace{-0.13cm}\raisebox{0.16cm}{\tiny $\mid$}\hspace{2pt}}}
\newcommand{\cost}{\mbox{\raisebox{0.12cm}{$\circ$}\hspace{-0.13cm}\raisebox{0.02cm}{\tiny $\mid$}\hspace{2pt}}}

\newtheorem{theoremm}{Theorem}[section]
\newtheorem{definitionn}[theoremm]{Definition}
\newtheorem{remarkk}[theoremm]{Remark}
\newtheorem{lemmaa}[theoremm]{Lemma}
\newtheorem{propositionn}[theoremm]{Proposition}
\newtheorem{conventionn}[theoremm]{Convention}
\newtheorem{examplee}[theoremm]{Example}
\newtheorem{exercisee}[theoremm]{Exercise}

\newenvironment{definition}{\begin{definitionn} \em}{ \end{definitionn}}
\newenvironment{exercise}{\begin{exercisee} \em}{ \end{exercisee}}
\newenvironment{remark}{\begin{remarkk} \em}{ \end{remarkk}}
\newenvironment{theorem}{\begin{theoremm}}{\end{theoremm}}
\newenvironment{lemma}{\begin{lemmaa}}{\end{lemmaa}}
\newenvironment{proposition}{\begin{propositionn} }{\end{propositionn}}

\newenvironment{proof}{ {\bf Proof.} }{\  $\Box$ \vspace{.1in} }
\newenvironment{example}{\begin{examplee} \em}{\end{examplee}}
\newenvironment{sublemma}[2]{\em \noindent{\bf Lemma \ref{#1}.#2} }{\vspace{5pt} }
\newenvironment{subproof}{\noindent{\bf Proof.} }{{\small  $\Box$} \vspace{5pt} }\newenvironment{subconvention}[2]{\em \noindent{\bf Convention \ref{#1}.#2}}{\vspace{5pt}}
\newenvironment{subassumption}[2]{\em \noindent{\bf Assumption \ref{#1}.#2}}{\vspace{5pt}}

\begin{document} 
\begin{frontmatter}
\title{From truth to computability I}
\author{Giorgi Japaridze
}
\address{Department of Computing Sciences\\ Villanova University\\ 800 Lancaster Avenue \\ Villanova, PA 19085, USA}
\ead{giorgi.japaridze@villanova.edu}
\begin{abstract}
\noindent  
The recently initiated approach called computability logic is a formal theory of interactive computation. It understands 
computational problems as games played by a machine against its environment, and uses  
logical formalism to describe valid principles of computability, with formulas representing computational problems and logical operators standing for operations on computational problems.
The concept of computability that lies under this approach is a     
generalization of Church-Turing computability from simple, two-step (question/answer, input/output) problems to problems of arbitrary degrees of interactivity. Restricting this concept to predicates, which are understood as computational problems of zero degree of interactivity,
yields exactly classical truth. This makes computability logic a generalization and refinement of classical logic. 

The foundational paper ``Introduction to computability logic" [Annals of Pure and Applied Logic 123 (2003), pp. 1-99]
was focused on semantics rather than syntax, and certain axiomatizability 
assertions in it were only stated as conjectures. The present contribution contains a verification of one of such conjectures: 
 a soundness and completeness proof for the deductive system $\predel$ 
which axiomatizes the most basic first-order fragment of computability logic called the finite-depth, elementary-base fragment. $\predel$ 
is a conservative extention of classical 
predicate calculus in the language which, along with all of the (appropriately generalized) logical operators of classical
logic, contains propositional connectives and quantifiers representing the so called choice operations. The atoms 
of this language are interpreted as elementary problems, i.e. predicates in the standard sense. Among 
the potential application areas for $\predel$ 
are the theory of interactive computation, constructive applied theories, knowledgebase systems, systems for resource-bound planning and action.

This paper is self-contained as it reintroduces all relevant definitions as well as main motivations. It is meant for a wide audience and does not assume that the reader has specialized knowledge in any particular subarea of logic or computer science.
\end{abstract}
\begin{keyword} 
\noindent Computability logic \sep Interactive computation \sep Game semantics \sep Linear logic \sep Constructive logics \sep Knowledge bases
\MSC primary: 03B47; secondary: 03F50; 03B70; 68Q10; 68T27; 68T30; 91A05
\end{keyword}
\end{frontmatter}

{\bf Acknowledgement:} This material is based upon work supported by the National Science Foundation under Grant No. 0208816.
\section{Introduction}\label{intr}

The question ``What 
can be computed?" is fundamental to theoretical computer science. The approach initiated recently in \cite{Jap03}, called {\em computability logic}, is about answering this question 
in a systematic way using logical formalism, with formulas understood as computational problems and logical operators as operations on computational problems. 

The collection of operators used in \cite{Jap03} to form the language of computability logic can be seen as  
a non-disjoint union of those of classical, intuitionistic and --- in a very generous sense --- linear logics, with the computational semantics of classical operators fully consistent with their standard meaning, and the semantics of the 
intuitionistic-logic and ``linear-logic" operators  formalizing the (so far rather abstract) computability and resource intuitions traditionally associated with those two logics. This collection  captures a set of most basic and natural operations on computational problems. But it generally remains, and will apparently always remain, open to different sorts of interesting extensions, depending on particular needs and taste. Some of such extensions are outlined in 
\cite{Japint}. Due to the fact that the language of computability logic has no clear-cut boundaries, every technical result in this area will deal with some fragment of that logic rather than the whole logic. The result presented in this paper concerns what in
\cite{Jap03} was called the {\em  finite-depth, elementary-base} fragment.  

This fragment is axiomatized as a rather unusual type of a deductive system called $\predel$. It is a conservative extension of classical first-order logic in a language obtained by incorporating into the formalism of the latter the 
``additive" and ``multiplicative" groups of to what we --- with strong reservations --- referred as ``linear-logic operators". 
The main technical result of this paper 
is a proof of soundness and completeness for $\predel$ with respect to computability semantics. A secondary result is a proof of decidability for the classical-quantifier-free (yet first-order) fragment of $\predel$. These proofs are given in Part 2. Part 1 is mainly devoted to a relatively brief (re)introduction to the relevant fragment and relevant aspects 
of computability logic, serving the purpose of keeping the paper self-contained both technically and motivationally. A more detailed and fundamental introduction 
to computability logic can be found in \cite{Jap03}. A considerably less technical and more compact --- yet comprehensive and  self-contained --- overview of computability logic  
is given in \cite{Japint}, reading which is most recommended for the first acquaintance with the subject and for 
a better appreciation of the import of the present results. The soundness and completeness of the propositional fragment {\bf CL1} of $\predel$ has been proven in \cite{Jap04}. 

Traditionally construed computational problems correspond to interfaces in transformational programs where the interaction between 
a system and its environment is simple and consists of two steps: querying the system and generating an answer. 
The computational problems that our approach deals with are more general in that the underlying interfaces
may have arbitrary complexity. Such problems and the corresponding computations can be called {\em interactive} as 
they model potentially long dialogues between the system and the environment. 
 From the technical point of view, computability logic is a game logic, because it defines interactive computations as games.
There is an extensive literature on ``game-style" models of computation in theoretical computer science: alternating Turing machines, interactive proof systems etc., that are typically only interesting in the context of computational complexity. Our approach, which is concerned with computability rather than  complexity and deals with deterministic rather than nondeterministic choices,
at present is only remotely related to that line of research, and the similarity is more terminological than conceptual. 
From the other, `games in logic' or `game semantics for linear logic' line, the closest to our present study appears to be Blass's  work [3], and less so some later studies by Abramsky, Jagadeesan (\cite{Abr94}), Hyland, Ong (\cite{Hay93}) and others. 
See \cite{Jap03} for discussions of how other approaches compare with ours.

There are considerable overlaps between the motivations and philosophies of linear (as well as intuitionistic) and computability logics, based on which \cite{Jap03} employed some linear-logic terminology. It should be pointed out, however, that computability logic is by no means {\em about} linear logic. Unlike most of the other game semantics approaches,  it is {\em not} an attempt to use games to construct good models for Girard's linear logic, Heyting's intuitionistic calculus or any other, already given popular
syntactic targets. Rather, computability logic evolves by the more and only natural scheme `from semantics to syntax': it views games as foundational entities in their own right, and then explores the logical principles validated by them. Its semantics, in turn, follows the scheme `from truth to computability'. It starts from the classical concept of truth and generalizes it to the more constructive, meaningful and useful concept of computability. As we are going to see, classical truth is nothing but a special case of computability; correspondingly, classical logic is nothing but a special fragment of computability logic and of $\predel$ in particular.  

The scope of the significance of our study is not limited to logic or theory of computing. As we will see later and more convincingly demonstrated in \cite{Jap03} and \cite{Japint}, some other application areas include constructive applied theories, knowledgebase systems, or resource-bound systems for planning and action. \vspace{20pt}

\begin{center}{\Large PART I}\vspace{0pt}\end{center} 
This part briefly reintroduces the subject and states the main results of the paper.

\section{Computational problems}\label{cpr}
The concept of computability on which the semantics of our logic is based is a natural but nontrivial generalization of Church-Turing computability from simple, two-step (question/answer, or input/output) problems to problems of arbitrary degrees and forms of interactivity where, in the course of interaction between the machine and the environment, input and output can be multiple and interlaced, perhaps taking place throughout the entire process of computation rather than just at the beginning (input) and the end (output) as this is the case with simple problems. Technically the concept is defined in terms of games: an interactive computational problem/task is a game between a machine and the environment, where dynamic input steps are called environment's moves, and output steps called machine's moves. 

The necessity in having a clear mathematical model of interactive computation hardly requires any justification: after all, most tasks that real computers and computer networks perform are truly interactive. And this sort of  tasks cannot always be reduced to simple series of (the well-studied and well-modeled) two-step tasks. E.g., interactive tasks involving multiple concurrent subtasks naturally generate situations/positions where both parties may have meaningful actions to take, and it may be up to the player whether to try to make a move or wait to see how things evolve, perhaps performing some vital computations while waiting and watching.\footnote{See Sections 3 and 15 of \cite{Jap03} for more detailed discussions and examples.} It is unclear whether the steps corresponding to such situations should be labeled as `machine-to-move' or `environment-to-move', which makes it impossible to 
break the whole process into consecutive pairs or alternately-labeled steps. 

Standard game-semantical approaches that understand players' strategies as functions from positions to moves\footnote{Often some additional restrictions are imposed on this sort of strategies. Say, in Abramsky's tradition, strategies only look 
at the other player's immediately preceeding moves.}  fail to capture the substance of interaction in full generality, as they essentially try to reduce interactive processes to simple chains of 
question/answer events. This is so because the 
`strategy=function' approach inherently only works when at every step of the play the player who is expected to make a move is uniquely determined. Let us call this sort of games \gj{strict}.\label{x58} Strictness is typically achieved by having  what in \cite{Ben01} is called 
\gj{procedural rules} or equivalent --- rules strictly regulating who and when should move, the most standard procedural rule being the one according to which players should take turns in alternating order. 

One of the main novel and distinguishing features of our games among the other concepts of games studied in the logical literature
--- including the one tackled by the author \cite{Jap97} earlier --- is the absence of procedural rules, based on which our games can be called \gj{free}.\label{x26} 
In these games, either player is free to make any move at any time. Instead of having procedural rules common for all games,
each particular game comes with its own what we call  
 \gj{structural rules}. These are rules telling what moves are \gj{legal}\label{x40} for a given player in a given position. Making an illegal move by a player is possible but it results in a loss for that player.  The difference between procedural and structural rules is not just terminological. Unlike the standard procedural rules that allow only one  player to move (at least move without penalty) 
in every given situation,  structural rules can be arbitrarily lenient. In particular, they do not necessarily have to
satisfy the condition that in every position at most one of the players may have legal moves. When, however, this condition still {\em is} satisfied, we essentially get the above-mentioned strict type of a game:  the structural rules of such a game 
can be thought of as procedural rules according to which the player that is expected to move in a given non-terminal position is the (now uniquely determined) one that has legal moves in that position. Strict games are thus special cases of our more general free games. The latters present a more adequate and apparently universal modeling tool for interactive tasks.  

Strategies for playing free games can and should no longer be defined as functions from positions to moves. In the next section we will define them as higher-level entities called play machines.

To define our games formally, we fix several classes of objects and dedicated (meta-) variables for them. By placing the common name for objects  between braces we denote the set of all objects of that type. Say, 
$\variables$ stands for the set of all variables. These objects are:

\begin{itemize}
\item \gj{Variables},\label{x68} with $\variables=\{v_0,v_1,v_2\}$. Letters $x,y,z$ will be used as metavariables for variables. 
\item \gj{Constants},\label{x8} with $\constants=\{0,1,2,\ldots\}$. Letter $c$ will be used as a metavariable for constants.
\item \gj{Terms},\label{x62} with $\terms=\variables\cup\constants$. Letter $t$ will be used as a metavariable for terms.
\item \gj{Valuations},\label{x67} defined as any functions of the type $\variables\rightarrow\constants$. 
A metavariable for valuations: 
$e$. Each valuation $e$ extends to a function of the type $\terms\rightarrow\constants$ by stipulating that, for every  constant $c$, $e(c)=c$.
\item \gj{Moves},\label{x42} defined as any finite strings over the standard keyboard alphabet
plus the symbol $\spadesuit$.\label{x13} 
Metavariables for moves: $\alpha,\beta$.
\item \gj{Players}, with $\players=\{\pp,\oo\}$. Here and from now on $\pp$ and $\oo$ are symbolic names for 
the machine and the environment, respectively. Letter $\xx$\label{x45} will range over players, with $\pneg\xx$ meaning ``$\xx$'s adversary", i.e. the player that is not $\xx$.
\item \gj{Labeled moves}, or \gj{labmoves}.\label{x39} They are defined as any moves prefixed with $\pp$ or $\oo$, with such a prefix (\gj{label}) indicating who has made the move. 
\item \gj{Runs},\label{x51} defined as any (finite or infinite) sequences of labmoves. Metavariables for runs: $\Gamma,\Delta$. 
\item \gj{Positions},\label{x48} defined as any finite runs. A metavariable for positions: $\Phi$. 
\end{itemize}
Runs and positions will often be delimited   with ``$\langle$" and ``$\rangle$", with $\emptyrun$\label{x93} thus denoting the \gj{empty run}. The meaning of an expression such as $\seq{\Phi,\xx\alpha,\Gamma}$ must be clear: this is the result of appending to the position $\Phi$ the position $\seq{\xx\alpha}$ and then the run $\Gamma$.

\begin{definition}\label{game}
 A {\bf game} is a pair $A=(\legal{A}{},\win{A}{})$, where:
\begin{itemize}
\item $\legal{A}{}$ is a function that sends each valuation $e$ to a set $\legal{A}{e}$ of runs, such that 
the following two conditions are satisfied:
\begin{description} 
\item[(a)] A run is in $\legal{A}{e}$ iff all of its nonempty finite initial segments are in $\legal{A}{e}$.
\item[(b)] No run containing the (whatever-labeled) move $\spadesuit$ is in 
$\legal{A}{e}$.
\end{description}
Elements of $\legal{A}{e}$ are called \gj{legal runs}\label{x41} of $A$ with respect to $e$, and all other runs called \gj{illegal}.\label{x29} In particular, if the last move of the shortest illegal initial segment of $\Gamma$ is $\xx$-labeled, then $\Gamma$ is said to be a {\bf $\xx$-illegal run}\label{x30} of $A$ with respect to $e$.
\item $\win{A}{}$  is a function of the type \{{\em valuations}\} $\times$ \{{\em runs}\} $\rightarrow$ \{{\em players}\} such that, writing $\win{A}{e}\seq{\Gamma}$ for $\win{A}{}(e,\Gamma)$,  the following condition is satisfied:
\begin{description}
\item[(c)]  If $\Gamma$ is a $\xx$-illegal run of $A$ with respect to $e$, then $\win{A}{e}\seq{\Gamma}=\pneg\xx$.
\end{description}
\end{itemize}
\end{definition}

To what we earlier referred as ``structural rules" are thus represented by the 
$\legal{}{}$ component of a game, and we call it the \gj{structure}\label{x59} of that game.  
The meaning of the $\win{}{}$ component, called the \gj{content},\label{x10} is that it tells us who has won 
a given run of the game. When $\win{A}{e}\seq{\Gamma}=\xx$, we say that $\Gamma$ is a \gj{$\xx$-won} (or \gj{won by $\xx$}) run of $A$ with respect to $e$. 

Just as predicates (their truth values) in classical logic
generally depend on how certain variables are interpreted, so do games: both the structure and the content of 
a game take valuation $e$ as a parameter. We will typically omit this parameter when it is irrelevant or clear from the context. 

Meaning by an  \gj{illegal move}\label{x28} a move adding which (with the corresponding label) to the given position makes it illegal,
condition (a) of Definition \ref{game} corresponds to the intuition that a run is legal iff no illegal moves have been made in it. This automatically makes the empty run $\emptyrun$ a legal run of every game. Our selection of the set of 
moves is very generous, and it is natural and technically very convenient to assume that certain moves will never be legal. According to condition (b), $\spadesuit$ is such a move. 
As for condition (c), it tells us that an illegal run is always lost by the player who has made the first illegal move. 

We say that a game $A$ \gj{depends on}\label{x11} a variable $x$ iff there 
are two valuations $e_1$ and $e_2$ that agree on all variables except $x$ such that either $\legal{A}{e_1}\not=\legal{A}{e_2}$ or $\win{A}{e_1}\not=\win{A}{e_2}$.

A game $A$ is said to be \gj{finitary}\label{x24} iff there is a finite set $\vec{x}$ of variables such that, for any two valuations
$e_1$ and $e_2$ that agree on all variables from $\vec{x}$, we have $\legal{A}{e_1}=\legal{A}{e_2}$ and 
$\win{A}{e_1}=\win{A}{e_2}$. Otherwise $A$ is \gj{infinitary}.\label{x31} 
One can easily show that for each finitary game $A$ there is a unique smallest set $\vec{x}$ that satisfies 
the above condition ---
in particular, the elements of this $\vec{x}$ are exactly the variables on which $A$ depends. A finitary game that depends on exactly $n$ variables is said to be \gj{$n$-ary}.\label{x1} 

A \gj{constant game}\label{x9} means  a 0-ary game. There is a natural operation, called \gj{instantiation},\label{x34} of the type 
$\valuations\times$\{{\em games}\}$\rightarrow$\{{\em constant games}\}. The result of applying this operation to $(e,A)$ is denoted \(e[A].\label{x92}\) 
Intuitively, $e[A]$ is the constant game obtained from $A$ by fixing the values of all variables to   
the constants assigned to them by $e$. Formally, game $e[A]$ is defined by stipulating that, for any valuation $e'$, 
$\legal{e[A]}{e'}=\legal{A}{e}$ and $\win{e[A]}{e'}=\win{A}{e}$. This makes $e'$ irrelevant, so that it can be omitted
and we can just write $\legal{e[A]}{}$ and $\win{e[A]}{}$. For any game $A$, these two expressions mean the same as
$\legal{A}{e}$ and $\win{A}{e}$.

Games whose $\legal{}{}$ component does not depend on the valuation parameter are said to be \gj{unistructural}.\label{x65} Constant games are thus special cases of unistructural games where the $\win{}{}$ component does not depend on valuation, either. Formally, a game $A$ is unistructural iff, for any two valuations $e_1$ and $e_2$, $\legal{A}{e_1}=\legal{A}{e_2}$.
 
We say that a game $A$ is \gj{finite-depth}\label{x25} iff there is a (smallest) integer $n$, called the \gj{depth of $A$},\label{x12} such that, for every valuation $e$ and run $\Gamma$ with $\Gamma\in\legal{A}{e}$, the length of $\Gamma$ does not exceed $n$. 
Games of depth $0$ are said to be \gj{elementary}.\label{x15} Thus, elementary games are games that  have no legal moves:
the empty run $\emptyrun$ is the only legal run of such games.  This automatically makes all elementary games unistructural.

Constant elementary games are said to be \gj{trivial}.\label{x63} Obviously there are exactly two trivial games. We denote them  
by the same symbols $\pp$ and $\oo$ as we use for the two players. In particular, $\pp$  is the (unique) trivial game with 
$\win{\twg}{}\emptyrun=\pp$, and $\oo$ is the (unique) trivial game with $\win{\tlg}{}\emptyrun=\oo$. 

Let us agree to understand classical \gj{predicates}\label{x49} --- in particular, predicates on $\{0,1,2,\ldots\}$ --- as functions from valuations to $\{\pp,\oo\}$ rather than, as more commonly done, functions from tuples of constants to $\{\pp,\oo\}$. Say, $x>y$ is the predicate that is true at a valuation $e$ 
(returns $\pp$ for it) iff $e(x)>e(y)$. This understanding of predicates is technically more convenient, and is also slightly more general as it captures infinite-arity predicates along with finite-arity ones. 
 
All elementary games have the same structure, so their trivial $\legal{}{}$ component can be ignored and each such game identified with its $\win{}{}$ component; furthermore, by setting the run parameter to its only relevant value
$\emptyrun$ in such games, $\win{}{}$ becomes a function of the type $\valuations\rightarrow\players$, i.e. exactly what we
call a predicate. We thus get a natural one-to-one correspondence between elementary games and predicates: every predicate $p$ can be thought of as the (unique) elementary game $A$ such that 
$\win{A}{e}\emptyrun=\pp$ iff $p$ is true at $e$; and vice versa: every elementary game $A$ thought of as the predicate 
$p$ that is true at $e$ iff $\win{A}{e}\emptyrun=\pp$. With this correspondence in mind, we will be using the terms ``predicate" and ``elementary game" as synonyms. So, computability logic understands each predicate $p$ as a
game of zero degree of interactivity,  (the only legal run $\emptyrun$ of) which is automatically won by the machine if $p$ is true, and lost if $p$ is false. This makes the classical concept of predicates a special case of our concept of computational problems; correspondingly, the classical concept of truth is going to be a special case of our concept of computability --- in particular, computability restricted to elementary games. 

The class of games in the above-defined sense is general enough to model anything that we would call a (two-player, two-outcome) interactive problem.  However, it is too general. There are games where the chances of a player to succeed essentially depend on the relative speed at which its adversary responds and, as it is not clear what particular speed would be natural to assume for the environment, we do not want to consider that sort of games meaningful computational problems. 
A simple example would be the game where all non-$\spadesuit$ moves are legal and that is won by the player who moves first. This is merely a contest of speed.  

Below we define a subclass of games called static. Intuitively, static games are games where speed is irrelevant: in order to succeed, only matters {\em what} to do (strategy) rather than {\em how fast} to do (speed). 

We say that a run $\Delta$ is a {\bf $\xx$-delay} of a run $\Gamma$ iff the following two conditions are satisfied:
\begin{itemize}
\item for each player $\xx'$, the subsequence of the $\xx'$-labeled moves of $\Delta$ is the same as that of $\Gamma$, and
\item 
for any $n,k\geq 1$, if the $n$th $\xx$-labeled move is made later than (is to the right of) the $k$th $\pneg\xx$-labeled move in $\Gamma$, then so is it in $\Delta$.
\end{itemize}

\noindent The above means that in  $\Delta$  each player has made the same sequence of moves as in $\Gamma$, only, in $\Delta$, $\xx$ might have been acting with some delay. 

\begin{definition}\label{static}
A game  $A$ is said to be \gj{static}\label{stg}
 iff, whenever $\win{A}{e}\seq{\Gamma}=\xx$ and 
$\Delta$ is a $\xx$-delay of $\Gamma$, we have $\win{A}{e}\seq{\Delta}=\xx$.
\end{definition}

Roughly, in a static game, if a player can succeed when acting fast, it will remain equally successful acting the same way but slowly. This releases the player from any pressure for time and allows it to select its own pace for the game. The following fact is a rather straightforward observation:

\begin{proposition}\label{predstat}
All elementary games are static.
\end{proposition}

One of the main theses on which computability logic relies philosophically is that the concept of static games is an adequate 
formal counterpart of our intuitive notion of ``pure", speed-independent computational problems. See Section 4 of 
\cite{Jap03} for a detailed discussion and examples in support of this thesis. 

Now we are ready to formally clarify what we mean by computational problems: we use the term ``(interactive) \gj{computational problem}" (or simply ``\gj{problem}") as a synonym of ``static game".

As shown in \cite{Jap03} (Proposition 4.8), all strict games are static. But not vice versa.
The class of static games is substantially more general, and is free of the limitations of strict games discussed earlier. 
The closure of the set of all predicates under the game operations that we will define in Section \ref{op} forms a natural class of static yet free games. Section 3 of \cite{Jap03} shows an example of a natural problem from this class that is impossible to model with strict games.

\section{Computability}\label{cp}
The definitions that we give in this section are semiformal and incomplete. All of the omitted technical details are however rather standard and can be easily restored by anyone familiar with Turing machines. If necessary, the corresponding detailed definitions can be found in Part II of \cite{Jap03}. 
  
The central point of our philosophy is to require that agent $\pp$ be implementable as a computer program, with effective and fully determined behavior. On the other hand, the behavior (including speed) of agent $\oo$, who represents a capricious user or the blind forces of nature, can be arbitrary. This intuition is captured by the model of interactive computation where $\pp$ is formalized as what we call HPM. 

An \gj{HPM}\label{x27} (\gj{hard-play machine}) ${\mathcal H}$ is a Turing machine with the capability of making moves. At any time, the current position of the game is fully visible to this machine, as well as it is fully informed about the valuation  with respect to which the outcome of the play will be evaluated. This effect is achieved by letting the machine have --- along with the ordinary read/write \gj{work tape} --- two additional read-only tapes: the \gj{valuation tape} and the 
\gj{run tape}. The former spells some valuation $e$ by listing constants in the lexicographic order of
the corresponding variables. Its contents remain unchanged throughout the work of the machine. As for the run tape, it serves as a dynamic input, spelling, at any time, the current position of the game.
Every time one of the players makes a move, that move (with the corresponding label) is automatically appended to the contents of the run tape. 

As always, the computation proceeds in discrete steps, also called \gj{clock cycles}. The technical details about how exactly ${\mathcal H}$ makes a 
move $\alpha$ are not very interesting, but for clarity let us say that this is done by constructing $\alpha$ in a certain section 
(say, the beginning) of the work tape and then entering one of the specially designated states called \gj{move states}. 
Thus, ${\mathcal H}$ can make at most one move per clock cycle. On the other hand, as we noted, there are no limitations to 
the relative speed of the environment, so the latter can make any finite number of moves per cycle. 
We assume that the run tape remains stable during a clock cycle and is updated only on a transition from one cycle to another. Again, there is flexibility in arranging 
 details regarding what happens if both of the players make moves ``simultaneously". For clarity, we assume that 
if, during a given cycle, ${\mathcal H}$ makes a move $\alpha$ and the environment makes moves $\beta_1,\ldots,\beta_n$, then the position spelled on the run tape throughout the next cycle will be the result of appending 
$\seq{\oo\beta_1,\ldots,\oo\beta_n,\pp\alpha}$ to the current position.

A \gj{configuration}\label{x7} of ${\mathcal H}$ is defined in the standard way: this is a full description of the 
(``current") state of the machine, the locations of its three scanning heads, 
and the contents of its tapes, with the exception that, in order to make finite descriptions of configurations possible, 
we do not formally include a description of the unchanging contents of the valuation tape as a part of configuration, but rather account for it in our definition of computation branches as this will be seen shortly. 
The \gj{initial configuration} is the configuration where ${\mathcal H}$ is in its initial state and the work and run tapes are empty. A configuration $C'$ is said to be an \gj{$e$-successor} of a configuration $C$ iff, when valuation $e$ is spelled on the valuation tape, $C'$ can legally follow $C$ in the standard (standard for multitape Turing machines) sense, based on the transition function of the machine and accounting for the possibility of the above-described nondeterministic updates 
of the contents of the run tape. An \gj{$e$-computation branch}\label{x5} of ${\mathcal H}$ is a sequence of configurations of ${\mathcal H}$ where the first
configuration is the   initial configuration and every other configuration is an $e$-successor of the previous one.
Thus, the set of all $e$-computation branches captures all possible scenarios corresponding to different behaviors by $\oo$.

Each $e$-computation branch $B$ of ${\mathcal H}$ incrementally spells (in the obvious sense) some run $\Gamma$ on the run tape, 
which we call the \gj{run spelled by $B$}.\label{x53} 
Then, for a game $A$, we write ${{\mathcal H}}\models_e A$\label{x94} (``${\mathcal H}$ \gj{wins $A$ on $e$}") iff, whenever 
$B$ is an $e$-computation branch of ${\mathcal H}$ and $\Gamma$ the run spelled by $B$, we have  
$\win{A}{e}\seq{\Gamma}=\pp$. And we write ${{\mathcal H}}\models A$ iff ${{\mathcal H}}\models_eA$ for every valuation $e$. The meaning of ${{\mathcal H}}\models A$ is that ${\mathcal H}$ \gj{wins}\label{x72} (\gj{computes}, \gj{solves}) $A$. Finally, we write $\models A$ and say that $A$ is 
\gj{winnable}\label{x71} (\gj{computable}, \gj{solvable}) iff there is an HPM ${\mathcal H}$ with ${{\mathcal H}}\models A$. 

The above ``hard-play" model of interactive computation seemingly strongly favors the environment in that the latter may 
be arbitrarily faster than the machine. What happens if we start limiting the speed of the environment? The answer is: 
{\em nothing} as far as computational problems are concerned. The model of computation  called 
EPM takes the idea if limiting the speed of the environment to the extreme, yet it yields the same class of computable problems. 

An \gj{EPM}\label{x18} (\gj{easy-play machine}) is defined in the same way as an HPM, with the only difference that now the environment can (but is not obligated to) make
a move only when the machine explicitly allows it to do so, the event that we call \gj{granting permission}.\label{x46} Technically permission is granted by entering one of the specially designated states called \gj{permission states}.\label{x47} The only requirement that the machine is expected to satisfy is that, as long as the adversary plays legal, the machine should grant permission every once in a while; how long that ``while" lasts, however, is totally up to the machine. This amounts to having full control over the speed of the adversary. 

The above intuition is formalized as follows. We say that an $e$-computation 
branch $B$ of a given EPM is \gj{fair}\label{x20} if permission is granted infinitely many times in $B$. A \gj{fair EPM}\label{x21} is an EPM  whose every $e$-computation branch (for every valuation $e$) is fair. For an EPM $\mathcal E$ and valuation $e$, we write 
${\mathcal E}\models_e A$ (``$\mathcal E$ \gj{wins $A$ on $e$}") iff, whenever 
$B$ is an $e$-computation branch of $\mathcal E$ and $\Gamma$ the run spelled by $B$, we have:
\begin{itemize}
\item  $\win{A}{e}\seq{\Gamma}=\pp$, and 
\item $B$ is fair unless $\Gamma$ is a $\oo$-illegal run of $A$ with respect to $e$.
\end{itemize}
Just as with HPMs, for an EPM $\mathcal E$, ${\mathcal E}\models A$ (``$\mathcal E$ \gj{wins}\label{x73} (\gj{computes}, \gj{solves}) $A$") means that 
${\mathcal E}\models_eA$ for every valuation $e$. 
Note that when we deal with fair EPMs, the second one of the above two conditions is always satisfied, and then the definition of $\models_e$ is literally the same as in the case of HPMs. 

\begin{remark}\label{assume}
When trying to show that a given EPM wins a given game, it is always perfectly safe to assume that the environment never makes an illegal move, for if it does, the machine automatically wins (unless the 
machine itself has made an illegal move earlier, in which case it does not matter what the environment did afterwards anyway, so that we may still assume that the environment did not make any illegal moves). 
Making such an assumption can often significantly simplify computability proofs.
\end{remark}

 The following fact, proven in \cite{Jap03} (Theorem 17.2), establishes equivalence between the two models for computational problems:

\begin{proposition}\label{eq}
For any static game $A$, the following statements are equivalent:\vspace{-5pt}
\[\begin{array}{ll}
\mbox{{\bf (i)}} & \mbox{there is an EPM that wins $A$;}\\
\mbox{{\bf (ii)}} & \mbox{there is an HPM that wins $A$;}\\
\mbox{{\bf (iii)}} & \mbox{there is a fair EPM that wins $A$.}\vspace{-5pt}
\end{array}\]
Moreover, there is an effective procedure that converts any EPM $($resp. HPM$)$ $\mathcal M$  into an HPM $($resp. fair EPM$)$ $\mathcal N$ such that, for every static game $A$ and valuation $e$,  ${\mathcal N}\models_e A$ whenever ${\mathcal M}\models_e A$.
\end{proposition} 

The philosophical significance of this proposition is that it reinforces the thesis according to which static games are games that allow us to make full abstraction from speed. Its technical importance is related to the fact that the EPM-model is much more convenient when it comes to describing strategies as we will have a chance to see in Part 2, and is a more direct and practical formal counterpart 
of our intuitive notion of what could be called \gj{interactive algorithm}.\label{x35} 

The two models act as natural complements to each other: we can meaningfully talk about the (uniquely determined) play between a given HPM and a given EPM, while this is impossible if both players are HPMs or both are EPMs. This fact will be essentially exploited in our completeness proof for logic $\predel$, where we describe an environment's strategy as an 
EPM and show that no HPM can win the given game against such an EPM. 

Let us agree on the following notation and terminology:
\begin{itemize} 
\item For a run $\Gamma$, $\rneg\Gamma$\label{x75} denotes the result of reversing all labels in $\Gamma$, i.e.
changing each labmove $\xx\alpha$ to $\pneg\xx\alpha$.
\item For a run $\Gamma$ and a computation branch $B$ of an HPM or EPM, we say that $B$ \gj{cospells}\label{x52} $\Gamma$ iff
$B$ spells $\rneg\Gamma$. 
\end{itemize}

Intuitively, when a given machine $\mathcal M$ plays as  $\oo$ (rather than $\pp$), then the run that is generated by  a given computation branch $B$ of $\mathcal M$ is the run cospelled (rather than spelled) by $B$, for the moves that $\mathcal M$ makes 
get the label $\oo$, and the moves that its adversary makes get the label $\pp$.
 
The following lemma will be used in our completeness proof for $\predel$. Its second clause assumes some standard 
encoding for play machines and their configurations. 

\begin{lemma}\label{lem}
Let ${\mathcal E}$ be a fair EPM.

(a) For any HPM ${\mathcal H}$ and any valuation $e$, there are a uniquely defined 
 $e$-computation branch $B_{{\mathcal E}}$ of ${\mathcal E}$ and a uniquely defined $e$-computation branch $B_{{\mathcal H}}$ of ${\mathcal H}$
--- which we respectively call {\bf the $({{\mathcal E}},e,{{\mathcal H}})$-branch} and {\bf the 
$({{\mathcal H}},e,{{\mathcal E}})$-branch}\label{x6}
--- such that the run spelled by $B_{{\mathcal H}}$ is the run cospelled by $B_{{\mathcal E}}$. 

(b) Suppose $e_0,e_1,e_2,\ldots$ are valuations such that the function $g$ defined by $g(c,i)=e_c(v_i)$ is effective. Then there is an effective function which takes $($the code of$)$ an arbitrary HPM ${\mathcal H}$ and arbitrary nonnegative integers $c,n$, and returns the $($code of the$)$ $n${\em th} configuration of the $({{\mathcal E}},e_c,{{\mathcal H}})$-branch. Similarly for 
the $({{\mathcal H}},e_c,{{\mathcal E}})$-branch.
\end{lemma}
 
When $e$, ${\mathcal H}$, ${\mathcal E}$, $B_{{\mathcal H}}$ are as in clause (a) of the above lemma, 
we call the run $\Gamma$ spelled by $B_{{\mathcal H}}$ {\bf the ${\mathcal H}$ vs ${\mathcal E}$ run on $e$};\label{x54}
then, if $A$ is a game with 
$\win{A}{e}\seq{\Gamma}=\pp$ (resp. $\win{A}{e}\seq{\Gamma}=\oo$), we say that ${\mathcal H}$ \gj{wins}\label{x74}
(resp. \gj{loses})\label{x744}
 \gj{$A$ against ${\mathcal E}$ on $e$}. 

A formal proof of Lemma \ref{lem} is given in \cite{Jap03} (Lemma 20.4),\footnote{Clause (b) of our lemma \ref{lem} is 
slightly stronger than the official formulation of the corresponding clause (c) of Lemma 20.4 of \cite{Jap03}. However, 
the proof given in \cite{Jap03} is just as good for our present strong formulation.} 
and we will not reproduce it here. Instead, the following intuitive explanation would suffice:

Assume ${\mathcal E}$ is a fair EPM, ${\mathcal H}$ is an arbitrary HPM and  $e$ an arbitrary valuation. 
The play that we are going to describe is the unique 
play generated when the two machines play against each other, with ${\mathcal H}$ in the role of $\pp$, ${\mathcal E}$ in the role of $\oo$, and valuation $e$ spelled on the valuation tapes of both machines. 
We can visualize this play as follows.
Most of the time during the play ${\mathcal H}$ remains inactive (sleeping); it is woken up only when ${\mathcal E}$ enters a permission state, on which event ${\mathcal H}$ makes a (one single) transition to its next computation step --- that may or may not result in making a move --- and goes back to sleep that will continue until ${\mathcal E}$ enters  a permission state again, and so on. From ${{\mathcal E}}$'s perspective, ${\mathcal H}$ acts as a patient adversary who makes one or zero move only when granted permission, just as the EPM-model assumes.  And from ${\mathcal H}$'s perspective, who, like a person under global anesthesia,  has no sense of time during its sleep and hence can think that the wake-up events that it calls the beginning of a clock cycle happen at a constant rate, ${\mathcal E}$ acts as an adversary who can make any finite number of moves during a clock cycle (i.e. while ${\mathcal H}$ was sleeping), just as the HPM-model assumes. This scenario uniquely determines an $e$-computation branch $B_{{\mathcal E}}$ of ${\mathcal E}$ that we call
the $({{\mathcal E}},e,{{\mathcal H}})$-branch, and an $e$-computation branch $B_{{\mathcal H}}$ of ${\mathcal H}$ that we call
the $({{\mathcal H}},e,{{\mathcal E}})$-branch. What we call the ${\mathcal H}$ vs ${\mathcal E}$ run on $e$ is the run generated 
in this play. In particular --- since we let ${\mathcal H}$ play in the role of $\pp$ --- this is the run spelled by $B_{{\mathcal H}}$. ${\mathcal E}$, who plays in the role of $\oo$, sees the same run, only it sees the labels of that run in negative colors. 
That is, $B_{{\mathcal E}}$ cospells rather than spells that run. This is exactly what clause (a) of Lemma \ref{lem} asserts.  
Now suppose $e_0,e_1,e_2,\ldots$ and $g$ are as in clause (b), and $e$ is one of the $e_c$. 
Then, using $g$, the contents of any initial segment of the valuation tape(s) can 
be effectively constructed from $c$. Therefore the work of either machine can be effectively traced up to any given computation step $n$, which implies clause (b). 


\section{Operations on computational problems}\label{op}
As noted, computability logic is an approach that uses logical formalism for specifying and studying interactive computational problems in a systematic way, understanding logical operators as operations on games/problems. It is time to define basic operations on games. It should be noted that even though our interests are focused on static games, the operations defined in this section are equally meaningful for non-static (\gj{dynamic}) games as well. So, we do not restrict the scope of those definitions to static games, and throughout the section we let the letters $A,B$ range over any games.
Here comes the first definition:

\begin{definition}\label{sov}
Let $A$ be any game, $x_1,\ldots,x_n$ ($n\geq 0$) pairwise distinct variables, and $t_1,\ldots,t_n$ any terms. For any valuation $e$, let $e^\circ$ denote the unique valuation that agrees with $e$ on all variables that are not among 
$x_1,\ldots,x_n$, such that, for each $x_i\in\{x_1,\ldots,x_n\}$, $e^\circ(x_i)=e(t_i)$. Then we define the game  
\(A[x_1/t_1,\ldots,x_n/t_n]\)
by stipulating that, for any valuation $e$, $\legal{A[x_1/t_1,\ldots,x_n/t_n]}{e}=\legal{A}{e^\circ}$ and
$\win{A[x_1/t_1,\ldots,x_n/t_n]}{e}=\win{A}{e^\circ}$; in other words, $e\bigl[A[x_1/t_1,\ldots,x_n/t_n]\bigr]=e^\circ \bigl[A\bigr]$.
\end{definition}

This operation, that we call \gj{substitution of variables},\label{x60} is a generalization of the standard operation of substitution of variables known from classical predicate logic. Intuitively, $A[x_1/t_1,\ldots,x_n/t_n]$ is the same as $A$, only with (the values of) variables $x_1,\ldots,x_n$ 
``read as" (the values of) $t_1,\ldots,t_n$, respectively. Each $t_i$ here can be either a variable or a constant. Remember from Section \ref{cpr} that when $t_i$ is a constant, $e(t_i)=t_i$. 

Example: If $A$ is the elementary game 
$x\times y> z+u$, then $A[x/z,z/6,u/y]$ would be the game $z\times y>6+y$.

Sometimes it is convenient to fix a certain tuple $(\tuple{x}{n})$ of pairwise distinct variables for a game $A$ throughout a 
context and write 
$A$ in the form $A(x_1,\ldots,x_n)$.\label{x90} We will refer to such a tuple $(\tuple{x}{n})$ as the \gj{attached tuple}\label{x4} of (the given representation of) $A$.  When doing so, we do not necessarily mean that $A(\tuple{x}{n})$ is an 
$n$-ary game and/or that $\tuple{x}{n}$ are exactly the variables on which this game depends.  Once $A$ is given with an attached tuple $(\tuple{x}{n})$, we will write $A(\tuple{t}{n})$ to mean the same as the more clumsy expression 
$A[x_1/t_1,\ldots,x_n/t_n]$. A similar notational practice is common in the literature for predicates. Thus, the above game 
$x\times y> z+u$ can be written as $A(x,z,u)$, in which case $A(z,6,y)$ will denote the game $z\times y>6+y$.

We have already seen two meanings of symbol $\gneg$: one was that of an operation on players (Section \ref{cpr}), and one that of an operation on runs (Section \ref{cp}). Here comes the third meaning of it --- that of an operation on games: 

\begin{definition}\label{neg}
The \gj{negation} $\gneg A$ of a game $A$ is defined by:
\begin{itemize}
\item $\legal{\gneg A}{e}=\{\Gamma\ |\ \rneg\Gamma\in\legal{A}{e}\}$.
\item $\win{\gneg A}{e}\seq{\Gamma}=\pneg\win{A}{e}\seq{\rneg\Gamma}$.
\end{itemize}
\end{definition}

Intuitively, $\gneg A$ is game $A$ with the roles of the two players switched: $\pp$'s moves or wins become $\oo$'s moves or wins, and vice versa.  For example, if {\em Chess} is the game of chess from the point of view of the white player, then $\gneg \mbox{\em Chess}$ would be the same game from the point of view of the black player. 

The operations $\mlc$ and $\mld$ produce parallel combinations of games. Playing $A_1\mlc\ldots\mlc A_n$ or $A_1\mld\ldots\mld A_n$ means playing the $n$ games concurrently. Both $A_1\mlc\ldots\mlc A_n$ and $A_1\mld\ldots\mld A_n$ have exactly the same structure (legal moves), and the only difference is in how the winner is determined: in order to win, in the former $\pp$ needs to win in each of the $n$ components, while in the latter winning in one of the components is  sufficient. To indicate that a given move is made in the $i$th component, the player should prefix it with the string ``$i.$". Any move that does not have one of such prefixes will be considered illegal. 

Here comes the formal definition. In it  
the notation 
\(\Gamma^{\gamma}\label{jun23}\) means the result of removing from $\Gamma$ all labeled moves 
except those of the form $\xx \gamma\alpha$ ($\xx\in\{\pp,\oo\}$), and then deleting the prefix ``$\gamma$" in the remaining moves, i.e. changing each such $\xx\gamma\alpha$ to $\xx\alpha$.

\begin{definition}\label{mc}
Let $A_1,\ldots,A_n$ ($n\geq 2$) be any games.\vspace{5pt}

The \gj{parallel} \gj{conjunction} $A_1\mlc\ldots\mlc A_n$ of $A_1,\ldots,A_n$\label{x76} 
is defined by:
\begin{itemize}
\item $\Gamma\in \legal{A_1\mlc\ldots\mlc A_n}{e}$ iff every move of $\Gamma$ has one of the prefixes $``1.",\ldots,``n."$ and, for each $i\in\{1,\ldots,n\}$, $\Gamma^{i.}\in\legal{A_i}{e}$. 
\item Whenever $\Gamma\in\legal{A_1\mlc\ldots\mlc A_n}{e}$, $\win{A_1\mlc\ldots\mlc A_n}{e}\seq{\Gamma}=\pp$ iff, for 
all $i\in\{1,\ldots,n\}$,
$\win{A_i}{e}\seq{\Gamma^{i.}}=\pp$.
\end{itemize}

The \gj{parallel} \gj{disjunction} $A_1\mld\ldots\mld A_n$\label{x77} of $A_1,\ldots,A_n$ 
is defined in exactly the same way, only with ``$\pp$" replaced by ``$\oo$". Equivalently, it can be defined 
by 
\(A_1\mld\ldots\mld A_n\ =_{def}\ \gneg(\gneg A_1\mlc\ldots\mlc \gneg A_n).\)
\end{definition}

The other operation from the same group --- the \gj{parallel} \gj{implication} $A\mli B$\label{x78} of games $A$ and $B$
 --- is defined by
\(A\mli B\ =_{def}\ (\gneg A)\mld B.\)
 
Intuitively, $A\mli B$ is the problem of \gj{reducing} $B$ ({\em consequent}) to $A$ ({\em antecedent}). That is, solving $A\mli B$ means solving $B$ having $A$ as a
\gj{computational resource}. Generally, computational resources are symmetric to computational tasks/problems: what is a problem for one player to solve, is a resource for the other player to use, and vice versa. 
Since in the antecedent of $A\mli B$ the roles of the players are switched, $A$ becomes a problem for $\oo$ 
to solve, and hence a resource that $\pp$ can use. Thus, our semantics of computational problems is, at the same time, a semantics of computational resources. As noted before, this offers a materialization of the abstract resource philosophy associated with linear logic \cite{Gir87}. We will see a couple of examples later supporting this claim. More elaborated  examples and discussions can be found in \cite{Jap03}, where, in Section 26,
the context of computational resources is further extended to informational and physical resources as well. \cite{Japint} also abounds with illustrative examples. 

On an intuitive level, our parallel operations can be related to the corresponding multiplicative operators of linear 
logic. The game-semantical approach to linear-logic-style connectives is not new in principle, even if it has rather stubbornly resisted a complete treatment within natural frameworks. What makes our understanding of ``multiplicatives" substantially different from other (\cite{Abr94,Bla92,Hay93} etc.) interpretations is that they are {\em free}, i.e. generate free games, even when applied to strict games. As an example, consider the game {\em Chess}$\mlc${\em Chess}. Assume an agent plays this two-board game over the Internet against two independent adversaries ---
adversary \#1 on board \#1 and adversary \#2 on board \#2 --- that, together, form the (one) environment for the agent. 
As we agreed, {\em Chess} is the game playing which means playing the game of chess white. Hence, in the initial position of
$\chess\mlc\chess$, only the agent has legal moves. But once such a move is made, say, on board \#1, the picture changes. Now both the agent and the environment have legal moves: the agent may make another opening move on board \#2, 
while the environment --- in particular adversary \#1 --- may make a reply move on board \#1. This is a situation where which player `is to move' is no longer strictly determined, so the next player to move will be the one who can or wants to act faster. A strict-game approach would impose some additional conditions uniquely determining the next player to move. Such conditions would most likely be artificial and not quite adequate, for the situation we are trying to model is a concurrent play on two boards against two independent adversaries, and we cannot or should not expect any coordination between their actions. Most of the compound tasks we perform in everyday life are free rather than strict, and so are most computer communication/interaction protocols. A strict understanding of $\mlc$ would essentially mean some sort of an 
(in a way interspersed but still) sequential rather than truly parallel/concurrent combination of tasks,  where no steps in one component would be allowed to be made until receiving a response in the other component, contrary to the very spirit of 
the idea of parallel/distributed computation. 

It is no accident that we use classical symbols for the above operations. As this is easy to see, the meanings of these operations, as well as the meanings of the so called  
blind quantifiers $\cla,\cle$ that will be defined shortly, are exactly classical when their scope is restricted to elementary games (in which case the compound games generated by these operations also remain elementary). This is 
what makes classical logic just a special fragment of the more general and expressive computability logic. Once the scope of the ``classical" propositional connectives is extended beyond elementary games, however, their behavior starts resembling that of the multiplicative operators of linear logic. E.g. the principle $A\mli A\mlc A$ generally fails for nonelementary games. Yet, this resemblance is rather shallow, and typically disappears as soon as we start considering longer and 
``deeper" formulas. See \cite{Jap03} for more about how computability logic relates to linear logic. We do not want to go into details of this discussion here because, as pointed out in Section 1, this work is everything but an attempt to find a justification for linear logic --- the popular logic  that is syntactically so appealing yet lacks a convincing  semantics. 

The next group of operations:  $\adc$, $\add$, $\ada$ and $\ade$ that we call {\em choice} operations, on the other hand, bear resemblance with the additive operators 
of linear logic. Based on their semantics, in more traditional terms they can be characterized as {\em constructive} versions of conjunction, disjunction, universal quantifier and existential quantifier, respectively. 

$\ada xA(x)$ is the game where, in the initial position, only $\oo$ has legal moves. Such a move consists in 
a choice of one of the elements of the universe of discourse. After $\oo$ makes a move $c\in\{0,1,\ldots\}$, 
the game continues (and the winner is determined) according to the rules of $A(c)$. If no initial move is made, $\pp$ is considered the 
winner as there was no particular (sub)problem specified by $\oo$ that $\pp$ failed to solve.
$A\adc B$ is similar, 
only here the choice is just made between ``left" (``1") and ``right" (``2"). $\ade$ and $\add$ are 
symmetric to $\ada$ and $\adc$, with
the only difference that now it is $\pp$ rather than $\oo$ who makes an initial move/choice. Here is the formal definition:

\begin{definition}\label{ad}
Let $A(x)$, $A_1,\ldots,A_n$ ($n\geq 2$) be any games.\vspace{5pt} 

The \gj{choice} \gj{conjunction} $A_1\adc\ldots\adc A_n$\label{x81} of $A_1,\ldots,A_n$ is defined by:
\begin{itemize}
\item $\legal{A_1\adc\ldots\adc A_n}{e}=\{\emptyrun\}\cup\{\seq{\oo i,\Delta}\ |\ i\in\{1,\ldots,n\},\ \Delta\in\legal{A_i}{e}\}$.
\item $\win{A_1\adc\ldots\adc A_n}{e}\seq{\Gamma}=\oo$ iff $\Gamma=\seq{\oo i,\Delta}$, where $i\in\{1,\ldots,n\}$ and $\win{A_i}{e}\seq{\Delta}=\oo$.
\end{itemize}

The \gj{choice} \gj{disjunction} $A_1\add\ldots\add A_n$\label{x82} of $A_1,\ldots,A_n$ is defined
in exactly the same way, only with ``$\pp$" instead of ``$\oo$". Equivalently, it can be defined by
\(A_1\add\ldots\add A_n\ =_{def}\ \gneg(\gneg A_1\adc\ldots\adc \gneg A_n).\)

The \gj{choice} \gj{universal quantification} $\ada xA(x)$\label{x83} of $A(x)$ is defined by:
\begin{itemize}
\item $\legal{\adai xA(x)}{e}=\{\emptyrun\}\cup\{\seq{\oo c,\Delta}\ |\ c\mbox{ is a constant, } \Delta\in\legal{A(c)}{e}\}$.
\item 
$\win{\adai xA(x)}{e}\seq{\Gamma}=\oo$ iff $\Gamma=\seq{\oo c,\Delta}$, where $c$ is a constant and $\win{A(c)}{e}\seq{\Delta}=\oo$.
\end{itemize}

The \gj{choice} \gj{existential quantification} $\ade xA(x)$\label{x84} of $A(x)$ is defined in 
 exactly the same way, only with ``$\pp$" instead of ``$\oo$". Equivalently, it can be defined by
\(\ade xA(x)\ =_{def}\ \gneg\ada x\gneg A(x).\)
\end{definition}

A few examples would help. The problem of computing a function $f$ can be specified as
\(\ada x\ade y\bigl(f(x)=y\bigr).\) 
 This is a game of depth 2, where the first legal move --- selecting a particular 
value $k$ for $x$ --- is by $\oo$. Making such a move brings the game down to $\ade y(f(k)=y)$. The second move --- selecting a value $n$ for $y$ --- is by $\pp$, after which the game continues (or rather stops) as $f(k)=n$. The latter is an elementary game won by $\pp$ iff $f(k)$ really equals $n$. Obviously $f$ is computable in the standard sense iff $\ada x\ade y\bigl( f(x)=y\bigr)$ is winnable, i.e. computable in our sense. 

Next, the problem of deciding a predicate $p(x)$ would be specified as
 \(\ada x\bigl(p(x)\add \gneg p(x)\bigr).\) 
This is the game where, after $\oo$ selects a value $k$ for $x$, the machine should reply by one of the moves 
1 or 2; the game will be considered won by the machine if $p(k)$ is true and the move 1 was made, or $p(k)$ is false and the choice was 2, so that decidability of $p(x)$ means nothing but existence of a machine that wins the game $\ada x\bigl(p(x)\add\gneg p(x)\bigr)$. 

To get a feel of $\mli$ as a problem reduction operation, let us consider reduction of the acceptance problem to the halting problem (the example borrowed from \cite{Jap03}). The halting problem can be expressed by 
\( \ada x\ada y \bigl(\mbox{\em Halts}(x,y) \add \gneg \mbox{\em Halts}(x,y)\bigr),\)
where $\mbox{\em Halts}(x,y)$ is the predicate 
``Turing machine  $x$ halts on input $y$". Similarly, the acceptance problem can be expressed by the formula
\(\ada x\ada y \bigl(\mbox{\em Accepts}(x,y) \add \gneg \mbox{\em Accepts}(x,y)\bigr),\)
 where $\mbox{\em Accepts}(x,y)$ is the predicate ``Turing machine $x$ accepts input $y$". While the acceptance problem is not decidable, it is algorithmically reducible to the halting problem. In particular, there is an HPM  that wins 

\(\ada x\ada y \bigl(\mbox{\em Halts}(x,y) \add \gneg \mbox{\em Halts}(x,y)\bigr)
\mli
\ada x\ada y \bigl(\mbox{\em Accepts}(x,y) \add \gneg \mbox{\em Accepts}(x,y)\bigr).\)

Here is a  strategy for solving this problem: Wait till the environment selects values $k$ and $n$ for $x$ and $y$ in the consequent (if such a selection is never made, the machine wins). Then specify the same values $k$ and $n$ for  $x$ and $y$ in
the antecedent (where the roles of the machine and the environment are switched), and see whether $\oo$ responds by 
1 or 2 there. If the response is 
1, simulate machine $k$ on input $n$ until it halts, and select, in the consequent, 1 or 2 depending on whether the simulation accepted or rejected. And if $\oo$'s response in the antecedent was 2, then select 2 in the consequent. 

We can see that what the machine did in the above strategy indeed was a reduction: it used an (external) solution 
to the halting problem to solve the acceptance problem. There are various natural concepts of reduction, and 
the sort of reduction captured by $\mli$, that we call {\bf linear reduction},\label{linred} is most basic among them.

One of the other, well-established, concepts of reduction is {\bf mapping reduction}:\label{june16} A predicate $p(x)$ is said to be mapping reducible to a predicate $q(x)$ iff there is an effective function $f$ such that, for any constant $c$, $p(c)$ is true iff
$q(f(c))$ is true. Using $A\mleq B$ as an abbreviation for $(A\mli B)\mlc (B\mli A)$, it is not hard to see that 
mapping reducibility of $p(x)$ to $q(x)$ means nothing but winnability of the game
\(\ada x\ade y\bigl(p(x)\mleq q(y)\bigr).\)
 
Notice that while standard approaches only allow us to talk about (a whatever sort of) reducibility as a {\em relation} between problems, in our approach reduction becomes an {\em operation} on problems, with reducibility as a relation simply meaning computability 
of the corresponding combination (such as \(\ada x\ade y\bigl(p(x)\mleq q(y)\bigr)\) or $A\mli B$) of problems. Similarly for other relations or properties such as the property of {\em decidability}. The latter becomes the operation of {\em deciding} if we define the problem of deciding 
a predicate (or any computational problem) $p(x)$ as the game $\ada x\bigl(p(x)\add\gneg p(x)\bigr)$. So, now we can meaningfully ask questions such as
``Is the linear reduction of the problem of deciding $p(x)$ to the problem of deciding $q(x)$ linearly reducible to the mapping reduction of $p(x)$ to $q(x)$?". This question would be equivalent to whether the following problem is (always) computable:
\begin{equation}\label{e1}  
\ada x\ade y\bigl(p(x)\mleq q(y)\bigr)\mli
\Bigl(\ada x\bigr(q(x)\add\gneg q(x)\bigr)\mli
\ada x\bigr(p(x)\add\gneg p(x)\bigr)\Bigr).
\end{equation}
This problem is indeed computable no matter what particular predicates $p(x)$ and $q(x)$ are, which means that mapping reduction is at least as strong as linear reduction.  Here is a strategy:
Wait till $\oo$ selects a value $k$ for $x$ in the consequent of the consequent of (\ref{e1}). Then specify the same value 
$k$ for $x$ in the antecedent of (\ref{e1}), and wait till $\oo$ replies there by selecting a value $n$ for $y$. Then select 
the same value $n$ for $x$ in the antecedent of the consequent of (\ref{e1}). $\oo$ will have to respond by 1 or 
2 in that component of the game. Repeat that very response in the consequent of the consequent of (\ref{e1}), and celebrate victory.

Expression (\ref{e1}) is a legal formula of the language of $\predel$ which, according to our main Theorem \ref{main}, is sound and complete with respect to computability semantics. So, had our ad hoc methods failed to find an answer (and this would certainly be the case if we dealt with a more complex problem), the existence of a successful algorithmic strategy could have been established by showing that (\ref{e1}) is provable in $\predel$. 
 Moreover, by clause (a) of Theorem \ref{main}, after finding an $\predel$-proof of (\ref{e1}), we would not only know that an algorithmic solution to (\ref{e1}) exists, but we would also be able to constructively obtain such a solution from the proof. On the other hand, the fact that linear reduction is not as strong as mapping reduction could be established by showing that $\predel$ does not prove 
\begin{equation}\label{e2}
\Bigl(\ada x\bigr(q(x)\add\gneg q(x)\bigr)\mli
\ada x\bigr(p(x)\add\gneg p(x)\bigr)\Bigr) \mli \ada x\ade y\bigl(p(x)\mleq q(y)\bigr).
\end{equation}
This negative fact, too, can be established effectively as, according to Theorem \ref{decid}, the relevant fragment of $\predel$  is decidable.
Our proof of the completeness part of Theorem \ref{main} would then offer a way how to construct particular predicates $p(x)$ and $q(x)$ for which 
(\ref{e2}) is not computable. 

These few examples must be sufficient to provide insights into the utility of computability logic and 
$\predel$ in particular for theory of computing: our logic offers a convenient tool for asking and answering questions in the above style (and beyond) in a systematic way, something that so far has been mostly done in an ad hoc manner or has been simply impossible to do. By iterating available operators, we can express and explore an infinite variety computational problems 
and relations between them,  only few of which may have special names established in the literature. 

The next, already mentioned group of operations is what we call ``blind quantifiers": $\cla$ and $\cle$,
with hardly any reasonably close counterparts in linear logic. For certain reasons, the operations $\cla x$ and $\cle x$ we only define for
games  called $x$-unistructural. A game $A$ is said to be \gj{$x$-unistructural}\label{x66} (or \gj{unistructural in $x$}) iff, for any two valuations $e_1$ and $e_2$ that agree on all variables except perhaps $x$, we have $\legal{A}{e_1}=\legal{A}{e_2}$. Intuitively, this is a game whose structure does not depend on $x$.  In fact whether we impose the $x$-unistructurality condition or not is 
irrelevant in our present case because this condition is automatically satisfied anyway: as shown in \cite{Jap03}, all games that can be expressed in the language of $\predel$ are unistructural, and obviously all unistructural games are also $x$-unistructural. 

\begin{definition}\label{bq}
Let $x$ be any variable and $A(x)$ any $x$-unistructural game.

The \gj{blind universal quantification $\cla x  A(x)$\label{x79} of $A(x)$} is defined by:

\begin{itemize}
\item $\legal{\clai xA(x)}{e}=\legal{A(x)}{e}$.
\item $\win{\clai xA(x)}{e}\seq{\Gamma}=\pp$ iff, for every constant $c$, $\win{A(c)}{e}\seq{\Gamma}=\pp$.
\end{itemize}

The \gj{blind existential quantification $\cle x  A(x)$\label{x80} of $A(x)$} is defined in exactly the same way, only with 
``$\oo$" instead of ``$\pp$". Equivalently, it can be defined by 
\(\cle xA(x)\ =_{def}\ \gneg\cla x\gneg A(x).\)
\end{definition} 

The meaning of $\cla xA(x)$ is similar to that of $\ada xA(x)$, with the difference that $\oo$ does not make a move specifying a particular value of $x$, so  that $\pp$ has to play blindly in a way that would be successful for any possible value of $x$. Alternatively, $\cla xA(x)$ can be thought of as the version of $\ada xA(x)$ where the particular value of $x$ selected by $\oo$ remains invisible to $\pp$.  This way, $\cla$ and $\cle$  produce games with {\em imperfect information}. 
Compare the problems
\(\ada x\bigl(\mbox{\em Even}(x)\add \mbox{\em Odd}(x)\bigr)\)
and
\(\cla x\bigl(\mbox{\em Even}(x)\add \mbox{\em Odd}(x)\bigr).\)
The former is an easy-to-compute problem of depth 2, while the latter is an incomputable problem 
of depth 1 with only by the machine to make a move --- select the true  disjunct, which is hardly possible to do as the value of $x$ remains unspecified. Some problems that depend on $x$ can be however solved having only partial information on  
$x$. For example, in order to tell whether  $x$ is even or odd, we do not really need to read the whole  
(decimal representation of) $x$ --- it would be sufficient to look at its last digit, i.e. know the value of \mbox{$(x \hspace{2pt}\mbox{\bf Mod}\hspace{2pt}10)$.}
Thus, the problem
\(\cla x\Bigl(\ade y\bigl(y=(x\hspace{2pt}\mbox{\bf Mod}\hspace{2pt}10)\bigr)\mli \bigl(\mbox{\em Even}(x)\add \mbox{\em Odd}(x)\bigr)\Bigr)\)
is computable, which is a more informative statement than if we had stated computability of the same problem 
with $\ada$ instead of $\cla$. 
As noted a while ago, the meanings of $\cla$ and $\cle$ are exactly classical  when applied to elementary games, which explains why we use the classical notation for them.

Another important group of operations comprises {\bf recurrence operations}. They come in different flavors
(see \cite{Japint}), perhaps the most interesting of which is what is called {\bf branching recurrence}\footnote{In \cite{Jap03} this operation is called {\em branching conjunction} and is denoted by $!$.} $\st$.\label{x19}  Intuitively $\st A$, as a resource, is $A$ that can be reused an arbitrary number of times. The same is true for the other sorts 
of recurrences, but there are several natural understandings of reusage, with $\st$ corresponding to the strongest form of it and the other recurrence operations corresponding to weaker concepts of reusage/recycling (and it is not clear which of the recurrence operations  
best corresponds to what the exponential operator $!$ of linear logic was meant to capture).  The operation $\intimpl$, called {\em weak reduction},\footnote{\cite{Jap03} uses the symbol $\Rightarrow$ for this operation.} is defined by $A\intimpl B = \st A\mli B$. This operation formalizes our weakest possible intuitive concept of reduction. The difference between $\mli$ and $\intimpl$ as reduction operations is that while in the former
every act of resource (antecedent) consumption is strictly accounted for, the latter allows uncontrolled usage of resources.
One of the conjectures stated in \cite{Jap03} is that we get exactly intuitionistic calculus when the intuitionistic 
implication is understood as $\intimpl$ and the other intuitionistic operators understood as the corresponding choice operations. Recurrence operators and weak reduction are not in the logical vocabulary of $\predel$, and hence we do not give formal definitions for them. Such definitions can be found in \cite{Jap03} and \cite{Japint}.
    
According to Theorem 14.1 of \cite{Jap03}, all of the operations that we discussed in this section preserve the static and unistructural properties of games. Taking into account that predicates as elementary games are static (Proposition \ref{predstat}) and obviously unistructural, their closure under those operations forms a natural class 
of unistructural computational problems. All of those operations except $\st$ and $\intimpl$ also preserve the finite-depth property. Hence the closure of the set of all elementary problems under substitution of variables, 
$\gneg$, $\mlc$, $\mld$, $\mli$, $\cla$, $\cle$, $\adc$, $\add$, $\ada$ and $\ade$ forms a natural 
class of finite-depth, unistructural computational problems. As we are going to see, this is exactly the class of problems expressible in the language of $\predel$.
Finally, as already noted more than once, the operations $\gneg$, $\mlc$, $\mld$, $\mli$, $\cla$, $\cle$ 
preserve the elementary property of games: they send predicates to predicates; and, when restricted to predicates, they coincide with the same-name classical operations. Of course, the same applies to the operation of substitution of variables, as well as the trivial games $\tlg$ and $\twg$ that can be understood as $0$-ary operations on games.    

One more game operation that we are going to look at is that of prefixation, which is somewhat reminiscent of 
the modal operator(s) of dynamic logic. This operation takes two arguments: a game $A$ and a position $\Phi$ that must 
be what we call a unilegal position of $A$ (otherwise the operation is undefined).  $\Gamma$ is said to be a \gj{unilegal run}\label{x64} (position if finite) of a game $A$ iff, for every valuation $e$, $\Gamma\in\legal{A}{e}$. 
As noted above, all games that we deal with in this paper are unistructural, and for such games obviously there is no difference between ``unilegal" and ``legal". 

\begin{definition}\label{prfx}
Assume $\Phi$ is a unilegal position of a game $A$. The \gj{$\Phi$-prefixation} of $A$, denoted $\seq{\Phi}A$, is defined as follows:\vspace{-5pt}
\begin{itemize}
\item $\legal{\seq{\Phi}A}{e}=\{\Gamma\ |\ \seq{\Phi,\Gamma}\in\legal{A}{e}\}$.
\item $\win{\seq{\Phi}A}{e}\seq{\Gamma}=\win{A}{e}\seq{\Phi,\Gamma}$.
\end{itemize} 
\end{definition}
  
Intuitively, $\seq{\Phi}A$ is the game playing which means playing $A$ starting (continuing) from position $\Phi$. 
That is, $\seq{\Phi}A$ is the game to which $A$ evolves (is ``brought down") after the (lab)moves of $\Phi$ have been made. We have already used this intuition when explaining the meaning of the choice operations. E.g., we said that after $\oo$ makes an initial move $c$, the game 
$\ada xA(x)$ continues as $A(c)$. What this meant was nothing but that $\seq{\oo c}\bigl(\ada xA(x)\bigr)=A(c)$. 
The following proposition summarizes this sort of a characterization of the choice operations, and extends it to the other operations, too. 
It tells us what the legal initial moves for a given game are, and to what game that game evolves after such a (uni)legal move is made. 

\begin{proposition}\label{ic}
In each of the following clauses, $e$ is any valuation, $\xx$ either player, and $\alpha,\beta$ any moves; 
in each subclause (b), the game on the left of the equation is assumed to be defined, $i\in\{1,\ldots,n\}$ and $c\in\{0,1,2,\ldots\}$.  
\begin{enumerate}
\item 
\begin{enumerate}
\item $\seq{\xx\alpha}\in\legal{\gneg A}{e}$ iff $\seq{\pneg\xx\alpha}\in\legal{A}{e}$;
\item $\seq{\xx\alpha}\gneg A=\gneg(\seq{\pneg\xx\alpha}A)$.
\end{enumerate}
\item 
\begin{enumerate}
\item $\seq{\xx\alpha}\in\legal{A_1\mlc\ldots\mlc A_n}{e}$ iff $\alpha=i.\beta$, where $i\in\{1,\ldots,n\}$ and $\seq{\xx\beta}\in\legal{A_i}{e}$;
\item $\seq{\xx i.\beta}(A_1\mlc\ldots\mlc A_n)=A_1\mlc\ldots\mlc A_{i-1}\mlc\seq{\xx\beta}A_i\mlc A_{i+1}\mlc\ldots\mlc A_n$.
\end{enumerate}
\item 
\begin{enumerate}
\item $\seq{\xx\alpha}\in\legal{A_1\mld\ldots\mld A_n}{e}$ iff $\alpha=i.\beta$, where $i\in\{1,\ldots,n\}$ and $\seq{\xx\beta}\in\legal{A_i}{e}$;
\item $\seq{\xx i.\beta}(A_1\mld\ldots\mld A_n)=A_1\mld\ldots\mld A_{i-1}\mld\seq{\xx\beta}A_i\mld A_{i+1}\mld\ldots\mld A_n$.
\end{enumerate}
\item 
\begin{enumerate}
\item $\seq{\xx\alpha}\in\legal{A\mli B}{e}$ iff $\alpha=i.\beta$, where 
$\left\{ \begin{array}{l} \mbox{$i=1$ and $\seq{\pneg\xx\beta}\in\legal{A}{e}$, 
or}\\
\mbox{$i=2$ and $\seq{\xx\beta}\in\legal{B}{e}$;}\end{array} \right.$
\item $\left\{ \begin{array}{l}
\seq{\xx 1.\beta}(A\mli B)=\seq{\pneg\xx\beta}A\mli B;\\
\seq{\xx 2.\beta}(A\mli B)=A\mli\seq{\xx\beta}B.
\end{array} \right.$
\end{enumerate}
\item
\begin{enumerate}
\item $\seq{\xx\alpha}\in\legal{A_1\adc\ldots\adc A_n}{e}$ iff $\xx=\oo$ and $\alpha=i\in\{1,\ldots,n\}$;
\item $\seq{\oo i}(A_1\adc\ldots\adc A_n)=A_i$.
\end{enumerate}
\item
\begin{enumerate}
\item $\seq{\xx\alpha}\in\legal{A_1\add\ldots\add A_n}{e}$ iff $\xx=\pp$ and $\alpha=i\in\{1,\ldots,n\}$;
\item $\seq{\pp i}(A_1\add\ldots\add A_n)=A_i$.
\end{enumerate}
\item
\begin{enumerate}
\item $\seq{\xx\alpha}\in\legal{\adai xA(x)}{e}$ iff $\xx=\oo$ and $\alpha=c\in\{0,1,2,\ldots\}$;
\item $\seq{\oo c}\ada xA(x)=A(c)$.
\end{enumerate}
\item
\begin{enumerate}
\item $\seq{\xx\alpha}\in\legal{\adei xA(x)}{e}$ iff $\xx=\pp$ and $\alpha=c\in\{0,1,2,\ldots\}$;
\item $\seq{\pp c}\ade xA(x)=A(c)$.
\end{enumerate}
\item
\begin{enumerate}
\item $\seq{\xx\alpha}\in\legal{\clai xA(x)}{e}$ iff $\seq{\xx\alpha}\in\legal{A(x)}{e}$;
\item $\seq{\xx\alpha}\cla xA(x)=\cla x\seq{\xx\alpha}A(x)$.
\end{enumerate}
\item 
\begin{enumerate}
\item $\seq{\xx\alpha}\in\legal{\clei xA(x)}{e}$ iff $\seq{\xx\alpha}\in\legal{A(x)}{e}$;
\item $\seq{\xx\alpha}\cle xA(x)=\cle x\seq{\xx\alpha}A(x)$.
\end{enumerate}

\end{enumerate}
\end{proposition}

The above fact is known from \cite{Jap03}. Its proof consists in just a routine analysis of the relevant definitions.

Just like this is the case with recurrence operations, the language of $\predel$ does not have any constructs corresponding to prefixation. However, this operation will be heavily exploited in our soundness and completeness proof for $\predel$ in Part 2. Generally, prefixation is very handy in visualizing a (unilegal) run of a given game $A$. In particular, every (sub)position $\Phi$ of such a run  can be represented by, or thought of as, the game $\seq{\Phi}A$. 

Here is an example. Remember game (\ref{e1}). Based on Proposition \ref{ic}, the run 
\(\seq{\oo 2.2.7,\pp 1.7,\oo 1.9,\pp 2.1.9,\oo 2.1.1,\pp 2.2.1}\)
is a (uni)legal run of that game, and to it 
corresponds the following sequence of games:\vspace{-5pt}
\begin{description}
\item[(i)] $\Bigl(\ada x\ade y\bigl(p(x)\mleq q(y)\bigr)\Bigr)\mli
\Bigl(\ada x\bigr(q(x)\add\gneg q(x)\bigr)\mli
\ada x\bigr(p(x)\add\gneg p(x)\bigr)\Bigr)$,

i.e. $\emptyrun$(\ref{e1});

\item[(ii)] $\Bigl(\ada x\ade y\bigl(p(x)\mleq q(y)\bigr)\Bigr)\mli
\Bigl(\ada x\bigr(q(x)\add\gneg q(x)\bigr)\mli
\bigr(P(7)\add\gneg p(7)\bigr)\Bigr)$,

i.e. $\seq{\oo 2.2.7}$(i), \ i.e. $\seq{\oo 2.2.7}$(\ref{e1});

\item[(iii)] $\Bigl(\ade y\bigl(p(7)\mleq q(y)\bigr)\Bigr)\mli
\Bigl(\ada x\bigr(q(x)\add\gneg q(x)\bigr)\mli
\bigr(p(7)\add\gneg p(7)\bigr)\Bigr)$,

i.e. $\seq{\pp 1.7}$(ii), \ i.e. $\seq{\oo 2.2.7,\pp 1.7}$(\ref{e1});

\item[(iv)] $\Bigl(p(7)\mleq q(9)\Bigr)\mli
\Bigl(\ada x\bigr(q(x)\add\gneg q(x)\bigr)\mli
\bigr(p(7)\add\gneg p(7)\bigr)\Bigr)$,

i.e. $\seq{\oo 1.9}$(iii), \ i.e. $\seq{\oo 2.2.7,\pp 1.7, \oo 1.9}$(\ref{e1});

\item[(v)] $\Bigl(p(7)\mleq q(9)\Bigr)\mli
\Bigl(\bigr(q(9)\add\gneg q(9)\bigr)\mli
\bigr(p(7)\add\gneg p(7)\bigr)\Bigr)$,

i.e. $\seq{\pp 2.1.9}$(iv), \ i.e. $\seq{\oo 2.2.7,\pp 1.7, \oo 1.9, \pp 2.1.9}$(\ref{e1});

\item[(vi)] $\Bigl(p(7)\mleq q(9)\Bigr)\mli
\Bigl(q(9)\mli
\bigr(p(7)\add\gneg p(7)\bigr)\Bigr)$,

i.e. $\seq{\oo 2.1.1}$(v), \ i.e. $\seq{\oo 2.2.7,\pp 1.7, \oo 1.9, \pp 2.1.9, \oo 2.1.1}$(\ref{e1});

\item[(vii)] $\Bigl(p(7)\mleq q(9)\Bigr)\mli
\Bigl(q(9)\mli p(7)\Bigr)$,

i.e. $\seq{\pp 2.2.1}$(vi), \ i.e. $\seq{\oo 2.2.7,\pp 1.7, \oo 1.9, \pp 2.1.9, \oo 2.1.1, \pp 2.2.1}$(\ref{e1}).\vspace{-5pt}
\end{description}

Player $\pp$ is the winner because the run hits a true elementary game. In this run $\pp$ has followed the winning strategy that we described for (\ref{e1}) earlier.

We finish this section by reproducing a fact proven in \cite{Jap03} (Proposition 21.3), according to which modus ponens preserves computability, and does so in a constructive sense:

\begin{proposition}\label{mp}
For any computational problems $A$ and $B$, if $\models A$ and \mbox{$\models A\mli B$,} then $\models B$. 
Moreover, there is an effective procedure that converts any two HPMs ${{\mathcal H}}_1$ and ${{\mathcal H}}_2$ into an HPM ${{\mathcal H}}_3$ such that, for any computational problems $A$, $B$ and any valuation $e$, whenever ${{\mathcal H}}_1\models_e A$ and ${{\mathcal H}}_2\models_e A\mli B$, we have ${{\mathcal H}}_3\models_e B$.
\end{proposition}

A similar closure property was proven in Section 21 of \cite{Jap03} with respect to the rules $A\mapsto \ada xA$ and $A\mapsto\st A$, with \({\mathcal P}\mapsto C\label{mpst}\) here and later meaning ``from premise(s) ${\mathcal P}$ conclude $C$".

\section{Logic $\predel$}\label{prdl}
By the {\em classical language} we mean the language of pure classical first-order logic with individual constants but without equality and functional symbols. We assume that the set of \gj{terms}\label{jun24} --- i.e. \gj{variables}\label{x69} and \gj{constants}\label{jun244} --- of this language is the same as the one we fixed in Section \ref{cpr}.  As always, each \gj{predicate letter}\label{x50} comes with a fixed \gj{arity}.\label{x2} An ($n$-ary\label{x203} non-logical) \gj{atom}\label{x3}\label{x202} is the expression 
$p(t_1,\ldots,t_n)$, where $p$ is an $n$-ary predicate letter and the $t_i$ are terms.

The language of $\predel$ extends the classical language by adding the operators $\adc,\add,\ada,\ade$ to its vocabulary. The names that we use for the logical operators of the language are the same as for the (same-symbolic-name) game operations defined in the previous section.  
Throughout the rest of this paper by a {\bf formula} we mean  a formula of this language. The definition is standard --- the set of formulas is the smallest set of expressions such that:
\begin{itemize}
\item Non-logical atoms and the {\bf logical atoms}\label{x201} $\twg$ and $\tlg$ are formulas;
\item If $F_1,\ldots,F_n$ ($n\geq 2$) are formulas, then so are $\gneg F_1$, $F_1\mlc\ldots\mlc F_n$, $F_1\mld\ldots\mld F_n$,
$F_1\mli F_2$, $F_1\adc\ldots\adc F_n$, $F_1\add\ldots\add F_n$;
\item If $F$ is a formula and $x$ is a variable, then $\cla xF$, $\cle xF$, $\ada xF$, 
$\ade x F$ are formulas.
\end{itemize}
 
The definitions of what a \gj{free} or \gj{bound} occurrence of a variable means are also standard, keeping in mind 
that now a variable can be bound by any of the four \gj{quantifiers} $\cla,\cle,\ada,\ade$. Every occurrence of a constant also counts as {\bf free}.
 By a \gj{free variable} of a formula $F$
we mean a variable that has free occurrences in $F$. Similarly, the \gj{free terms}\label{jun2444} of $F$ are its free variables plus the constants occurring in $F$. As known, {\bf classical validity} of a formula 
of the classical language that contains constants means the same as validity of the same formula with its constants understood as free variables. So, for a reader more used to the version of classical logic where variables are the only terms, it is perfectly safe to think of constants as free variables.  

In the previous section, substitution of variables was defined as an operation on games. Here we define
a ``similar" operation on formulas called \gj{substitution of terms}\label{x61}.
Suppose $F$ is a formula, $t_1,\ldots,t_n$ are pairwise distinct terms, and $t'_1,\ldots,t'_n$ are any terms. Then
\(F[t_1/t'_1,\ldots,t_n/t'_n]\)
stands for the result of simultaneously substituting in $F$ all free occurrences of $t_1,\ldots,t_n$ by $t'_1,\ldots,t'_n$, respectively. 

In concordance with a similar notational practice established in Section \ref{op} for games, sometimes we represent a formula $F$ as $F(t_1,\ldots,t_n)$\label{x91} where the $t_i$ are pairwise distinct terms. In the context defined by such a representation, $F(t'_1,\ldots,t'_n)$ will mean the same as $F[t_1/t'_1,\ldots,t_n/t'_n]$. Our disambiguating convention is that 
the context is set by the expression that was used earlier. That is, when we first mention $F(t_1,\ldots,t_n)$ and only after that 
use the expression $F(t'_1,\ldots,t'_n)$, the latter should be understood as $F[t_1/t'_1,\ldots,t_n/t'_n]$ rather 
than the former understood as $F[t'_1/t_1,\ldots,t'_n/t_n]$. 
It should be noted that, when representing $F$ as $F(t_1,\ldots,t_n)$,
we do not necessarily mean that $t_1,\ldots,t_n$ are exactly the free terms of $F$. 
  
An \gj{interpretation}\label{x36} is a function $^*$ that sends each $n$-ary predicate letter $p$ to an elementary game $p^*=A(x_1,\ldots,x_n)$ with an attached $n$-tuple of (pairwise distinct) variables. This assignment extends to formulas by commuting with all operations. That is:
Where $p$ and $A(x_1,\ldots,x_n)$ are as above and $t_1,\ldots,t_n$ are any terms, $\bigl(p(t_1,\ldots,t_n)\bigr)^*=A(t_1,\ldots,t_n)$; $\tlg^*=\tlg$; \ $(\gneg F)^*=\gneg(F^*)$; \ $(F_1\adc\ldots\adc F_k)^*=F_{1}^{*}\adc\ldots\adc F_{k}^{*}$; \ $(\cla xF)^*=\cla x(F^*)$; \ etc.
For a predicate letter $p$, we will say ``\hspace{2pt}\gj{$^*$ interprets $p$ as $A$}" to mean that $p^*=A$. Similarly, for a formula $F$, if $F^*=A$, we say that 
\gj{$^*$ interprets $F$ as $A$}. 

For a formula $F$, an interpretation $^*$ is said to be \gj{$F$-admissible}\label{x37} iff, for any $n$-ary 
predicate letter $p$, the game $A(x_1,\ldots,x_n)$ assigned to $p$ by $^*$ does not depend on any variables that are not among $x_1,\ldots,x_n$ but occur in $F$. We need this condition to avoid possible collisions of variables.

\begin{definition}\label{val}
A formula $F$ is said to be \gj{valid} iff $\models F^*$ for every $F$-admissible interpretation $^*$.
\end{definition}

To axiomatize the set of valid formulas, we need some technical preliminaries. Understanding $F\mli G$ as an abbreviation for $\gneg F\mld G$, a \gj{positive}\label{x43} (resp. \gj{negative}) \gj{occurrence} of a subformula is
one that is in the scope of an even (resp. odd) number of occurrences of $\gneg$.
A \gj{surface occurrence}\label{x44} of a subformula is an occurrence that is 
not in the scope of any choice operators. A formula not containing choice operators --- i.e. a formula of the classical language --- is said to be \gj{elementary}.\label{x16} The 
\gj{elementarization}\label{x17} of a formula $F$ is the result of replacing
in $F$ all surface occurrences of subformulas of the form  $G_1\add\ldots\add G_n$ or $\ade xG$ by $\tlg$ and all surface occurrences of subformulas  
of the form  $G_1\adc\ldots\adc G_n$ or $\ada xG$ by $\twg$. A formula is said to be \gj{stable}\label{x57} iff its elementarization is classically valid. Otherwise it is \gj{instable}.\label{x32}

\begin{definition}\label{pred}
Logic $\predel$ is given by the following rules:  
\begin{description}
\item[\ \ A.\ ]  $\vec{H}\mapsto F$, where $F$ is stable and $\vec{H}$ is 
a set of formulas satisfying the following conditions:
\begin{description}
\item[(i)]  Whenever $F$ has a positive (resp. negative)  surface occurrence of a subformula $G_1\adc\ldots\adc G_n$ (resp. $G_1\add \ldots\add G_n$), for each 
$i\in\{1,\ldots,n\}$, $\vec{H}$ contains the result of replacing that occurrence in $F$ by $G_i$;
\item[(ii)] Whenever $F$ has a positive (resp. negative) surface occurrence of a subformula $\ada x G(x)$ (resp. $\ade xG(x)$), $\vec{H}$ contains the result of replacing that occurrence in $F$ by $G(y)$ for some variable $y$ not occurring in $F$.
\end{description}
\item[\ \ B1.]  $F'\mapsto F$, where $F'$ is the result of replacing in $F$ a negative (resp. positive) surface occurrence of a subformula $G_1\adc\ldots\adc G_n$ (resp. $G_1\add\ldots\add G_n$) by $G_i$ for some $i\in\{1,\ldots, n\}$.
\item[\ \ B2.]  $F'\mapsto F$, where $F'$ is the result of replacing in $F$ a negative (resp. positive) surface occurrence of a subformula $\ada xG(x)$ (resp. $\ade xG(x)$) by $G(t)$ for some term $t$ such that (if $t$ is a variable) 
neither the above occurrence of 
$\ada xG(x)$ (resp. $\ade xG(x)$) in $F$ nor any of the free occurrences of $x$ in $G$ are in the scope of $\cla t$,
$\cle t$, $\ada t$ or $\ade t$. 
\end{description}
\end{definition}

Axioms are not explicitly stated, but note that the set $\vec{H}$ of premises of Rule {\bf A} can be empty, in which case the conclusion $F$ of that rule acts as an axiom. Even though this may not be immediately obvious, $\predel$ essentially {\em is} a (refined sort of) Gentzen-style system. Consider, for example, Rule {\bf B1}. It is very similar to the 
additive-disjunction-introduction rule of linear logic. The only difference is that while linear logic requires that 
$G_i$ be a $\mld$- (multiplicative) disjunct of the premise, $\predel$ allows it to be any positive surface occurrence. This is what 
the calculus of structures \cite{gug} calls {\em deep inference}, as opposed to the {\em shallow inference} of linear logic. 
Natural semantics appear to naturally call for this sort of inference. Yet the traditional Gentzen-style axiomatizations   
for classical logic do not use it. To the question ``why only shallow inference?" classical logic has a simple answer:
``because it is sufficient" (for G\"{o}del's completeness). Linear logic, however, may not have a very good answer to this or similar questions. 

In the following examples and exercise, $p$ and $q$ are unary predicate letters.

\begin{example}\label{ex1}
The following is a $\predel$-proof of $\ada x\ade y\bigl(p(x)\mld\gneg p(y)\bigr)$:\vspace{3pt}\\
1. $p(z)\mld\gneg p(z)$ \ (from $\{\}$ by Rule {\bf A});\\
2. $\ade y\bigl(p(z)\mld\gneg p(y)\bigr)$ \ (from $1$ by Rule {\bf B2});\\
3. $\ada x\ade y\bigl(p(x)\mld\gneg p(y)\bigr)$ \ (from $\{2\}$ by Rule {\bf A}). 
\end{example}

  \begin{example}\label{ex2}
While  $\cle y\cla x\bigl(p(x)\mld\gneg p(y)\bigr)$ is a classically valid elementary formula and hence derivable in 
$\predel$ by Rule {\bf A} from the empty set of premises, $\predel$ does not prove its ``constructive version"
$\ade y\ada x\bigl(p(x)\mld\gneg p(y)\bigr)$. Indeed, the latter is instable, so it could only be derived by Rule {\bf B2} from the premise $\ada x\bigl(p(x)\mld\gneg p(t)\bigr)$ for some term $t$ different from $x$. Rules {\bf B1} or {\bf B2} are not applicable to $\ada x\bigl(p(x)\mld\gneg p(t)\bigr)$, so this formula 
could only be derived by Rule {\bf A}. Then its (single) premise should be $p(z)\mld\gneg p(t)$ for some variable $z$ different from $t$. But   $p(z)\mld\gneg p(t)$ is now an instable formula not containing any choice operators, so it cannot be derived by any of the rules of $\predel$.
\end{example}

\begin{exercise}\label{ex3}
With $\mbox{\em Logic}\vdash F$\label{vdsh} (resp. $\mbox{\em Logic}\not\vdash F$) here and later meaning ``$F$ is provable (resp. not provable) in {\em Logic}", verify that:\vspace{3pt}\\
1. $\predel\vdash \cla x\ p(x)\ \mli\ \ada x\ p(x)$;\\
2. $\predel\not\vdash \ada x\ p(x)\ \mli\ \cla x\ p(x)$;\\
3. $\predel$ proves formula (\ref{e1}) from Section \ref{op};\\
4. $\predel$ does not prove formula (\ref{e2}) from Section \ref{op}.
\end{exercise}

From the definition of $\predel$ it is clear that if $F$ is an elementary formula, then the only way to prove $F$ in $\predel$ is to derive it by Rule {\bf A} from the empty set of premises. In particular, this rule will be applicable when $F$ is stable, which for an elementary $F$ means nothing but that $F$ is classically valid. And vice versa: every classically valid formula is an elementary formula derivable in $\predel$ by Rule {\bf A} from the empty set of premises. Thus we have:

\begin{proposition}\label{clas}
The $\adc,\add,\ada,\ade$-free fragment of $\predel$ is exactly classical logic.
\end{proposition}

This is what we should have expected for, as noted in Section \ref{op}, 
when restricted to elementary problems --- and elementary formulas are exactly the ones that represent such problems --- 
the meanings of all non-choice operators of $\predel$ are exactly classical.

Another natural fragment of $\predel$ is the one obtained by forbidding the blind operators in its language. This is still a first-order logic as it contains the constructive quantifiers $\ada$ and $\ade$. So, the following theorem, that will be proven later in Section \ref{s6}, may come as a pleasant surprise:
 
\begin{theorem}\label{decid}
The $\cla,\cle$-free fragment of $\predel$ is decidable.
\end{theorem}

Of course $\predel$ in its full language cannot be decidable as it contains classical logic. However, 
taking into account that classical validity and hence stability of a formula is recursively enumerable, the following 
fact can be immediately seen from the way $\predel$ is defined:

\begin{proposition}\label{re}
$\predel$ is recursively enumerable.
\end{proposition} 

Here comes our main theorem, according to which $\predel$ precisely describes the set of all valid principles of computability.
This theorem is just a combination of Propositions \ref{sound} and \ref{unic} proven in Part 2.

\begin{theorem}\label{main}
$\predel\vdash F$ iff $F$ is valid $($any formula $F$ $)$.
Furthermore:

a) There is an effective procedure that takes an $\predel$-proof of a formula $F$ and 
constructs an HPM that wins $F^*$ for every $F$-admissible interpretation $^*$.

b) If $\predel\not\vdash F$, then $F^*$ is not computable  for some $F$-admissible interpretation 
$^*$ that interprets atoms as finitary predicates of arithmetical complexity\footnote{See page \pageref{x555} for an explanation of what ``arithmetical complexity $\Delta_2$" means.} $\Delta_2$.
\end{theorem}

$\predel$ is a fragment of the logic {\bf FD}\label{x22} introduced in \cite{Jap03}. The language of {\bf FD} is more expressive\footnote{Ignoring the minor detail that constants were not allowed in {\bf FD}.} in that it has an additional sort of letters called {\em general letters}. Unlike our predicate  letters (called {\em elementary letters} in \cite{Jap03}) that can only be interpreted as elementary games, general letters can be interpreted as any computational problems. $\predel$ is obtained from {\bf FD} by mechanically deleting 
the last two Rules {\bf C} and {\bf D}. Those two rules introduce general letters that are alien to the language we now consider.  
Once a general letter is introduced, it never disappears in any later formulas of an {\bf FD}-proof. Based on this observation, a formula in our present sense is provable in {\bf FD} iff it is provable in $\predel$, so that {\bf FD} is a conservative extension of $\predel$. It was conjectured in \cite{Jap03} (Conjecture 25.4) that {\bf FD} is sound and complete with respect to computability semantics. Our Theorem \ref{main} signifies a successful verification of that conjecture restricted 
to the general-letter-free fragment of {\bf FD}. This fragment is called {\bf elementary-base} as it only has elementary 
letters, i.e. all atoms of it represent elementary problems. The fragment of computability logic that {\bf FD} is conjectured to axiomatize, in turn, is called {\bf finite-depth} as all of its logical operators represent game operations that preserve 
the finite-depth property of games. Hence the fragment of computability logic captured by $\predel$ was called in \cite{Jap03} the 
{\em finite-depth, elementary-base} fragment.  

The language of {\bf FD}, in turn, is just a fragment of the bigger language introduced in \cite{Jap03} for computability logic, called the {\em universal language}. 
The latter is the extension of the former by adding the operators $\st$, $\cost$ ($\cost=\gneg\st\gneg$) and $\intimpl$ to it. \cite {Japint} further augments the official language of computability logic with a few other natural operators. 
Along with the above-mentioned Conjecture 25.4 regarding the soundness and completeness of {\bf FD}, there were two other major conjectures stated in \cite{Jap03} regarding the universal language: Conjecture 24.4 and Conjecture 26.2. A positive verification
of those two conjectures restricted to the language of $\predel$ is also among the immediate consequences of our Theorem \ref{main}.

When restricted to the language of $\predel$, Conjecture 24.4 of \cite{Jap03} sounds as follows:

\begin{quote}{\em If a formula $F$ is not valid, then $F^*$ is not computable for some $F$-admissible interpretation $^*$ that
interprets every atom as a finitary predicate.}\footnote{The original formulation of Conjecture 24.4 imposes three more restrictions on the problems through which atoms are interpreted: those problems can be chosen to also be 
determined (see \cite{Jap03} for a definition), strict (in the sense that in every position at most one of the players has legal moves) and unistructural. These conditions can be omitted in our case as they are automatically satisfied for elementary games.}   
\end{quote}

The significance of this conjecture is related to the fact that showing non-validity of a given formula by appealing to interpretations that interpret atoms as infinitary predicates generally would seriously weaken such a non-validity statement. E.g., if game $p^*$ depends on infinitely many variables, 
then  $p^*\add\gneg p^*$ may be incomputable just due to the fact that the machine would never be able to finish reading all the relevant information from the valuation tape necessary to determine whether $p^*$ is true or false. On the other hand, once we restrict our considerations only to interpretations that interpret atoms as finitary predicates, the non-validity statement for $p\add\gneg p$ is indeed highly informative: the failure to solve $p^*\add\gneg p^*$ in such a case signifies fundamental limitations of algorithmic methods rather than just impossibility to obtain all the necessary external information. A positive solution to Conjecture 24.4 of \cite{Jap03} restricted to the language of $\predel$ is contained in clause (b) of our Theorem \ref{main}.
 
As for Conjecture 26.2, it was about equivalence between validity and another version of this notion called {\em uniform validity}.
If we disabbreviate ``\hspace{1pt}$\models F^*$\hspace{1pt}" as ``\hspace{1pt}$\exists\ {{\mathcal H}}\ \bigl({{\mathcal H}}\models F^*\bigr)$\hspace{1pt}" 
(with $^*$ ranging over $F$-admissible interpretations and ${\mathcal H}$ over HPMs), then validity of $F$ in the sense of Definition \ref{val} can be written as  \mbox{``\hspace{1pt}$\forall\ ^*\ \exists\ {{\mathcal H}}\ \bigl({{\mathcal H}}\models F^*\bigr)$\hspace{1pt}".} Reversing the order of quantification yields the following stronger property of uniform validity:

\begin{definition}\label{uv}
A formula $F$ is said to be \gj{uniformly valid} iff there is an HPM ${\mathcal H}$ such that, for every $F$-admissible interpretation $^*$, \ ${{\mathcal H}}\models F^*$.
\end{definition}

Intuitively, uniform validity means existence of an interpretation-independent solution: since no information regarding interpretation $^*$ comes as a part of input to our play machines, the above HPM ${\mathcal H}$ with $\forall ^*({{\mathcal H}}\models F^*)$ will have to play in some standard, uniform way that would be successful for any possible $^*$. 
 
The term ``uniform" is borrowed from \cite{Abr94} as this understanding of validity in its spirit is close to that in Abramsky and Jagadeesan's tradition. The concepts of validity in Lorenzen's \cite{Lor59} tradition, or in the sense of Japaridze \cite{Jap00,Jap02a}, also belong to this category. Common to those uniform-validity-style notions is that validity there 
is not defined as being ``always true" (true=winnable) as this is the case with the classical understanding 
of this concept; in those approaches the concept of truth is often simply absent, and validity is treated as a basic concept 
in its own rights. 
As for (simply) validity, it is closer to validities in the sense of Blass \cite{Bla92} or Japaridze \cite{Jap97}, and presents a direct generalization of the corresponding classical concept in that it indeed means being true (winnable) in every particular setting.  

Which of our two versions of validity is more interesting depends on the motivational standpoint. 
It is validity rather than uniform validity that tells us what can be computed in principle. So, a 
computability-theoretician would focus on validity. Mathematically, non-validity is generally by an order of magnitude more informative --- and correspondingly harder to prove --- than non-uniform-validity. Say, the non-validity of $p\add\gneg p$,
with the above-quoted and now successfully verified 
Conjecture 24.4 of \cite{Jap03} in mind, means existence of solvable-in-principle yet algorithmically unsolvable problems
 --- the fact 
that became known to the mankind only as late as in the 20th century. 
As for 
the non-uniform-validity of $p\add\gneg p$, it is trivial: of course there is no way to choose one of the two disjuncts that would be true for all possible values of $p$ because, as the Stone Age intellectuals were probably aware, some $p$ are true and some are false.  

On the other hand, it is uniform validity rather than validity that is of interest in more applied areas of computer science such as knowledgebase systems (see Section \ref{sappl}) or resourcebase and planning systems (see Section 26 of \cite{Jap03} or Section 8 of \cite{Japint}). 
In such applications we want a logic on which a universal problem-solving machine can be based. Such a machine would or should be able to solve problems represented by formulas of $\predel$ without any specific knowledge of the meaning of 
their atoms, i.e. without knowledge of the actual interpretation. Remembering what was said about the intuitive meaning of uniform validity, this concept is exactly what fits the bill. 

Anyway, the good news, signifying a successful verification of Conjecture 26.2 of \cite{Jap03} restricted to the language of $\predel$, is that the two concepts of validity yield the same logic. If $F$ is uniformly valid, then it is automatically also valid, as uniform validity is stronger than validity. Suppose now $F$ is valid. Then, by the completeness part of Theorem \ref{main},  $\predel \vdash F$. But then, according to the `furthermore' clause (a) of the same theorem, $F$ is uniformly valid.  
Thus, where --- in accordance to our present convention --- ``formula" means formula of the language of $\predel$, we have:

\begin{theorem}\label{valuval}
A formula is valid if and only if it is uniformly valid.
\end{theorem}

In many contexts, such as the one of the following section, the above theorem allows us to talk about ``{\em the} semantics" of $\predel$ without being specific regarding which of the two possible underlying concepts of validity we have in mind.

\section{$\predel$-based applied systems}\label{sappl}

As demonstrated in Section \ref{op}, the language of $\predel$ presents a convenient formalism for specifying and studying computational problems and relations between them. Its axiomatization provides a systematic way to answer not only the 
question `{\em what} can be computed' but  --- in view of clause
(a) of Theorem \ref{main} --- also `{\em how} can be computed'.  Our approach brings logic and theory of computing closer together, and its general theoretical importance is obvious. The property of computability is at least as interesting as the property of (classical) truth. Moreover, as we saw, computability is also more general than truth: the latter is nothing but the former restricted to formulas of classical logic, i.e. elementary formulas. Thus, studying the logic of computability makes at least as much sense as studying the logic of truth. 
The latter --- classical logic --- is well-studied and  well-explored. The former, however, has never received the treatment it naturally deserves.

The significance of our study is not limited to the theory of computation or pure logic. The fact that $\predel$
is a conservative extension of classical logic makes the former a reasonable and appealing alternative to the latter in every aspect of its applications. In particular, there are good reasons to try to base applied theories --- such as, say, Peano  arithmetic --- on $\predel$ instead of just classical logic. From axioms of such a theory we would require to be ``true" in our sense, i.e. represent (under the fixed, ``standard" interpretation/model) computable problems, and from its rules of inference require to preserve the property of computability.
One of the particular ways to construct such theories is to treat the theorems of $\predel$ as logical axioms and use modus ponens as the only logical rule of inference.\footnote{One could show that including 
some other standard logical rules such as quantification rules along with modus ponens 
generally would not increase the deductive power of the theory as long as non-logical axioms are (re)written in a proper manner.} 
All of the non-logical axioms of the old, classical-logic-based version of the theory are true elementary formulas and hence computable in our sense, so they can be automatically included into the new set of non-logical axioms. To those could be added new, more constructive and informative axioms that involve choice operators, which would allow us to delete some or most 
of the old axioms that have become no longer independent. A new, computability-preserving inference rule that could be included in the $\predel$-based arithmetic  is the \gj{constructive rule of induction}:\label{cri} 
\[\ada x\bigl(F(x)\mli F(x+1)\bigr), \ F(0) \ \mapsto \ \ada xF(x).\]
       
No old information whatsoever would be lost when following this path. On the other hand we would get a much more expressive, constructive and computationally meaningful theory. All theorems of such a theory would be computable problems in our sense. E.g., provability of 
$\ada x\ade y\hspace{2pt} p(x,y)$ --- as opposed to $\cla x\cle y\hspace{2pt} p(x,y)$ --- 
would imply that, for every $x$, a $y$ with $p(x,y)$ not only exists, but can be algorithmically found. 
Moreover, such an algorithm itself can be effectively constructed from a proof of the formula and algorithmic solutions 
(winning HPMs) to the problems represented by the non-logical axioms of the theory. This would be guaranteed 
by clause (a) of theorem \ref{main}, Proposition \ref{mp} and similar facts regarding any additional, non-logical rules of inference if such are present, such as the above constructive rule of induction. 

Looks like our approach materializes what the constructivists have been calling for, yet without unsettling the classically-minded: whatever we could say or do by the means that classical logic offered, we can automatically still 
say and do, with the only difference that now many things we can say and do in a much more informative and constructive way. Our way of constructivization of theories is conservative and hence peaceful. This contrasts with  many other attempts to constructivize theories, that typically try to replace classical logic by weaker logics with ``constructive" syntactic features (often in a not very clear sense) --- such as intuitionistic calculus ---
yielding loss of information and causing the frustration of those who see nothing wrong with the classical way of reasoning.

From the purely logical point of view, it could be especially interesting to study applied theories in the 
$\cla,\cle$-free sublanguage of the language of $\predel$. Let us use {\bf CA} to denote the version of arithmetic  based on the $\cla,\cle$-free fragment of $\predel$. Of course, it would be more accurate to use the indefinite article ``{\em a}" instead of ``{\em the}" here, for we are not very specific about what the axioms of {\bf CA} should be. Let us just say that {\bf CA}
has some ``standard" collection of basic axioms characterizing $=$, $+$, $\times$ and the successor function, and 
includes the above constructive rule of induction. For the traditional, classical-logic based version of arithmetic we use the standard name {\bf PA}. Due to the big difference between the underlying logics of {\bf CA} and {\bf PA} --- enough to remember that one is decidable and the other is not --- 
{\bf CA} might have some new and interesting features. Could we obtain a reasonably expressive 
yet decidable theory this way?\footnote{Even if the set of non-logical axioms of {\bf CA} is chosen finite and the rule of induction is not included, the fact that the underlying logic is decidable does not imply the decidability 
of {\bf CA} itself, for the deduction theorem for $\predel$-based theories would work in a way rather different from 
how it works for classical-logic-based theories.} Generally, how strong a theory (whether decidable or not) could we get and what would be the fundamental limitations to its deductive power? 
Would {\bf CA} still be able to numerically represent all decidable predicates and functions as {\bf PA} does? How much of its own metatheory would {\bf CA} be able to formalize? 
One can show that the property of computability of the problems represented by formulas of {\bf CA}  can be expressed 
in the language of {\bf CA}, so that {\bf CA},
unlike {\bf PA}, would be able to talk about its own ``truth". One can also show that {\bf PA} can constructively prove that everything provable in {\bf CA} is true and hence {\bf CA} is consistent. What are the limitations of the
deductive strength of {\bf CA} that make it impossible to reproduce the same proof? If there are none, then what happens 
to G\"{o}del's incompleteness theorems in the context of {\bf CA}? How about provability logic in general, which has been so well-studied for {\bf PA} (see \cite{Jap98})? These are a few examples of the many intriguing questions naturally arising in this new framework and calling for answers. 

$\predel$ can as well be of high interest in applied areas of computer science such as AI. The point is  that 
the language of $\predel$, being a specification language for computational problems, is, at the same tame, a coherent and comprehensive query and knowledge specification language --- something that the language of classical logic fails to be. Where {\em Age}$(x,y)$ is the predicate ``Person $x$ is $y$ years old", the knowledge represented by the formula $\cla x\cle y\mbox{\em Age}(x,y)$ is knowledge of the almost tautological fact that all people have their age. However, 
how to express (the stronger) knowledge of every person's actual age, which is more likely to be of relevance in a knowledgebase system? In classical logic we cannot do this, and this limited expressive power precludes classical logic from serving as a satisfactory logic of knowledgebase systems,  
that all the time deal with the necessity to distinguish  between just truth and the system's actual ability to know/find/tell what is true. Within the framework of traditional approaches, classical logic needs to be extended (say, by adding to it epistemic modalities and the like) to obtain a more or less suitable logic of knowledgebases. In our case, however, the situation 
is much more nice: there is no need to have separate languages and logics for {\em theories} on one hand 
and {\em knowledgebases} on the other hand: the same logic $\predel$, with its standard semantics, can be successfully used in both cases, without the need to modify/extend/adjust it. Back to our example,  knowledge of everyone's actual age 
can be expressed by $\ada x\ade y\mbox{\em Age}(x,y)$. Obviously the ability of an agent to solve this problem means its ability to correctly tell each person's age. Within the framework of computability logic, the concept of the \gj{knowledge}\label{x38} of an agent can formally be defined as the set of the queries that the agent can actually solve. The word ``\gj{query}" here is a synonym of what we call ``problem" (game), and we prefer to use the former in this 
new context only because it is more common in the database and knowledgebase systems lingo.  

Let us look at the query intuitions associated with our game semantics. Every formula whose main operator is $\add$ or $\ade$ can be thought of as a question asked by the user (environment). E.g., 
\(\mbox{
{\em Male}$(${\em Dana}$)\hspace{2pt} \add\hspace{2pt} ${\em Female}$(${\em Dana})}\)
 is the question ``Is Dana male or female?". Solving this problem, by our semantics for $\add$, means correctly telling the gender of Dana. Formulas whose main operator is $\adc$ or $\ada$, on the other hand, represent questions asked by the system. E.g., 
\(\ada x\bigl(\mbox{\em Male}(x) \add\mbox{\em Female}(x)\bigr)\)
 is the question ``Whose gender do you want me to tell you?". The user's response can be ``Dana", which will bring the game down to the above user-asked question regarding the gender of Dana. Just as we noted when discussing computational problems, the language of $\predel$ allows us to form queries of arbitrary complexities and degrees of interactivity. Negation turns queries into counterqueries; parallel operators generate parallel queries where both the user and the system can have simultaneous questions and counterquestions, with $\mli$ acting as a query reduction operator; and blind quantifiers generate imperfect-information queries.  Let us look at  
\begin{equation}\label{diet}
\cla x\Bigl(\ade y\mbox{\em Age}(x,y)\mlc \bigl(\mbox{\em Male}(x)\add\mbox{\em Female} (x)\bigr)\mli \ade z\mbox{\em BestDiet}(z,x)\Bigr),
\end{equation}
where  {\em BestDiet}$(z,x)$ is the predicate ``$z$ is the best diet for $x$". The ability of the knowledgebase system to solve this query means its ability to determine the best diet for any person, provided that the system is told that person's age and gender --- that is, its ability to reduce the `best diet' problem to the `age and gender' problem. That $x$ is quantified with $\cla$ rather than $\ada$ means that the system does not need to be told who the person really is. 
The following is a possible legal scenario of interaction over this query. The system is waiting till the user specifies, in the antecedent, the age and gender of $x$ (without having explicitly specified the value of $x$). Our semantics automatically makes the system successful (winner) if the user fails to respond to either of 
those two counterqueries. Once  responses in the antecedent are received, the system selects a diet for $x$. The system has 
been successful if the diet it selected is really the best diet for $x$ as long as the user has told it the true age and gender of $x$.   

Most of the real information systems are interactive, and this makes our logic, which is designed to be a logic of interactive tasks, a well-suited formal framework for them and an appealing alternative to the more traditional frameworks. Imagine a medical diagnostics system. What we would like the system to do is to tell us, for any patient $x$, the diagnosis  $y$ for $x$. That is, to solve the query 
{\em for all $x$} $\ade y${\em Diagnosis}$(x,y)$. If here we understand `{\em for all}\hspace{1pt}' as $\cla$, the problem has no solution: an abstract $x$ cannot be diagnosed even by God. With `{\em for all}\hspace{1pt}' understood as 
$\ada$, the query does have a solution in principle. But 
diagnosing a patient just based on his/her identity would require having all the relevant medical information regarding that patient, which in a real knowledgebase system is unlikely to be the case. Most likely, the query that the system 
 solves would look like 
$\cla x(Q(x)\mli \ade y${\em Diagnosis}$(x,y))$, where $Q(x)$ is a  (counter)query with questions regarding $x$'s symptoms, blood pressure, cholesterol level, reaction to various drugs, etc. (one of such questions could be $\ade z(x=z)$, effectively turning the main quantifier $\cla x$ into $\ada x$). Most likely $Q(x)$ would not be just a $\mlc$-conjunction of such questions as this was the case with the antecedent of (\ref{diet}), but rather it would have a more complex structure, where what questions are asked could depend on the answers that the user gave to previous questions, yielding a long dialogue with a series of interspersed moves by both parties. 
 
A more familiar to each of us real-life example is the automated bank  
account information system. You dial the bank-by-phone number to inquire about your balance. 
But the query that the system solves is not really as simple as $\ade x${\em MyBalance}$(x)$. If this was the case, then you would be told your balance right after dialing the number. Rather, you will have to go through quite a dialogue, with all sorts of questions regarding your preferences, account type and number, secret PIN or even mother's maiden name.

The style of the above  examples and the terminology employed to explain the associated intuitions
are somewhat different from those that we saw in Section \ref{op} when discussing computational problems and operations on them, or at the beginning of the present section when discussing $\predel$-based applied theories. But notice that 
the underlying formal semantics remains the same: whether we talk about valid principles of computability, constructive applied theories, or knowledgebase systems --- in each case we deal with the same (language of) $\predel$ with its standard semantics. Using the same logic $\predel$ in all these cases is possible only due to Theorem \ref{valuval} though. The reason for the failure of the principle $p\add\gneg p$ in the context of computability theory is that the corresponding problem may have no algorithmic solution. That is, $p\add\gneg p$ is not {\em valid}. The reason for the failure of the same principle in the context of knowledgebase systems is much simpler. An intelligent system may fail to solve the problem 
{\em Male}({\em Dana})$\add\gneg${\em Male}({\em Dana}) not because the latter has no algorithmic solution (of course it has one), but simply because the system does not possess sufficient knowledge to determine Dana's gender. In particular, the system with empty non-logical (but perfect logical) knowledge would not be able to solve  
$p\add\gneg p$ because it is not {\em uniformly valid}. According to Theorem \ref{valuval}, however, validity and uniform validity are equivalent. Hence, the logic of computability, which  is about what can be computed in principle, is the same as the logic of knowledgebase systems, which is about what can be actually solved
by knowledge-based agents. 

The point to be made here is that our approach brings together applied theories and knowledgebase systems, traditionally studied by different clans of researchers with different motivations, 
visions and methods. Every computability-logic-based applied theory automatically {\em is} a knowledgebase system, and vice versa. Knowledgebase systems can be axiomatized in exactly the same way as we would axiomatize arithmetic. The set of non-logical axioms of such a system may include atomic formulas representing factual knowledge, such as {\em Father}$(${\em Bob,Jane}$)$ (``Bob is Jane's father"); it can include nonatomic elementary formulas representing general knowledge, 
such as 
\(\cla x\bigl(x\times(y+1)=(x\times y)+x\bigr)\) 
or
 \(\cla x\bigl(\cle y\mbox{\em Father}(x,y)\mli\mbox{\em Male}(x)\bigr);\) 
and it can include nonelementary formulas such as 
\(\ada x\ada y\ade z(z= x\times y)\) 
or
\(\ada x\ade y\mbox{\em Age}(x,y),\)
expressing the ability of the system to compute the $\times$ function or its knowledge of (ability to tell) everyone's age.
These axioms would represent what can be called the {\em explicit knowledge} of the system --- the basic set of problems/queries that the system is able to solve. And the set of theorems of such a system would represent its
overall --- perhaps what can be called {\em implicit} --- knowledge. Each theorem would be a query that the system,
with $\predel$ built into it, is actually capable of solving:
as we noted when discussing $\predel$-based applied theories, a solution to the problem/query expressed by a formula $F$ can be automatically obtained from a proof of $F$ and known solutions to the non-logical axioms. Furthermore, 
one can show that it is not even necessary for the knowledgebase system to know  actual solutions (winning HPMs) for 
its axioms.  Rather, it would suffice to have unlimited access to machines or other knowledgebase systems (\gj{external computational/informational resources}) that solve those axioms. ``Unlimited access" here means the possibility to query 
(play against) those resources any finite number of times and perhaps in parallel. There is no need for the system to know how exactly those external resources do their 
job as long as they do it right. The system would still be able to dynamically solve any theorem $F$, even if 
no longer able to construct an actual HPM that solves $F$.

Extending the meaning of the term ``resource" to physical resources as well, computability-logic-based knowledgebase systems can be further generalized to resourcebase  systems and systems for resource-bound planning and action.
See Section 26 of \cite{Jap03} for a discussion and illustrations. A more elaborated discussion of applied systems based on computability logic is given in Section 8 of \cite{Japint}.\vspace{20pt}

\begin{center}{\Large PART 2}\vspace{0pt}\end{center}  
This part can be considered a technical appendix to Part 1. It is exclusively devoted to proofs of our two main results: Theorem \ref{main} (Sections \ref{sprlm}-\ref{s5}) and Theorem \ref{decid} (Section \ref{s6}).

\section{Preliminaries}\label{sprlm}

The concept of admissible interpretation can be naturally extended from formulas to sets of formulas: For a set $S$ of formulas, an \gj{$S$-admissible} interpretation is an interpretation that is $F$-admissible for each $F\in S$. To simplify things, we will assume throughout the rest of this paper that all the formulas we deal with are from some fixed set $S$, and by ``interpretation" we will always mean $S$-admissible interpretation. 
 
Reiterating and extending our earlier conventions, in what follows $E,F,G,$ $H,I,J,K$ will be exclusively used as a metavariable for formulas, $\alpha,\beta$ for moves, $^*$,$^\star$ for interpretations, $x,y,z,s,u,w$ for variables, $a,b,c,d$ for constants, 
$t$ for terms and $e,f$ for valuations. 

The following lemma, on which our reasoning will often rely implicitly, is just a straightforward observation:

\begin{lemma}\label{struct}
For any formula $F(x_1,\ldots,x_n)$, the set $\legal{(F(t_1,\ldots,t_n))^*}{e}$ does not depend on $e$, $^*$ or $t_1,\ldots,t_n$.
\end{lemma}

With the above fact in mind and in accordance with our conventions from Section \ref{cpr}, we will usually omit the parameter $e$ in the expression $\legal{F^*}{e}$, as well as omit ``with respect to $e$" in the phrase ``legal run of $F^*$ with respect to $e$". Remember also from Section \ref{cpr} that $e$ can as well be omitted in the expression $\win{A}{e}$ when $A$ is a constant game and hence $e$ is irrelevant.

\begin{lemma}\label{new2}
Suppose $x$ is a variable occurring in a formula $F$. Then, for any  interpretation $^*$, 
 constant $c$ and subformula $G$ of $F$,  $(G[x/c])^*=G^*[x/c]$.  
\end{lemma} 

\begin{proof} Assume $x$ occurs in $F$. Pick an arbitrary interpretation $^*$, constant $c$ and subformula $G$ of $F$. Our goal statement 
\((G[x/c])^*=G^*[x/c]\) can be proven by induction on the complexity of $G$. We will only outline the proof scheme. Verification of details can be done by a routine analysis of the relevant definitions, which we lazily omit and just say something like ``it is easy to see that..."

Assume $G$ is an $n$-ary nonlogical atom $p(t_1,\ldots,t_n)$ (the case of logical atoms $\tlg,\twg$ is trivial). Let $p^* = A(x_1,\ldots,x_n)$. 

First consider the case when $x$ is not among $t_1,\ldots,t_n$. Then $G[x/c]=G$, so it would be sufficient to show that $G^*=G^*[x/c]$. But indeed, by our convention, $^*$ is $F$-admissible; since $x$ occurs in $F$, according to the definition of $F$-admissible interpretation, either $A(x_1,\ldots,x_n)$ does not depend on $x$, or $x$ is among $x_1,\ldots,x_n$. In either case it can be seen that $A(t_1,\ldots,t_n)$ does not depend on $x$. Hence $A(t_1,\ldots,t_n)[x/c]=A(t_1,\ldots,t_n)$, i.e. $G^*[x/c]=G^*$.

Next consider the case when $x$ is among $t_1,\ldots,t_n$. For convenience of visualization, we may assume that $t_1=\ldots=t_i=x$ and all $t_j$ with $i<j\leq n$ are different from $x$. Then $G[x/c]=p(c,\ldots,c,t_{i+1},\ldots,t_n)$ and hence $(G[x/c])^*=A(c,\ldots,c,t_{i+1},\ldots,t_n)$. It is not hard to verify that $A(c,\ldots,c,t_{i+1},\ldots,t_n)=
A(t_1,\ldots,t_n)[x/c]$, so that we get $(G[x/c])^*=G^*[x/c]$.  This competes our proof of the basis case of induction.

For the inductive step, let us consider the case when $G=H_1\mlc H_2$ as an example. The following equation is based on the obvious fact that substitution of terms commutes with $\mlc$:
\begin{equation}\label{dec61}\bigl((H_1\mlc H_2)[x/c]\bigr)^* =   
\bigl((H_1[x/c])\mlc (H_2[x/c])\bigr)^*.\end{equation}
Next, the operation $^*$ also commutes with $\mlc$, so that we have
\[\bigl((H_1[x/c])\mlc (H_2[x/c])\bigr)^* =
(H_1[x/c])^*\mlc (H_2[x/c])^*.\]
By the induction hypothesis, $(H_1[x/c])^*=H_{1}^{*}[x/c]$ and $(H_2[x/c])^*=H_{2}^{*}[x/c]$, so we have
\[(H_1[x/c])^*\mlc (H_2[x/c])^*=(H_{1}^{*}[x/c])\mlc(H_{2}^{*}[x/c]).\]
Since the game operation of substitution of variables obviously commutes with $\mlc$, we have
\[(H_{1}^{*}[x/c])\mlc(H_{2}^{*}[x/c])=(H_{1}^{*}\mlc H_{2}^{*})[x/c].\]
Finally, again because $^*$ commutes with $\mlc$, we have
\begin{equation}\label{dec62}
(H_{1}^{*}\mlc H_{2}^{*})[x/c]=(H_{1}\mlc H_{2})^*[x/c].\end{equation}
The chain of equations from (\ref{dec61}) to (\ref{dec62}) yields 
\(\bigl((H_1\mlc H_2)[x/c]\bigr)^* = (H_{1}\mlc H_{2})^*[x/c],\) i.e.
$(G[x/c])^*=G^*[x/c]$.

The cases with the other propositional connectives will be handled in a similar way, based on the fact that the three operations: $^*$, $[x/c]$ (as an operation on formulas) and $[x/c]$ (as an operation on problems)
commute with $\gneg,\mld,\mli,\adc,\add$ just as they commute with $\mlc$. Moreover, those three operations commute with
$Qy$ as well, where $Q$ is any of the four quantifiers and $y$ is a variable different from $x$, so the case $G=
QyH$ with $y\not=x$ can also be handled in a way similar to the way we handled the case $G=H_1\mlc H_2$. 

The only remaining case is $G=QxH$ (one can see that $[x/c]$ does not commute with $Qx$). Obviously we have $( QxH)[x/c]=QxH$, so that 
\begin{equation}\label{dec63}\bigl((QxH)[x/c]\bigr)^*=(QxH)^*.\end{equation}
The operation $^*$ commutes with $Qx$, and therefore 
\[(QxH)^*=Qx(H^*).\]
$Qx(H^*)$ obviously does not depend on $x$, which easily implies 
\[Qx(H^*)=\bigl(Qx(H^*)\bigr)[x/c].\]
Again by the fact that $^*$ commutes with $Qx$, we have
\begin{equation}\label{dec64}\bigl(Qx(H^*)\bigr)[x/c]=(QxH\bigr)^*[x/c].\end{equation}
The chain of equations from (\ref{dec63}) to (\ref{dec64}) yields
\(\bigl((QxH)[x/c]\bigr)^*=(QxH\bigr)^*[x/c],\) i.e. $(G[x/c])^*=G^*[x/c]$.
\end{proof} 

By a \gj{perfect interpretation}\label{pi} we mean an interpretation that interprets any $n$-ary predicate letter $p$ as a finitary predicate
 $A(x_1,\ldots,x_n)$ that does not depend on any variables others than $x_1,\ldots,x_n$. Every perfect interpretation $^*$ 
is nothing but a model in the classical sense (\gj{classical model})\label{jun25} with domain $\{\mbox{\em constants}\}$ --- the model that interprets each constant $c$ as the element $c$ of the domain and interprets each $n$-ary predicate letter $p$ with $p^*=A(x_1,\ldots,x_n)$ as the predicate $A(x_1,\ldots,x_n)$. Such a predicate $A(x_1,\ldots,x_n)$ 
is generally $\leq n$-ary in our sense but can be thought of as exactly $n$-ary under   
the more traditional understanding of $n$-ary predicates as sets of $n$-tuples of objects of the domain (the understanding that we slightly revised in Section \ref{cpr}). 
By a {\bf closed}\label{clfrm} formula we mean a formula not containing free occurrences of variables.  

A straightforward induction based on a routine analysis of relevant definitions reveals that:

\begin{lemma}\label{jan23}
For any formula $F$ and perfect interpretation $^*$, the game $F^*$ $($is finitary and$)$ does not depend on any variables 
that do not occur free in $F$; hence, if $F$ is closed, $F^*$ is a constant game. 
\end{lemma}

With the above fact in mind and in accordance with our conventions, as long as $F$ is closed and $^*$ is perfect, we can always safely omit the valuation parameter $e$ in $\win{F^*}{e}$ and simply write $\win{F^*}{}$ as this is done in Lemma \ref{cl} below. 

Remembering the observations made in Section \ref{op} about the classical behavior 
of our operations $\tlg$, $\twg$, $\gneg$, $\mlc$, $\mld$, $\mli$, $\cla$, $\cle$, we obviously have: 

\begin{lemma}\label{cl}
For any closed elementary formula $F$ and perfect interpretation $^*$,  $\win{F^*}{}\emptyrun=\pp$ iff
$F$ is true in $^*$ understood as a classical model.
\end{lemma}  

Based on the above fact, for a closed elementary formula $F$ and perfect interpretation $^*$, the phrases ``$F$ is true in $^*$" and ``$\win{F^*}{}\emptyrun=\pp$" will be used interchangeably. Remember also from Section \ref{cpr} that, for a predicate $A$, another way to say
``$\win{A}{e}\emptyrun=\pp$" or ``$A$ is true at $e$" is to say ``$e[A]$ is true".
 
Let $^*$ be an arbitrary interpretation and $e$ an arbitrary valuation. The 
\gj{perfect interpretation induced by $(^*,e)$}\label{x99} is the interpretation $^\star$ such that, for every $n$-ary predicate letter $p$ with $p^*=A(x_1,\ldots,x_n)$, we have $p^\star=
A'(x_1,\ldots,x_n)$, where $A'(x_1,\ldots,x_n)$ is the unique game such that, for any tuple $c_1,\ldots,c_n$ of constants, $A'(c_1,\ldots,c_n)=e[A(c_1,\ldots,c_n)]$. This means nothing but that $A'(x_1,\ldots,x_n)$  is the 
predicate such that $A'(c_1,\ldots,c_n)$ is true (at whatever valuation) iff $A(c_1,\ldots,c_n)$ is true at $e$. 
Note that while $A(c_1,\ldots,c_n)$ may depend on some hidden variables,
$A'(c_1,\ldots,c_n)$ is a constant game.

For a formula $F$, we will be using the notation  
\(\mbox{$\elz{F}$}\label{x95}\)
 for the elementarization of $F$. 
The following two lemmas can be verified by straightforward induction on the complexity of $F$:

\begin{lemma}\label{new1}
For any formula $F$, interpretation $^*$ and valuation $e$, $\win{F^*}{e}\emptyrun=\win{\elzi{F}^*}{e}\emptyrun$.
\end{lemma}

\begin{lemma}\label{simpll}
Suppose $^\star$ is the perfect interpretation induced by $(^*,e)$, and $F$ is a closed elementary\footnote{In fact the lemma holds for any closed formula, but for our purposes the elementary case is sufficient.} formula. Then  $e[F^*]=F^\star$.
\end{lemma}

A valuation $f$ is said to be \gj{finite} iff there is a finite set $\vec{x}$ of variables such that for every variable $y\not\in\vec{x}$, $f(y)=0$. A \gj{representation} of a finite valuation $f$ is a set $\{x_1/c_1,\ldots,x_n/c_n\}$, where $x_1,\ldots,x_n$ are pairwise distinct variables such that each variable $x$ with $f(x)\not=0$ is among $x_1,\ldots,x_n$,
and  $c_1,\ldots,c_n$ are constants with $f(x_1)=c_1,\ldots,f(x_n)=c_n$. We will say that such a set 
$\{x_1/c_1,\ldots,x_n/c_n\}$ \gj{represents} $f$. By abuse of terminology, we will often identify a representation of a given finite valuation with that valuation itself. 

Where $f$ is a valuation and $F$ is a formula, $fF$ will denote the result of substituting in $F$ every free occurrence of every variable $x$ by the constant $f(x)$. That is, $fF=F[x_1/f(x_1),\ldots,x_n/f(x_n)]$, where $x_1,\ldots,x_n$ are all the free variables of $F$. Thus, $fF$ is always a closed formula. Generally, we say that $G$ is an \gj{instance}\label{x33} of $F$ iff $G=fF$ for some valuation $f$.

We say that a valuation $f$ is \gj{$F$-distinctive}\label{dst} iff for any free terms $t_1$ and $t_2$ of $F$, as long as $t_1\not=t_2$, we have $f(t_1)\not=f(t_2)$.  

\begin{lemma}\label{agr}
For any formula $F$, interpretation $^*$, and valuations $e$ and $f$ that agree on all free variables of $F$, we have  
$e[F^*]=e[(fF)^*]$.
\end{lemma}

\begin{proof} Assume $F$, $^*$, $e$, $f$ are as above. Let $x_1,\ldots,x_n$ be all the free variables of $F$, and 
let \(c_1=e(x_1)=f(x_1),\ \ldots,\ c_n=e(x_n)=f(x_n).\) 
Obviously we have 
\begin{equation}\label{q1}e[F^*]=e\bigl[F^*[x_1/c_1,\ldots,x_n/c_n]\bigr].\end{equation}
Observe that $F^*[x_1/c_1,\ldots,x_n/c_n]=(\ldots((F^*[x_1/c_1])[x_2/c_2])\ldots)[x_n/c_n]$, and similarly for ``$F$" instead of ``$F^*$". Therefore, applying Lemma \ref{new2} $n$ times, we get \(F^*[x_1/c_1,\ldots,x_n/c_n]=(F[x_1/c_1,\ldots,x_n/c_n])^*\) and hence 
\begin{equation}\label{jun255}e\bigl[F^*[x_1/c_1,\ldots,x_n/c_n]\bigr]=e\bigl[(F[x_1/c_1,\ldots,x_n/c_n])^*\bigr] .\end{equation}
But $F[x_1/c_1,\ldots,x_n/c_n]$ is nothing but $fF$, so we have $(F[x_1/c_1,\ldots,x_n/c_n])^*=(fF)^*$ and hence 
\begin{equation}\label {q3}e\bigl[(F[x_1/c_1,\ldots,x_n/c_n])^*\bigr]=e[(fF)^*].\end{equation}
Equations (\ref{q1}), (\ref{jun255}) and (\ref{q3}) yield $e[F^*]=e[(fF)^*]$.
\end{proof} 

Now we define a function that, for a formula $F$ and a surface occurrence $O$ in $F$, returns a string $\alpha$ called 
the \gj{$F$-specification of $O$},\label{x300} which is said to \gj{$F$-specify $O$}. In particular:\vspace{-5pt}
\begin{itemize}
\item The occurrence of $F$ in itself is $F$-specified by 
the empty string.
\item If $F$ is $\gneg G$, $\cla xG$ or $\cle x G$,  then an occurrence that happens to be in $G$ is $F$-specified by the same string that $G$-specifies that occurrence.
\item If  $F$ is $G_1\mlc \ldots \mlc G_n$,  $G_1\mld \ldots \mld G_n$ or $G_1\mli G_2$, then an occurrence that happens to be in $G_i$ is $F$-specified by $i.\alpha$, where 
$\alpha$ is the $G_i$-specification of that occurrence.\vspace{-5pt}
\end{itemize}

Example: The second occurrence of $p\add q$ in $F=G\mld (p\add q)\mld\gneg(p\mli \cle x(G\mlc (p\add q)))$ is $F$-specified by the string ``$3.2.2.$".

With Lemma \ref{new2} in mind and based on Proposition \ref{ic}, the following lemma can be easily verified by induction on the complexity of $F$, the routine details of which we omit: 

\begin{lemma}\label{splem} For every formula $F$, move $\alpha$ and interpretation $^*$:
\begin{description}
\item[(a)] $\seq{\oo\alpha}\in\legal{F^*}{}$ iff one of the following two conditions holds:
\begin{enumerate}
\item $\alpha=\beta i$, where $\beta$ is the $F$-specification 
of a positive $($resp. negative$)$ surface occurrence of a subformula $G_1\adc\ldots\adc G_n$ $($resp. $G_1\add\ldots\add G_n$$)$ 
and $i\in\{1,\ldots,n\}$. In this case  $\seq{\oo\alpha}F^*=H^*$, where $H$ is the result of substituting in $F$ the above occurrence by $G_i$. 
\item $\alpha=\beta c$, where $\beta$ is the $F$-specification 
of a positive $($resp. negative$)$ surface occurrence of a subformula $\ada xG(x)$ $($resp. $\ade xG(x)$$)$ 
and $c\in\{\mbox{constants}\}$. In this case  $\seq{\oo\alpha}F^*=H^*$, where $H$ is the result of substituting in $F$ the above occurrence by $G(c)$. 
\end{enumerate}

\item[(b)] $\seq{\pp\alpha}\in\legal{F^*}{}$ iff one of the following two conditions holds:
\begin{enumerate}
\item $\alpha=\beta i$, where $\beta$ is the $F$-specification 
of a negative $($resp. positive$)$ surface occurrence of a subformula $G_1\adc\ldots\adc G_n$ $($resp. $G_1\add\ldots\add G_n$$)$ 
and $i\in\{1,\ldots,n\}$. In this case  $\seq{\pp\alpha}F^*=H^*$, where $H$ is the result of substituting in $F$ the above occurrence by $G_i$. 
\item $\alpha=\beta c$, where $\beta$ is the $F$-specification 
of a negative $($resp. positive$)$ surface occurrence of a subformula $\ada xG(x)$ $($resp. $\ade xG(x)$$)$ 
and $c\in\{\mbox{constants}\}$. In this case  $\seq{\pp\alpha}F^*=H^*$, where $H$ is the result of substituting in $F$ the above occurrence by $G(c)$. 
\end{enumerate}
\end{description}
\end{lemma}

\section{Soundness of $\predel$}\label{s3}

\begin{proposition}\label{sound}
 If $\predel\vdash F$, then $F$ is valid {\em (}any formula $F${\em )}. 
Moreover, there is an effective procedure that takes an $\predel$-proof of a formula $F$ and returns an HPM ${\mathcal H}$ such that, for all $^*$, \  ${{\mathcal H}}\models F^*$.
\end{proposition}

\begin{proof} Assume $\predel\vdash F$. Let us fix a particular $\predel$-proof of $F$. We will be referring to at as ``the proof", and referring to the formulas occurring in the proof as ``proof formulas". We assume that this is a sequence (rather than tree) of formulas without repetitions, and that every proof formula comes with a fixed {\em justification} --- a record indicating by which rule and from what premises the formula was derived. 

It would be sufficient to describe an effective way of constructing an EPM ${\mathcal E}$ with `for all $^*$, ${{\mathcal E}}\models F^*$'. By Proposition \ref{eq}, such an EPM ${\mathcal E}$ can then be effectively converted into an HPM ${\mathcal H}$ with 
`for all $^*$, ${{\mathcal H}}\models F^*$'.

We construct the EPM ${\mathcal E}$, that will play in the role of $\pp$, as follows. At the beginning, this machine creates two records on its work tape: $E$ to hold a formula, and $f$ to hold (a representation of) a finite valuation. $E$ is initialized to $F$, and $f$ initialized to $\{x_1/c_1,\ldots,x_q/c_q\}$, where $x_1,\ldots,x_q$ are all the free variables of $F$ and, for each $1\leq i\leq q$, $c_i$ is the value assigned to $x_i$ by the valuation spelled on the valuation tape. After the initialization step, the machine follows the  algorithm LOOP described below.\vspace{10pt}
  
\noindent{\bf Procedure} LOOP: {\bf While} $E$ is a proof formula, {\bf do} one of the following, depending on which of 
the three rules was used (last) to derive $E$ in the proof:

\begin{description}
\item[Case of Rule A:] Keep granting permission until the adversary makes a move  
$\alpha$ that satisfies the conditions of one of the following two subcases, and then act as the corresponding subcase 
prescribes: 
\begin{description}
\item[Subcase (i):] $\alpha=\beta i$, where $\beta$ \ $E$-specifies a positive (resp. negative) surface occurrence of a subformula $G_1\adc\ldots\adc G_n$ (resp. $G_1\add\ldots\add G_n$) and $i\in\{1,\ldots,n\}$. Let $H$ be  
the result of substituting in $E$ the above occurrence by $G_i$. Then update $E$ to $H$, and update $f$ by deleting in it all pairs $u/d$ where $u$ is not a free variable of $H$.
\item[Subcase (ii):] $\alpha=\beta c$, where $\beta$ \ $E$-specifies a positive (resp. negative) surface occurrence of a subformula $\ada xG(x)$ (resp. $\ade x G(x)$) and $c\in\constants$. Let $H$ be the premise\footnote{If there are many such premises, select the lexicographically smallest one. The presence of more than one such premise, however, signifies that the proof has some (easy-to-get-rid-of) redundancies, and we may safely assume that this is not the case.} of $E$ that is  
the result of substituting in $E$ the above occurrence by $G(y)$, where $y$ does not occur in $E$. Then update 
$E$ to $H$, and update $f$ to $f\cup\{y/c\}$ (unless $x$ did not really have free occurrences in $G(x)$, in which case 
$f$ should stay the same as it was). 
\end{description}
\item[Case of Rule B1:] Let $H$ be the premise of $E$ in the proof. $H$ is the result of substituting, in $E$, a certain negative (resp. positive) surface occurrence of a subformula $G_1\adc\ldots\adc G_n$ (resp. $G_1\add\ldots\add G_n$) by $G_i$ for some $i\in\{1,\ldots,n\}$. 
Let $\beta$ be the $E$-specification of that occurrence. Then make the move $\beta i$, update $E$ to $H$, and 
update $f$ by deleting in it all pairs $u/d$ where $u$ is not a free variable of $H$.
\item[Case of Rule B2:] Let $H$ be the premise of $E$ in the proof. $H$ is the result of substituting, in $E$, a certain negative (resp. positive) surface occurrence of a subformula $\ada xG(x)$ (resp. $\ade xG(x)$) by $G(t)$ for some term $t$ such that (if $t$ is a variable) neither the above occurrence of 
$\ada xG(x)$ (resp. $\ade xG(x)$) in $F$ nor any of the free occurrences of $x$ in $G(x)$ are in the scope of $\cla t$,
$\cle t$, $\ada t$ or $\ade t$. 
 Let $\beta$ be the $E$-specification of the above occurrence of $\ada xG(x)$ (resp. $\ade xG(x)$). Let $c=f(t)$ if $t$ is either a free variable of $E$ or a constant,\footnote{Remember that when $t$ is a constant, $f(t)=t$.} and 
$c=0$ otherwise. Then make the move $\beta c$, update $E$ to $H$, and --- if $t$ is a variable --- update $f$ to $f\cup\{t/c\}$ (unless $x$ did not really have free occurrences in $G(x)$, in which case 
$f$ should stay the same as it was). 
\end{description}\vspace{8pt}

It is obvious that (the description of) ${\mathcal E}$ can be constructed effectively from the $\predel$-proof of $F$. What we need to do now is to show that ${\mathcal E}$ wins $F^*$ for every $^*$.
In doing so, we will assume that ${{\mathcal E}}$'s adversary never makes illegal moves. By Remark \ref{assume}, making such an assumption is perfectly legitimate. 

Pick an arbitrary interpretation $^*$, an arbitrary valuation $e$ and an arbitrary $e$-computation branch $B$ of ${\mathcal E}$. Fix $\Gamma$ as the run spelled by $B$. Consider the work of ${\mathcal E}$ in $B$. For each $i\geq 1$ such that LOOP makes at least $i$ iterations in $B$, let $E_i$ and $f_i$ denote the values of the records $E$ and $f$ at the beginning of the $i$th iteration of LOOP, 
and $K_i$ denote $f_iE_i$. Thus, $E_1=F$ and, by Lemma \ref{agr}, 
$e[F^*]=e[K_{1}^{*}]$. Our goal is to show that $B$ is fair and   
$\win{F^*}{e}\seq{\Gamma}=\pp$, 
i.e. $\win{K_{1}^{*}}{e}\seq{\Gamma}=\pp$.  

Evidently $E_{i+1}$ 
is always one of the premises of $E_i$ in the proof, so that LOOP is iterated only a finite number of times. For the same reason, the value of record $E$ is always a proof formula (incidentally,
this means that the {\bf while} condition of LOOP is always satisfied, so that the reason why LOOP is only iterated a finite number of times is simply that one of the iterations never terminates).
Fix $l$ as the number of iterations of LOOP. 
The $l$th iteration deals with the case of 
Rule {\bf A}, for otherwise there would be a next iteration. This implies that 
\begin{equation}\label{e71} E_l\ \mbox{\em is stable.}\end{equation}

For each $i$ with $1\leq i\leq l$, let $\Theta_i$ be the sequence of the moves made by the players by the beginning of the $i$th iteration of LOOP, where the moves made by ${\mathcal E}$ are $\pp$-labeled and the moves made by its adversary $\oo$-labeled.
\begin{equation}\label{e71.5} 
\begin{array}{l}\mbox{\em For each $i$ with $1\leq i\leq l$, we have $\Theta_i\in\legal{K_{1}^{*}}{}$ and $\seq{\Theta_i}K_{1}^{*}=K_{i}^{*}$.}
\end{array} 
\end{equation}

This statement can be proven by induction on $i$. The basis case with $i=1$ is trivial. Now consider an arbitrary $i$ 
with $1\leq i < l$. By the induction hypothesis, $\Theta_i\in\legal{K_{1}^{*}}{}$ and $\seq{\Theta_i}K_{1}^{*}=K_{i}^{*}$. 
If the $i$th iteration of LOOP deals with the case of Rule {\bf B1} or {\bf B2}, then exactly one move $\alpha$ is made
 during that iteration, and this move is by the machine, so that $\Theta_{i+1}=\seq{\Theta_i,\pp\alpha}$. A simple analysis of the corresponding steps of our algorithm, in conjunction with Lemma \ref{splem}(b), can convince us that 
$\seq{\pp\alpha}\in\legal{K_{i}^{*}}{}$ and $\seq{\pp\alpha}K_{i}^{*}=K_{i+1}^{*}$. With the equalities $\Theta_{i+1}=\seq{\Theta_i,\pp\alpha}$ and $K_{i}^{*}=\seq{\Theta_i}K_{1}^{*}$ in mind,
the former then implies $\Theta_{i+1}\in\legal{K_{1}^{*}}{}$ and the latter implies 
$\seq{\Theta_{i+1}}K_{1}^{*}=K^{*}_{i+1}$. Suppose now the $i$th iteration of LOOP deals with the case of Rule {\bf A}. Then the machine does not make a move. This means that $\oo$ makes a move $\alpha$, for otherwise we would have $i=l$.
 Our assumption that $\oo$ never makes illegal moves here means nothing but that  
$\seq{\Theta_i,\oo\alpha}\in\legal{K_{1}^{*}}{}$ and therefore (as $K^{*}_{i}=\seq{\Theta_i}K_{1}^{*}$)
$\seq{\oo\alpha}\in\legal{K_{i}^{*}}{}$.
Applying Lemma \ref{splem}(a) to the fact that $\seq{\oo\alpha}\in\legal{K_{i}^{*}}{}$  and analyzing the corresponding steps of our algorithm, it is easy to see that $\Theta_{i+1}=\seq{\Theta_i,\oo\alpha}$ and $\seq{\oo\alpha}K_{i}^{*}=K_{i+1}^{*}$.
Hence $\Theta_{i+1}\in\legal{K_{1}^{*}}{}$ and $\seq{\Theta_{i+1}}K_{1}^{*}=K^{*}_{i+1}$. Statement (\ref{e71.5})
is proven.
 
\begin{equation}\label{rr} \Gamma=\Theta_l.\end{equation}

Indeed. Since the $l$th iteration of LOOP deals with the case of Rule {\bf A}, ${\mathcal E}$ does not make any moves during that iteration. We claim that $\oo$ does not make any moves either, so that run $\Gamma$ that is generated in the play is exactly $\Theta_l$. To verify this claim, suppose, for a contradiction, that during the $l$th iteration of LOOP $\oo$ makes a move $\alpha$. As we assume that $\oo$ plays legal, we should have  $\seq{\Theta_l,\oo\alpha}\in\legal{K_{1}^{*}}{}$. 
In view of (\ref{e71.5}), this means that $\seq{\oo\alpha}\in\legal{K_{l}^{*}}{}$.
From Lemma \ref{splem}(a), just as this was observed in the proof of (\ref{e71.5}), 
it is obvious that then $\alpha$  would satisfy the conditions of either Subcase {\bf (i)} or Subcase {\bf (ii)}, and 
then there would be an $(l+1)$th iteration, which, however, is not the case. Statement (\ref{rr}) is proven.
 
The fact that the last iteration of LOOP deals with the case of Rule {\bf A} and $\oo$ does not make any moves during that iteration guarantees that ${\mathcal E}$ will grant permission infinitely many times during that iteration, so that branch $B$ is fair. Thus, in order to complete our proof of Proposition \ref{sound}, what remains to show is that $\win{K_{1}^{*}}{e}\seq{\Gamma}=\pp$.

According to (\ref{e71.5}), $\Theta_l$ is a legal position of $K_{1}^{*}$ and 
$\seq{\Theta_l}K_{1}^{*}=K_{l}^{*}$. This implies that $\win{K_{1}^{*}}{e}\seq{\Theta_l}=\win{K^{*}_{l}}{e}\emptyrun$. But, by 
(\ref{rr}), $\Theta_l=\Gamma$. Hence  
\begin{equation}\label{e75} 
\begin{array}{l}\mbox{\em $\win{K_{1}^{*}}{e}\seq{\Gamma}=\win{K_{l}^{*}}{e}\emptyrun $.}
\end{array} 
\end{equation}

Suppose, for a contradiction, that 
$\win{K_{1}^{*}}{e}\seq{\Gamma}\not=\pp$. Then, by (\ref{e75}),  $\win{K_{l}^{*}}{e}\emptyrun\not=\pp$, whence,  according to Lemma \ref{new1}, $\win{\elzi{K_l}^*}{e}\emptyrun\not=\pp$. Then Lemma \ref{simpll} implies that $\win{\elzi{K_l}^\star}{}\emptyrun\not=\pp$, 
where $^\star$ is the perfect interpretation induced by $(^*,e)$. That is, $\elz{K_l}$ is false in $^\star$ understood as a classical model. But this is impossible because, by (\ref{e71}), $\elz{E_l}$ is classically valid and hence $\elz{K_l}$,
which is an instance of $\elz{E_l}$, is true in all classical models.
 \end{proof}

\section{Completeness of $\predel$}\label{s5}

\begin{lemma}\label{newest}
Let $t$ be any term, $F(t)$ any formula, and $t'$ any term that does not occur in $F(t)$. Then 
$\predel\vdash F(t)$ iff $\predel\vdash F(t')$. 
\end{lemma}
\begin{proof} This lemma can be proven by induction on the lengths of $\predel$-derivations. The step corresponding to Rule {\bf A} will rely on a similar fact known from classical logic. The routine details of this induction are left to the reader.\end{proof}

In our completeness proof for $\predel$ we will employ the complementary  logic $\copredel$,\label{x23} whose language is the same as that of $\predel$ and which is given by the following rules:\vspace{-5pt}

\begin{description} 
\item[\ \ A.\ ]  $\vec{H}\mapsto F$, where $F$ is instable and $\vec{H}$ is a 
set of formulas satisfying the following conditions:
\begin{description}
\item[(i)]  Whenever $F$ has a negative (resp. positive)  surface occurrence of a subformula $G_1\adc\ldots\adc G_n$ (resp. $G_1\add \ldots\add G_n$), for each 
$i\in\{1,\ldots,n\}$, $\vec{H}$ contains the result of replacing that occurrence in $F$ by $G_i$;
\item[(ii)] Whenever $F$ has a negative (resp. positive) surface occurrence of a subformula $\ada x G(x)$ (resp. $\ade xG(x)$), $\vec{H}$ contains the result of replacing that occurrence in $F$ by $G(y)$, where $y$ is a variable that does not occur in $F$.
\item[(iii)] Whenever $F$ has a negative (resp. positive) surface occurrence of a subformula $\ada x G(x)$ (resp. $\ade xG(x)$)
and $t$ is a free term of $F$,  $\vec{H}$ contains the result of replacing in $F$ the above occurrence 
of $\ada xG(x)$ (resp. $\ade xG(x)$)  by $G(y)$ and\footnote{``and" = ``and replacing in the resulting formula".}
 all free occurrences of $t$ by $y$, where $y$ is a variable that does not occur in $F$.

\end{description}
\item[\ \ B1.]  $F'\mapsto F$, where $F'$ is the result of replacing in $F$ a positive (resp. negative) surface occurrence of a subformula $G_1\adc\ldots\adc G_n$ (resp. $G_1\add\ldots\add G_n$) by $G_i$ for some $i\in\{1,\ldots, n\}$.
\item[\ \ B2.]  $F'\mapsto F$, where $F'$ is the result of replacing in $F$ a positive (resp. negative) surface occurrence of a subformula $\ada xG(x)$ (resp. $\ade xG(x)$) by $G(y)$, where $y$ is a variable that does not occur in $F$.  
\end{description}

\begin{lemma}\label{cnl} 
If $\predel\not\vdash F$, then $\copredel\vdash F$ \ $($any formula $F$ $)$.
\end{lemma}

\begin{proof} We prove this lemma by induction on the complexity of $F$. Assume $\predel\not\vdash F$. There are two cases to consider:

{\em Case 1:} $F$ is stable. Then one of the following two subcases must hold (otherwise $F$ would be $\predel$-derivable by Rule {\bf A}): 

{\em Subcase 1.1:} There is a $\predel$-unprovable formula $H$ that is the result of replacing in 
$F$ some positive (resp. negative) surface occurrence of a subformula $G_1\adc\ldots\adc G_n$ (resp. $G_1\add \ldots\add G_n$) by $G_i$ for some $i\in\{1,\ldots ,n\}$. By the induction hypothesis $\copredel\vdash H$, whence, by Rule {\bf B1}, $\copredel\vdash F$.

{\em Subcase 1.2:} There is a $\predel$-unprovable formula $H$ that is the result of replacing in 
$F$ some positive (resp. negative) surface occurrence of a subformula $\ada xG(x)$ (resp. $\ade xG(x)$)  by $G(y)$,
where $y$ is a variable that does not occur in $F$. By the induction hypothesis $\copredel\vdash H$, whence, by Rule {\bf B2}, $\copredel\vdash F$.

{\em Case 2:} $F$ is instable.  Let $\vec{H}$ be a minimal set of formulas satisfying the three conditions (i)-(iii) 
of Rule {\bf A} of $\copredel$. We claim that 

\begin{equation}\label{apr10}
\mbox{\em None of the elements of $\vec{H}$ is $\predel$-provable.}\end{equation}

To show this, consider an arbitrary element $H$ of $\vec{H}$. One of the following three subcases must hold:

{\em Subcase 2.1:} $H$ is the result of replacing in $F$ a negative (resp. positive)  surface occurrence of a subformula $G_1\adc\ldots\adc G_n$ (resp. $G_1\add \ldots\add G_n$) by $G_i$ for some $1\leq i\leq n$. If $\predel\vdash H$, then
$F$ would be $\predel$-derivable from $H$ by Rule {\bf B1}, which is a contradiction. 

{\em Subcase 2.2:} $H$ is the result of replacing in $F$ a negative (resp. positive)  surface occurrence of a subformula $\ada xG(x)$ (resp. $\ade xG(x)$) by $G(y)$ for some $y$ not occurring in $F$. Just as in the previous subcase, 
$\predel\vdash H$ is impossible, for otherwise, by Rule {\bf B2}, we would have $\predel\vdash F$. 

{\em Subcase 2.3:} $H$ is the result of replacing in $F$  a negative (resp. positive) surface occurrence of a subformula $\ada x G(x)$ (resp. $\ade xG(x)$)
 by $G(y)$ and all free occurrences of some term $t$ by $y$, where $y$ is a variable that does not occur in $F$.  
Notice that then $F[t/y]$ follows  follows from $H$ by Rule {\bf B2} of $\predel$. So, if $\predel\vdash H$, then 
$\predel\vdash F[t/y]$, and therefore, by Lemma \ref{newest}, $\predel\vdash F$. Again a contradiction, and (\ref{apr10}) is thus proven.
 
Applying the induction hypothesis to (\ref{apr10}), we conclude that each element of $\vec{H}$ is $\copredel$-provable, whence, by Rule {\bf A}, $\copredel\vdash F$. \end{proof}

Remember that a (finitary) predicate $A$ is said to be of \gj{complexity $\Sigma_2$}\label{x55} iff it is (``can be written as") $\cle y\cla z B$ for some decidable predicate $B$; and $A$ is of \gj{complexity $\Delta_2$}\label{x555} iff both $A$ and $\gneg A$ are of complexity $\Sigma_2$. The rest of this section is devoted to a proof of the following proposition:

\begin{proposition}\label{unic} 
 If $\predel\not\vdash F$, then $F$ is not valid {\em (}any formula $F${\em )}.

In particular, if $\predel\not\vdash F$, then $F^*$ is not computable for some interpretation $^*$ that interprets atoms as finitary predicates of complexity $\Delta_2$. 
\end{proposition}

\begin{proof} Assume $\predel\not\vdash F$. Then, by Lemma \ref{cnl}, $\copredel\vdash F$. Let us 
fix a $\copredel$-proof for $F$, call it ``the proof" and call the formulas occurring in the proof ``proof formulas". Our conventions about what a proof means are the same as in Section \ref{s3}. In particular, we assume that the proof has no repetitions: every proof formula appears in it exactly once. Based on the proof, we are going to construct the fair EPM ${\mathcal E}$ which will be shown to be such that no HPM ${\mathcal H}$ wins 
$F^*$ against ${\mathcal E}$ on $e_c$ for an appropriately selected interpretation $^*$ (which does not depend on ${\mathcal H}$) and valuation $e_c$. Our selection of such $^*$ and $e_c$ will be based on a diagonalization-style idea. 

Let us agree for the rest of this section that  $x_1,\ldots,x_q$ are all the $($pairwise distinct$)$ free variables of $F$, and that\vspace{7pt}

\begin{subconvention}{unic}{1} \ \\
a) $e$ always means the $($arbitrary but fixed$)$ valuation spelled on the valuation tape of ${\mathcal E}$;\\
b) $B$ always stands for an $($arbitrary but fixed$)$ $e$-computation branch of ${\mathcal E}$.
\end{subconvention}\vspace{0pt}

The work of ${\mathcal E}$ consists of three stages, that we call the preinitialization, initialization and postinitialization stages.
During the {\em preinitialization stage}, ${\mathcal E}$ checks whether $e$ is $F$-distinctive (see page \pageref{dst}). If $e$ passes the test for 
$F$-distinctiveness,
${\mathcal E}$ goes to the initialization stage. Otherwise ${\mathcal E}$ simply goes into an infinite loop in a permission state to formally ensure fairness, thus forever remaining in the preinitialization stage.
During the {\em initialization stage}, ${\mathcal E}$ creates two records: $E$ to hold a formula, and $f$ to hold (a description of) a finite valuation. ${\mathcal E}$ initializes $E$ to $F$ and $f$ to $\{x_1/e(x_1),\ldots,x_q/e(x_q)\}$,  and goes to the postinitialization stage.
During the {\em postinitialization stage}, ${\mathcal E}$ simply follows the following procedure:\vspace{10pt}

\noindent{\bf Procedure} LOOP: {\bf While} $E$ is a proof formula and $f$ is an $E$-distinctive valuation, 
{\bf do} one of the following, depending on which of the three rules was used (last) to derive $E$ in the proof:

\begin{description}
\item[Case of Rule A:] Keep granting permission until the adversary makes a move  
$\alpha$. Then act depending on which of the following four subcases applies:
\begin{description}
\item[Subcase (i):] $\alpha=\beta i$, where $\beta$ \ $E$-specifies a negative (resp. positive) surface occurrence of a subformula $G_1\adc\ldots\adc G_n$ (resp. $G_1\add\ldots\add G_n$) and $i\in\{1,\ldots,n\}$. Let $H$ be  
the result of substituting in $E$ the above occurrence by $G_i$. Then update $E$ to $H$, and update $f$ by deleting in it all pairs $x/d$ such that $x$ is not a free variable of $H$.
\item[Subcase (ii):] $\alpha=\beta c$, where $\beta$ \ $E$-specifies a negative (resp. positive) surface occurrence of a subformula $\ada xG(x)$ (resp. $\ade x G(x)$) and $c$ is a constant not occurring in $fE$. Let $H$ be the 
premise\footnote{As in Section \ref{s3}, if there are many such premises, select the lexicographically smallest one.}
 of $E$ that is  
the result of substituting  in $E$ the above occurrence by $G(y)$, where $y$ is a variable that does not occur in $E$. Then update 
$E$ to $H$, and update 
$f$ to $f\cup\{y/c\}$ (unless $x$ did not really have free occurrences in $G(x)$, in which case $f$ should stay as it was).
\item[Subcase (iii):] $\alpha=\beta c$, where $\beta$ \ $E$-specifies a negative (resp. positive) surface occurrence of a subformula $\ada xG(x)$ (resp. $\ade x G(x)$) and $c$ is a constant that occurs in $fE$. Let $t$ be the free term of $E$ with $f(t)=c$. Let $H$ be the premise\footnote{Again, select the lexicographically smallest one if there are many such premises.} of $E$ that is  
the result of substituting in $E$ the above occurrence of $\ada xG(x)$ (resp. $\ade xG(x)$) by  $G(y)$ and all free occurrences of $t$ by $y$, where $y$ is a variable that does not occur in $E$. Then update 
$E$ to $H$; update
$f$ to $f\cup\{y/c\}$ if $t$ is a constant, and to $(f-\{t/c\})\cup\{y/c\}$ if $t$ is a variable. 
\item[Subcase (iv):] $\alpha$ does not satisfy any of the above conditions (i)-(iii). Then go into an infinite loop in a permission state.
\end{description}
\item[Case of Rule B1:] Let $H$ be the premise of $E$ in the proof. $H$ is the result of substituting, in $E$, a certain positive (resp. negative) surface occurrence of a subformula $G_1\adc\ldots\adc G_n$ (resp. $G_1\add\ldots\add G_n$) by $G_i$ for some $i\in\{1,\ldots,n\}$. 
Let $\beta$ be the $E$-specification of that occurrence. Then make the move $\beta i$, update $E$ to $H$, and update $f$ by deleting in it all pairs 
$x/d$ such that $x$ is not a free variable of $H$. 
\item[Case of Rule B2:] Let $H$ be the premise of $E$ in the proof. $H$ is the result of substituting, in $E$, a certain positive (resp. negative) surface occurrence of a subformula $\ada xG(x)$ (resp. $\ade xG(x)$) by $G(y)$ for some variable $y$ not occurring in $F$. Let $\beta$ be the $E$-specification of that occurrence. Let $c$ be the smallest constant not occurring in $fE$. Then make the move $\beta c$, update $E$ to $H$, and update $f$ to $f\cup\{y/c\}$ (unless $x$ did not really have free occurrences in $G(x)$, in which case $f$ should stay as it was).
\end{description}

\begin{sublemma}{unic}{2} Suppose $e$ is $F$-distinctive. For each $i\geq 1$ such that LOOP is iterated at least $i$ times in $B$, let $E_i$ and $f_i$ be the values of $E$ and $f$ at the beginning of the $i${\em th} iteration. Then, for each such $i$ 
{\em (}in clauses $($a$)$-$($c$)${\em )}, we have:
\begin{description}
\item[(a)] $E_i$ is a proof formula.
\item[(b)] $f_i$ is an $E_i$-distinctive valuation.
\item[(c)] As long as $i>1$, $E_{i}$ is a premise of $E_{i-1}$ in the proof.
\item[(d)] LOOP is iterated a finite, nonzero number of times in $B$.
\item[(e)] Where $l$ is the number of iterations of LOOP in $B$, $E_l$ is derived by Rule {\bf A} and hence is instable.
\item[(f)] $B$ is a fair branch.
\end{description}
\end{sublemma}

\begin{subproof} Clauses (a)-(c) are obvious from the description of LOOP. Formally they can be verified by  straightforward induction on $i$. Note that clauses (a) and (b) imply that the {\bf while} condition of LOOP is always satisfied. 

In view of the assumption of the lemma regarding $e$, $e$ will pass the test for $F$-distinctiveness during the preinitialization stage, so LOOP will be iterated at least once. And clause (c) implies that the number of iterations of 
LOOP cannot be infinite --- in particular, cannot exceed the number of proof formulas. This proves clause (d).

For the remaining two clauses, assume $l\geq 1$ is the number of iterations of LOOP in $B$. 
As $E_l$ is a proof formula, it should be derived by one of the three rules of $\copredel$. Among 
those rules, only Rule {\bf A} is possible, for otherwise, as it is easy to see, we would have a next iteration of LOOP. Thus, clause (e) holds.

For clause (f), we want to show that ${\mathcal E}$ will grant permission infinitely many times --- in particular, it will do 
so during the $l$th iteration of LOOP. By clause (e), the $l$th iteration of LOOP deals with the case of Rule {\bf A}.
What ${\mathcal E}$ does during that iteration is that it keeps granting permission until the adversary responds by a move. If 
such a response is never made, permission will be granted infinitely many times. Suppose now the adversary makes a 
move $\alpha$.
$\alpha$ cannot be a move that satisfies the conditions of one of the Subcases (i)-(iii), for then we would have an 
$(l+1)$th iteration of LOOP. Thus, we deal with Subcase (iv), in which, again, ${\mathcal E}$ will grant permission infinitely many times.
\end{subproof}

\begin{sublemma}{unic}{3} ${\mathcal E}$ is fair.\end{sublemma}

\begin{subproof} Keeping in mind that $e$ and $B$ are arbitrary (Convention \ref{unic}.1), all we need to show is that 
 $B$ is fair, i.e. permission will be granted infinitely many times in $B$. By Lemma \ref{unic}.2(f), if
$e$ is $F$-distinctive, then $B$ is fair. And 
if $e$ is not $F$-distinctive, then the fairness of $B$ can be directly seen from the description of the preinitialization stage. \end{subproof}

As mentioned earlier, we are going to use ${\mathcal E}$ as an environment's strategy, so that we will be interested in runs cospelled rather than spelled by computation branches of ${\mathcal E}$. This means that when analyzing how such runs are generated,
we should assume that the moves made by ${\mathcal E}$ get the label $\oo$ rather than $\pp$, and the moves made by its adversary get the label $\pp$ rather than $\oo$.

For the rest of this section, let us agree on the following:\vspace{7pt}

\begin{subconvention}{unic}{4} Suppose $e$ is $F$-distinctive so that, according to Lemma \ref{unic}.2(d), LOOP makes a  finite, nonzero number of iterations in $B$. Then:
\begin{itemize}
\item $l$ will denote the number of iterations of LOOP in $B$, so that the $l${\em th} iteration is the last iteration.
\item $E_i$ $($where $1\leq i\leq l$$)$ will denote the value of record $E$ at the beginning of the $i${\em th}
iteration of LOOP in $B$.
\item $f_i$ $($where $1\leq i\leq l$$)$ will denote the value of record $f$ at the beginning of the $i${\em th}
iteration of LOOP in $B$.
\item $K_i$ $($where $1\leq i\leq l$$)$ will stand for $f_iE_i$.
\item $\Theta_i$ $($where $1\leq i\leq l$$)$ will stand for the sequence of the moves made by the players --- in their normal order --- by the beginning of the $i${\em th} iteration of LOOP in $B$, where the moves made by ${\mathcal E}$ are $\oo$-labeled and the moves made by its adversary $\pp$-labeled.
\end{itemize}
\end{subconvention}

\begin{sublemma}{unic}{5} Suppose $e$ is $F$-distinctive. Then, for every $i$ with $1\leq i\leq l$ and 
every interpretation $^*$, we have
$\Theta_i\in\legal{K_{1}^{*}}{}$ and $\seq{\Theta_i}K_{1}^{*}=K_{i}^{*}$.
\end{sublemma}

\begin{subproof} Assume $e$ is $F$-distinctive. We proceed by induction on $i$. The basis case with $i=1$ is trivial
taking into account that $\Theta_1=\emptyrun$. 
Now consider an arbitrary $i$ with $1\leq i <l$. By the induction hypothesis, $\Theta_i\in\legal{K_{1}^{*}}{}$ and $\seq{\Theta_i}K_{1}^{*}=K_{i}^{*}$.

Suppose the $i$th iteration of LOOP in
$B$ deals with the case of Rule {\bf A}. Then ${\mathcal E}$ does not make a move during this iteration. Since $i$ is not the last iteration, the adversary should make a move $\alpha$ that satisfies the conditions of 
one of the Subcases (i)-(iii), and then we will have $\Theta_{i+1}=\seq{\Theta_i,\pp\alpha}$.  Analyzing how $E_i$ and $f_i$ are updated to $E_{i+1}$ and $f_{i+1}$ in this case, in view of Lemma \ref{splem}(b) it is easy to see that then 
$\seq{\pp\alpha}\in\legal{K_{i}^{*}}{}$ and $\seq{\pp\alpha}K_{i}^{*}=K_{i+1}^{*}$, whence, with the equalities 
 $K_{i}^{*}=\seq{\Theta_i}K_{1}^{*}$ and $\Theta_{i+1}=\seq{\Theta_i,\pp\alpha}$ in mind, we have 
$\Theta_{i+1}\in\legal{K_{1}^{*}}{}$ and $\seq{\Theta_{i+1}}K_{1}^{*}=K_{i+1}^{*}$.

Suppose now the $i$th iteration of LOOP deals with the case of one of the Rules {\bf B1} or {\bf B2}. Then the adversary does not 
move during this iteration. ${\mathcal E}$ makes a one single move $\alpha$ so that $\Theta_{i+1}=\seq{\Theta_i,\oo\alpha}$. Again, analyzing what kind of a move this $\alpha$ is and how $E_i$ and $f_i$ are updated to $E_{i+1}$ and $f_{i+1}$, in view of Lemma \ref{splem}(a) we can easily see that $\seq{\oo\alpha}\in\legal{K_{i}^{*}}{}$ and $\seq{\oo\alpha}K_{i}^{*}=K_{i+1}^{*}$, whence, with the equalities 
 $K_{i}^{*}=\seq{\Theta_i}K_{1}^{*}$ and $\Theta_{i+1}=\seq{\Theta_i,\oo\alpha}$ in mind, we have 
$\Theta_{i+1}\in\legal{K_{1}^{*}}{}$ and $\seq{\Theta_{i+1}}K_{1}^{*}=K_{i+1}^{*}$. 
\end{subproof}

\begin{sublemma}{unic}{6} Suppose $e$ is $F$-distinctive, and 
$\Gamma$ is the run cospelled by $B$. Then, for any interpretation $^*$ with $\win{K_{l}^{*}}{e}\emptyrun=\oo$, we have $\win{F^*}{e}\seq{\Gamma}=\oo$.
\end{sublemma}

\begin{subproof} Assume $e$ is $F$-distinctive, $B$ cospells $\Gamma$ and $\win{K_{l}^{*}}{e}\emptyrun=\oo$. By Lemma  \ref{unic}.5, $\Theta_l\in\legal{K_{1}^{*}}{}$ and $\seq{\Theta_l}K_{1}^{*}=K_{l}^{*}$. Since 
$\win{K_{l}^{*}}{e}\emptyrun=\oo$, we then have  $\win{\seq{\Theta_l}K_{1}^{*}}{e}\emptyrun=\oo$, whence $\win{K_{1}^{*}}{e}\seq{\Theta_l}=\oo$, i.e. $\win{e[K_{1}^{*}]}{}\seq{\Theta_l}=\oo$, i.e. $\win{e[(f_1E_1)^*]}{}\seq{\Theta_l}=\oo$. Then, remembering from the description of the initialization stage that $f_1$ agrees with $e$ on all free variables of $F$ and $E_1=F$, Lemma \ref{agr} yields
$\win{e[F^{*}]}{}\seq{\Theta_l}=\oo$, i.e.
\begin{equation}\label{nov25}
\win{F^*}{e}\seq{\Theta_l}=\oo.
\end{equation}
Back to $\Gamma$. Obviously $\Theta_l$ is an initial segment of $\Gamma$. Since $E_l$ is derived by Rule {\bf A} (Lemma \ref{unic}.2(e)), the $l$th iteration of
LOOP deals with Case of Rule {\bf A}. So, ${\mathcal E}$ does not move during this iteration. If its adversary does not make moves either,
then $\Theta_l=\Gamma$ and, by (\ref{nov25}), $\win{F^*}{e}\seq{\Gamma}=\oo$. Suppose now the adversary makes a move $\alpha$ during the $l$th iteration. $\alpha$ cannot be a move that satisfies the conditions of one of the 
Subcases (i)-(iii), for otherwise there would be an $(l+1)$th iteration of LOOP. But if none of those three conditions is satisfied, then it can be seen from Lemma \ref{splem}(b) that we must have $\seq{\pp\alpha}\not\in\legal{K_{l}^{*}}{}$. Consequently,
by Lemma \ref{unic}.5, $\seq{\pp\alpha}\not\in\legal{\seq{\Theta_l}K_{1}^{*}}{}$, whence 
$\seq{\Theta_l,\pp\alpha}\not\in\legal{K_{1}^{*}}{}$, whence, in view of Lemmas \ref{struct} and \ref{new2},    $\seq{\Theta_l,\pp\alpha}\not\in\legal{F^{*}}{}$. But 
$\seq{\Theta_l,\pp\alpha}$ is an initial segment of $\Gamma$, which makes $\Gamma$ a $\pp$-illegal and hence $\oo$-won run of $e[F^*]$. \end{subproof}

To proceed with our proof of Proposition \ref{unic}, we need to agree on some additional terminology. In the following convention, when using set-theoretic notation such as $c\in\vec{c}$,  
we identify a tuple $\vec{c}$ of constants with the set of the constants that appear in $\vec{c}$.\vspace{7pt}
 
\begin{subconvention}{unic}{7} Suppose $\vec{a}=(a_1,\ldots,a_r)$ and $\vec{b}=(b_1,\ldots,b_r)$ are two $r$-tuples of pairwise distinct constants. 
Let $(a'_1,\ldots,a'_m)$ be the result of deleting in $\vec{a}$ all constants that are in $\vec{b}$. Similarly, let 
$(b'_1,\ldots,b'_m)$ be the result of deleting in $\vec{b}$ all constants that are in $\vec{a}$.  
We define the \gj{$(\vec{a},\vec{b})$-permutation}\label{x400} as the function $\hbar:\ \constants\rightarrow\constants$ such that, for every constant $c$, we have:
\begin{itemize}
\item If $c\not\in(\vec{a}\cup\vec{b})$, then $\hbar c=c$.
\item If $c=b_i\in\vec{b}$ $($$1\leq i\leq r$$)$, then $\hbar c=a_i$.
\item If $c=a'_j\in(\vec{a}-\vec{b})$ $($$1\leq j\leq m$$)$, then $\hbar c=b'_j$.
\end{itemize} 
\end{subconvention}

The following statement is obvious:\vspace{7pt}

\begin{sublemma}{unic}{8} For any tuples $\vec{a}=(a_1,\ldots,a_r)$ and $\vec{b}=(b_1,\ldots,b_r)$ of pairwise distinct constants, the $(\vec{a},\vec{b})$-permutation is an effective, bijective function from $\constants$ to $\constants$.
\end{sublemma}

For the rest of this section we assume that:\vspace{7pt}

\begin{subassumption}{unic}{9} \begin{itemize}
\item $H_1,\ldots,H_k$ are all the instable proof formulas. 
\item $G_1(x_{1}^{1},\ldots,x_{r_1}^{1}),\ldots,G_k(x_{1}^{k},\ldots,x_{r_k}^{k})$ are the elementarizations of $H_1,\ldots,H_k$, respectively, where, for each $1\leq i\leq k$, we assume that 
$x_{1}^{i},\ldots,x_{r_i}^{i}$ are all the $($pairwise distinct$)$ free variables of $G_i(x_{1}^{i},\ldots,x_{r_i}^{i})$.\vspace{-5pt} \end{itemize}
\end{subassumption}

By a \gj{$\Delta_2$-interpretation}\label{x56} we mean an interpretation that interprets each predicate letter as a (finitary) predicate of complexity $\Delta_2$. 

By G\"{o}del's completeness theorem for classical predicate calculus --- in particular, the version of the proof of that theorem as given in Section 72 of \cite{Kle52} --- for every formula $G(w_1,\ldots,w_r)$ of the classical language that is not (classically) valid and 
whose free variables are exactly $w_1,\ldots,w_r$, there is a classical model with domain $\{0,1,2,\ldots\}$ 
 and an $r$-tuple $a_1,\ldots,a_r$ of pairwise distinct objects of the domain such that, in that model,
\begin{itemize}
\item every predicate letter is interpreted as a predicate of complexity $\Delta_2$;
\item $G(a_1,\ldots,a_r)$ is false.
\end{itemize}

\noindent Such a model is nothing but what we would call a perfect (see page \pageref{pi}) $\Delta_2$-interpretation.
Based on the above fact and taking into account that each of the $G_i(x_{1}^{i},\ldots,x_{r_i}^{i})$ ($1\leq i\leq k$) is
a classically non-valid  elementary formula, we fix the following perfect $\Delta_2$-interpretations and tuples of constants:\vspace{7pt}

\begin{subassumption}{unic}{10} 
For each $1\leq i\leq k$, 
\begin{itemize}
\item $^{\star i}$ is a perfect $\Delta_2$-interpretation and 
\item $\vec{a^i}=(a_{1}^{i},\ldots,a_{r_i}^{i})$ are pairwise distinct constants such that
$G_i(a_{1}^{i},\ldots,a_{r_i}^{i})$ is false in $^{\star i}$. 
\end{itemize}
\end{subassumption}

For each $1\leq i\leq k$ and each $n$-ary predicate letter $p$,  let  

\begin{center}$A_{i}^{p}(u_1,\ldots,u_{n})\ =\ p^{\star i}$\end{center}

(of course, it is legitimate to assume that the attached tuple of each $p^{\star i}$ comes from the same pool 
$u_1,u_2,\ldots$ of variables).   

Let us fix an effective encoding of tuples of pairwise distinct constants. We assume that every such tuple has exactly one code, and vice versa: every $c_0\in\{0,1,\ldots\}$ is the code of exactly one tuple of pairwise distinct constants.

For each $1\leq i\leq k$ and each $n$-ary predicate letter $p$, we define the predicate
\begin{center}$B_{i}^{p}(u_0,u_1,\ldots,u_{n})$\end{center}
by stipulating that, for any $c_0,\ldots,c_{n}$,  
$B_{i}^{p}(c_0,\ldots,c_{n})$ is true iff $c_0$ is the code of an $r_i$-tuple $\vec{b}$ of pairwise distinct constants and, where $\hbar$ is the $(\vec{a^i},\vec{b})$-permutation, 
$A_{i}^{p}(\hbar c_1,\ldots, \hbar c_{n})$ is true. 

Since $\hbar$ is an effective function and the complexity of  $A_{i}^{p}$ is $\Delta_2$, we obviously have:\vspace{10pt}

\begin{sublemma}{unic}{11} For any $n$-ary predicate letter $p$ and any $1\leq i\leq k$, the complexity of $B_{i}^{p}(u_0,u_1,\ldots,u_n)$ is $\Delta_2$.
\end{sublemma}

Remember that $x_1,\ldots,x_q$ are all the free variables of $F$. We also select and fix an arbitrary variable $s$ that does not occur in $F$. And we fix a constant $d_0$ such that no constant occurring in $F$ is greater than $d_0$.

For each constant $c$, we define the valuation $e_c$\label{x500} by stipulating  that:
\begin{itemize}
\item $e_c(s)=c$;
\item $e_c(x_1)=d_{0}+1$;\ \ \ \ldots;\ \ \ $e_c(x_q)=d_{0}+q$;
\item for any other variable $z$, \ $e_c(z)=0$.
\end{itemize}

Notice that:\vspace{7pt}

\begin{sublemma}{unic}{12} \\
a) For any constant $c$, \ $e_c$ is an $F$-distinctive valuation.\\
b) The function $g$ defined by $g(c,i)=e_c(v_i)$ is effective.\end{sublemma}

We fix the list 
\({{\mathcal H}}_0,\ {{\mathcal H}}_1,\ {{\mathcal H}}_2,\ \ldots\)
of all HPMs arranged according to the lexicographic order of their (standardized) descriptions. 

According to Lemma \ref{unic}.3, ${\mathcal E}$ is fair. Hence, for every HPM ${{\mathcal H}}$ and valuation $f$, the 
$({{\mathcal E}},f,{{\mathcal H}})$-branch (see Lemma \ref{lem}) is defined. 

For each constant $c$, we define:
\begin{itemize}
\item $B_c$\label{x600} as the $({{\mathcal E}},e_c,{{\mathcal H}}_c)$-branch,\footnote{Not to confuse with the predicates $B^{p}_{i}$.} and
\item $\Gamma_c$\label{x700} as the ${{\mathcal H}}_c$ vs ${\mathcal E}$ run on $e_c$, i.e. the run cospelled by $B_c$.
\end{itemize}

Note that, by Lemmas \ref{unic}.12(a) and
\ref{unic}.2(d), LOOP is iterated a finite, nonzero number of times in $B_c$.

Next, where $1\leq i\leq k$, we define the predicate $\Last_i(x,x')$\label{x800}  
by stipulating that, for any constants $c,c'$,
\begin{itemize}
\item $\Last_i(c,c')$ is true iff we have:
\begin{itemize}
\item The value of record $E$ in the last iteration of LOOP in $B_c$ is $H_i$; 
\item $c'$ is the code of $\vec{b}$, where $\vec{b}=b_1,\ldots,b_{r_i}$ are the constants assigned to the variables $x_{1}^{i},\ldots,x_{r_i}^{i}$ by the value of record $f$ in the last iteration of LOOP in $B_c$. Note that, in view of Lemma \ref{unic}.2(b), $b_1,\ldots,b_{r_i}$ are pairwise distinct.
\end{itemize}
\end{itemize}  

\begin{sublemma}{unic}{13} For each $1\leq i\leq k$, the predicate {\em $\Last_i(x,x')$} has complexity $\Delta_2$.
\end{sublemma} 

\begin{subproof} Updates of records $E$ and $f$ generally may take several computation steps. Let us call such steps (configurations of ${\mathcal E}$) --- together with the steps within the preinitialization and initialization stages --- {\em transitional}, and call all other steps {\em non-transitional}. Thus, it is the non-transitional configurations in which records $E$ and $f$ have definite values, with the former being a proof formula and the latter being a finite valuation. For each $1\leq i\leq k$, let $K_i(y,x,x')$ be the predicate such that 
$K_i(n,c,c')$ is true iff the $n$th configuration of $B_c$ is non-transitional, 
the value of record $E$ in that configuration is $H_i$, and $c'$ is the code of $\vec{b}$, where $\vec{b}=b_1,\ldots,b_{r_i}$ are the constants assigned to the variables $x_{1}^{i},\ldots,x_{r_i}^{i}$ by the value 
of record $f$ in the $n$th configuration. 
In view of Lemmas \ref{lem}(b) and \ref{unic}.12(b),
it is not hard to see that $K_i$ is a decidable predicate. We know that the values of records $E$ and $f$ should stabilize at some computation step $m$ 
of $B_c$ and never change afterwards. In particular, such an $m$ is the first configuration of the last iteration of LOOP in $B_c$. With this fact in mind and some little thought, we can find that  $\Last_i(x,x')=\cle z\cla y\bigl(y\geq z\mli K_i(y,x,x')\bigr)$ and  $\gneg\Last_i(x,x')=\cle z\cla y\bigl(y\geq z\mli \gneg K_i(y,x,x')\bigr)$. This means that 
$\Last_i(x,x')$ has complexity $\Delta_2$. \end{subproof}

For any $n$-ary predicate letter $p$ and any $1\leq i\leq k$, we now define the predicate $C_{i}^{p}(s,u_1,\ldots,u_{n})$
by
\[C_{i}^{p}(s,u_1,\ldots,u_{n}) = \cle u_0\bigl(\Last_{i}(s,u_0)\mlc  B_{i}^{p}(u_0,u_1,\ldots,u_{n})\bigr).\]
 
For each $n$-ary predicate letter $p$, we define the predicate $D^p(u_1,\ldots,u_{n})$ by  
\[D^p(u_1,\ldots,u_{n})\ =\ C_{1}^{p}(s,u_1,\ldots,u_{n})\mld\ldots\mld C_{k}^{p}(s,u_1,\ldots,u_{n}).\]
(Notice that $D^p(u_1,\ldots,u_{n})$ is generally $n+1$-ary rather than $n$-ary, with $s$ being a hidden variable on which it depends.)\vspace{7pt}

\begin{sublemma}{unic}{14} The predicate $D^p(u_1,\ldots,u_{n})$ has complexity $\Delta_2$ {\em (}any $n$-ary predicate letter $p${\em )}.
\end{sublemma} 

\begin{subproof} Disjunction preserves $\Delta_2$-complexity. So, in order to show that the predicate $D^p(u_1,\ldots,u_{n})$ is
of complexity $\Delta_2$, it would be sufficient to verify that each disjunct  $C_{i}^{p}(s,u_1,\ldots,u_{n})$ ($1\leq i\leq k$) of it has complexity 
$\Delta_2$. From Lemmas \ref{unic}.11 and \ref{unic}.13, together with the fact that $\mlc$ and $\cle$ preserve 
$\Sigma_2$-complexity, it is obvious that   $C_{i}^{p}(s,u_1,\ldots,u_{n})$ is of complexity $\Sigma_2$. Thus, what remains to show is that $\gneg C_{i}^{p}(s,u_1,\ldots,u_{n})$ is also of complexity $\Sigma_2$. We claim that 
\begin{equation}\label{jun29}
\begin{array}{ll}
\gneg C_{i}^{p}(s,u_1,\ldots,u_{n})= & 
\cle u_0\Bigl(\mld\{\Last_{j}(s,u_0)\ |\ j\not=i,\ j\in\{1,\ldots,k\}\}\\
 & \mld
\bigl(\Last_{i}(s,u_0)\mlc\gneg B_{i}^{p}(u_0,u_1,\ldots,u_{n})\bigr)\Bigr)
\end{array}
\end{equation} 
\noindent ($\mld S$ means the $\mld$-disjunction of the elements of $S$, understood as $\tlg$ when $S$ is empty). 
This claim can be verified based on the meanings of the predicates $C_{i}^{p}$ and $\Last_i$, and the observation 
that, for every (value of) $s$, there is exactly one $j\in\{1,\ldots,k\}$ and exactly one (value of) $u_0$ such that $\Last_{j}(s,u_0)$ is true. Details of this verification are left to the reader. 

Now, from Lemmas \ref{unic}.11 and \ref{unic}.13, together with the fact that $\mlc$, $\mld$ and $\cle$ preserve 
$\Sigma_2$-complexity, (\ref{jun29}) allows us to conclude that $\gneg C_{i}^{p}(s,u_1,\ldots,u_{n})$ is indeed of complexity $\Sigma_2$. \end{subproof}

Now we define the interpretation $^*$ by stipulating that, for each $n$-ary predicate letter $p$, 
\begin{center}$p^* \ = \ D^p(u_1,\ldots,u_{n}).$\end{center} 
Lemma \ref{unic}.14 then means that $^*$ is a $\Delta_2$-interpretation. The fact that variable $s$ does not occur in $F$ guarantees that this interpretation is $F$-admissible. What remains to show is that no HPM wins 
$F^*$. We are going to do this by proving that each ${{\mathcal H}}_c$ loses $F^*$ against ${\mathcal E}$ on $e_c$.\vspace{7pt}

\begin{sublemma}{unic}{15} Assume the following:
\begin{enumerate}
\item $c,c'\in\constants$ and $i\in\{1,\ldots,k\}$ are such that {\em $\Last_i(c,c')$} is true;
\item $\vec{b}=(b_1,\ldots,b_{r_i})$ is the tuple of pairwise distinct constants encoded by $c'$;
\item $\hbar$ is the $(\vec{a^i},\vec{b})$-permutation.
\end{enumerate}
Then, for any elementary formula $J(z_1,\ldots,z_n)$ whose free variables are exactly $z_1,\ldots,z_n$ and any 
constants 
$c_1,\ldots,c_n$, $e_c\bigl[\bigl(J(c_1,\ldots,c_n)\bigr)^*\bigr]=\bigl(J(\hbar c_1,\ldots,\hbar c_n)\bigr)^{\star i}$.
\end{sublemma}

\begin{subproof}  Assume the conditions of the lemma are satisfied, and $J(z_1,\ldots,z_n)$ is an elementary formula 
whose free variables are exactly $z_1,\ldots,z_n$. We prove the lemma by induction on the complexity of $J(z_1,\ldots,z_n)$.

For the basis of induction, we need to consider the case when $J(z_1,\ldots,z_n)$ is atomic. The cases when it is $\tlg$ or $\twg$ are trivial, so
suppose $J(z_1,\ldots,z_n)$ is a non-logical atom. For simplicity of representation and obviously without loss of generality, we may assume that $J(z_1,\ldots,z_n)=p(z_1,\ldots,z_n)$, where $p$ is an $n$-ary predicate letter. Then 
\(\bigl(J(c_1,\ldots,c_n)\bigr)^*=D^p(c_1,\ldots,c_n).\)
In turn, 
\(D^p(c_1,\ldots,c_n)=C_{1}^{p}(s,c_1,\ldots,c_{n})\mld\ldots\mld C_{k}^{p}(s,c_1,\ldots,c_{n}).\)
Thus, 
\begin{equation}\label{y1}e_c\bigl[\bigl(J(c_1,\ldots,c_n)\bigr)^*\bigr]=e_c[C_{1}^{p}(s,c_1,\ldots,c_{n})\mld\ldots\mld C_{k}^{p}(s,c_1,\ldots,c_{n})].\end{equation}
We obviously have 
\[\begin{array}{l}e_c[C_{1}^{p}(s,c_1,\ldots,c_{n})\mld\ldots\mld C_{k}^{p}(s,c_1,\ldots,c_{n})]=\\
C_{1}^{p}(c,c_1,\ldots,c_{n})\mld\ldots\mld C_{k}^{p}(c,c_1,\ldots,c_{n}).\end{array}\]
According to assumption (1) of the lemma, $\Last_i(c,c')$ is true. As noted earlier in the proof of Lemma \ref{unic}.14, 
$i$ and $c'$ are unique values for which $\Last_i(c,c')$ is true. Each component $C_{j}^{p}(c,c_1,\ldots,c_{n})$ in the above disjunction contains (under $\cle u_0$) the conjunct $\Last_j(c,u_0)$ which is thus false when 
$j\not=i$, and hence each such disjunct $C_{j}^{p}(c,c_1,\ldots,c_{n})$ can be deleted. So, 
\[\begin{array}{l}C_{1}^{p}(c,c_1,\ldots,c_{n})\mld\ldots\mld C_{k}^{p}(c,c_1,\ldots,c_{n})=C_{i}^{p}(c,c_1,\ldots,c_{n})
=\\
\cle u_0\bigl(\Last_{i}(c,u_0)\mlc  B_{i}^{p}(u_0,c_1,\ldots,c_{n})\bigr).\end{array}\]
Since $c'$ is the only constant for which 
$\Last_i(c,c')$ is true, $\cle u_0\bigl(\Last_{i}(c,u_0)\mlc  B_{i}^{p}(u_0,c_1,\ldots,c_{n})\bigr)$ can be equivalently rewritten as $B_{i}^{p}(c',c_1,\ldots,c_{n})$. Thus,  
\[\cle u_0(\Last_{i}(c,u_0)\mlc  B_{i}^{p}(u_0,c_1,\ldots,c_{n})\bigr)=B_{i}^{p}(c',c_1,\ldots,c_{n}).\]
In turn, based on assumptions (2) and (3) of the lemma,
\[B_{i}^{p}(c',c_1,\ldots,c_{n})=A_{i}^{p}(\hbar c_1\ldots,\hbar c_{n}).\]
Finally, notice that 
\begin{equation}\label{y2}A_{i}^{p}(\hbar c_1\ldots,\hbar c_{n})=\bigl(p(\hbar c_1\ldots,\hbar c_{n})\bigr)^{\star i}=\bigl(J(\hbar c_1\ldots,\hbar c_{n})\bigr)^{\star i}.\end{equation}
From the chain of equations from (\ref{y1}) to (\ref{y2}) we get 
$e_c\bigl[\bigl(J(c_1,\ldots,c_n)\bigr)^*\bigr]=\bigl(J(\hbar c_1\ldots,\hbar c_{n})\bigr)^{\star i}$, which completes our proof of the basis case of induction.

For the inductive step, we will only consider the case when the main operator of $J(z_1,\ldots,z_n)$ is $\cle$. The case with $\cla$ is similar, and the cases with $\gneg,\mlc,\mld,\mli$ are simpler or straightforward.

So, \ assume \ $J(z_1,\ldots,z_n)=\cle z_0 I(z_0,z_1,\ldots,z_n)$. \ Then \ 
$\bigl(J(c_1,\ldots,c_n)\bigr)^*=$ $\bigl(\cle z_0I(z_0,c_1,\ldots,c_n)\bigr)^*=\cle z_0 \Bigl(\bigl(I(z_0,c_1,\ldots,c_n)\bigr)^*\Bigr)$
and 
\(\bigl(J(\hbar c_1,\ldots,\hbar c_n)\bigr)^{\star i}=\bigl(\cle z_0I(z_0,\hbar c_1,\ldots,\hbar c_n)\bigr)^{\star i}=\cle z_0 \Bigl(\bigl(I(z_0,\hbar c_1,\ldots,\hbar c_n)\bigr)^{\star i}\Bigr).\)
Thus, we need to show that $e_c\Bigl[\cle z_0 \Bigl(\bigl(I(z_0,c_1,\ldots,c_n)\bigr)^*\Bigr)\Bigr]=\cle z_0 \Bigl(\bigl(I(z_0,\hbar c_1,\ldots,\hbar c_n)\bigr)^{\star i}\Bigr)$. In other words, show that $e_c\Bigl[\cle z_0 \Bigl(\bigl(I(z_0,c_1,\ldots,c_n)\bigr)^*\Bigr)\Bigr]$ is true iff $\cle z_0 \Bigl(\bigl(I(z_0,\hbar c_1,\ldots,\hbar c_n)\bigr)^{\star i}\Bigr)$ is so.
In what follows we implicitly rely on Lemma \ref{new2}.
Suppose $e_c\Bigl[\cle z_0 \Bigl(\bigl(I(z_0,c_1,$ $\ldots,c_n)\bigr)^*\Bigr)\Bigr]$   is true.  This means that there is a constant $c_0$ such that 
$e_c\bigl[\bigl(I(c_0,c_1,$ $\ldots,c_n)\bigr)^*\bigr]$ is true. Then, by the induction hypothesis,
 $\bigl(I(\hbar c_0,\hbar c_1,\ldots,\hbar c_n)\bigr)^{\star i}$ is true. In turn, this implies that $\cle z_0 \Bigl(\bigl(I(z_0,\hbar c_1,\ldots,\hbar c_n)\bigr)^{\star i}\Bigr)$ is true.
Now suppose  $\cle z_0 \Bigl(\bigl(I(z_0,\hbar c_1,\ldots,\hbar c_n)\bigr)^{\star i}\Bigr)$ is true. This means that for some constant $d$,
$\bigl(I(d,\hbar c_1,\ldots,\hbar c_n)\bigr)^{\star i}$ is true. Since $\hbar$ is a bijection (Lemma \ref{unic}.8), there is 
a constant $c_0$ with $\hbar c_0=d$. Then, by the induction hypothesis, $e_c\bigl[\bigl(I(c_0,c_1,\ldots,c_n)\bigr)^{*}\bigr]$ is true.
Consequently, $e_c\Bigl[\cle z_0\Bigl(\bigl(I(z_0,c_1,\ldots,c_n)\bigr)^{*}\Bigr)\Bigr]$ is true.
\end{subproof}

\begin{sublemma}{unic}{16} ${{\mathcal H}}_c$ does not win $F^*$ against ${\mathcal E}$ on $e_c$ {\em (}any constant $c${\em )}.
\end{sublemma}

\begin{subproof} Fix an arbitrary $c$. 
In what follows we rely on our Convention \ref{unic}.1 with $e_c$ and $B_c$ in the roles of $e$ and $B$, respectively. 
That is, in the present context $e_c$ and $B_c$ should be understood as synonyms of to what the earlier parts of the present section referred as $e$ and $B$.
The fact that $e_c$ is $F$-distinctive (Lemma \ref{unic}.12) allows us to use the notation established in Convention \ref{unic}.4. According to Lemma \ref{unic}.2(d), LOOP is iterated a finite (and nonzero) number of times --- 
in particular, $l$ times in $B_c$. Then, by clauses (a) and (e) of Lemma \ref{unic}.2, $E_l$ is an instable proof formula. Hence
$E_l=H_i$ for one (and exactly one as we assume that the $\copredel$-proof of $F$ has no repetitions) of the $i$ 
with $1\leq i\leq k$. Fix this $i$. 
Consider $f_l$ --- the value of record $f$ at the beginning of the last iteration of LOOP in $B_c$. Let $\vec{b}=(b_1,\ldots,b_{r_i})$ be the values returned for $x_1,\ldots,x_{r_i}$ by $f_l$, and let $c'$ be the code 
of $\vec{b}$. So, $\Last_i(c,c')$ is true. 
Let $\hbar$ be the $(\vec{a^i},\vec{b})$-permutation. Thus, 
the three conditions of Lemma \ref{unic}.15 are satisfied. Then, according to that lemma,
$e_c\bigl[\bigl(G_i(b_1,\ldots,b_{r_i})\bigr)^*\bigr]=\bigl(G_i(\hbar b_1,\ldots,\hbar b_{r_i})\bigr)^{\star i}.$
But remembering the meaning of $\hbar$, we have 
\(\hbar b_1=a_{1}^{i},\ \ \ldots,\ \ \hbar b_{r_i}=a_{r_i}^{i}.\)
Thus, 
$e_c[\bigl(G_i(b_1,\ldots,b_{r_i})\bigr)^*]$ has the same truth value as 
$\bigl(G_i(a_{1}^{i},\ldots,a_{r_i}^{i})\bigr)^{\star i}$.
According to Assumption \ref{unic}.10, the latter is false. Then so is the former, which can be expressed by writing
\begin{equation}\label{t1}
\win{(G_i(b_1,\ldots,b_{r_i}))^*}{e_c}\emptyrun=\oo.\end{equation}

We have $E_l=H_i$ and hence $K_l=f_lH_i$. This obviously implies that $\elz{K_l}=f_l\elz{H_i}$. In turn, by Assumption 
\ref{unic}.9,  $\elz{H_i}=G_i(x_{1}^{i},\ldots,x_{r_i}^{i})$. And we also have $f_lG_i(x_{1}^{i},\ldots,x_{r_i}^{i})=
G_i(b_{1}^{i},\ldots,b_{r_i}^{i})$. Thus, $\elz{K_l}=G_i(b_{1}^{i},\ldots,b_{r_i}^{i})$. By (\ref{t1}), we then get 
$\win{\elzi{K_l}^*}{e_c}\emptyrun=\oo$. This, by Lemma \ref{new1}, implies 
$\win{K_{l}^{*}}{e_c}\emptyrun=\oo$. Then, by Lemma \ref{unic}.6, we get $\win{F^*}{e_c}\seq{\Gamma_c}=\oo$. Thus $\Gamma_c$, which is the ${{\mathcal H}}_c$ vs ${\mathcal E}$ run on $e_c$, is a lost (by ${{\mathcal H}}_c$) run of $F^*$ with respect to $e_c$, which means that ${{\mathcal H}}_c$ does not win $F^*$ against ${\mathcal E}$ on $e_c$.
\end{subproof}

Lemma \ref{unic}.16 essentially completes our proof of Proposition \ref{unic}: 
that ${{\mathcal H}}_c$ does not win $F^*$ against ${\mathcal E}$ on $e_c$ clearly means that it simply does not win $F^*$.
But every HPM is ${{\mathcal H}}_c$ for some $c$. Hence, no HPM wins $F^*$, and $F$ is not valid. \end{proof}

\section{Decidability of the $\forall,\exists$-free fragment of $\predel$}\label{s6}

This section is devoted to a proof of Theorem \ref{decid}. 
Let $F$ be an arbitrary formula that does not contain blind quantifiers. The decidability of the question $\predel\vdash F$ can be shown by induction on the complexity of $F$.  

$F$ is provable iff it is derivable from some provable formulas by one of the Rules {\bf A}, {\bf B1}, or {\bf B2}. We define a procedure that tests, as described below, each of these
three possibilities. If one of those three tests succeeds, the procedure returns 
``yes", otherwise returns ``no". \vspace{5pt}

{\bf Testing Rule A:} This routine has the following three steps. The whole test is considered to have succeeded iff 
each of those three steps succeed.

{\em Step 1:} Check whether $F$ is stable, i.e. whether $\elz{F}$ is classically valid. Note that the latter does not contain any quantifiers. 
The question of classical validity of a quantifier-free formula is, of course, decidable. Thus, this step takes only a finite amount of time. If $F$ is stable, the step has succeeded. Otherwise it has failed. 

{\em Step 2:} For each positive (resp. negative) surface occurrence of a subformula $G_1\adc\ldots\adc G_n$ (resp. 
$G_1\add\ldots\add G_n$) in $F$ and each $1\leq i\leq n$, see if $H$ is provable, where $H$ the result of replacing 
in $F$ the above occurrence by $G_i$. Just like $F$, $H$ does not contain blind quantifiers, and its 
complexity is lower than that of $F$. Hence, by the induction hypothesis, testing whether $\predel\vdash H$ takes a finite amount of time. Obviously there is only a finite number of such $H$s to test, so the whole Step 2 takes a finite amount of time. If all of such $H$s turn out to be provable, then the step has succeeded. Otherwise it has failed.

{\em Step 3:} For each positive (resp. negative) surface occurrence of a subformula $\ada xG(x)$ (resp. 
$\ade xG(x)$) in $F$, see if $H$ is provable, where $H$ the result of replacing 
in $F$ the above occurrence by $G(y)$, where $y$ is the smallest (in the lexicographic order) variable not occurring in 
$F$. As in the previous step, $H$ is $\cla,\cle$-free and its complexity is lower than that of $F$, whence, by the induction hypothesis, testing whether $\predel\vdash H$ takes a finite amount of time. Also, again there is only a finite number of such $H$s to test, so the whole Step 3 takes a finite amount of time. If all of such $H$s turn out to be provable, then the step has succeeded. Otherwise it has failed.

Before we describe how the other rules are tested, let us verify that $F$ is derivable by Rule {\bf A} iff each of the above three steps (and hence the whole test) succeeds. 

Assume all three steps succeed. Success of Step 1 means that $F$ is stable. And success of Steps 2 and 3 obviously means that there is $\vec{H}$ that satisfies conditions {\bf (i)} and {\bf (ii)} of Rule {\bf A}. Hence $F$ is derivable from that $\vec{H}$ by Rule {\bf A}.

Now assume one of the three steps fails. We want to show that then one of the conditions of Rule {\bf A} is violated for $F$ as a possible conclusion of that rule. Indeed: Failure of Step 1 means that the condition of stability of $F$ is violated.
Failure of Step 2 obviously means that there is no set $\vec{H}$ of formulas that would satisfy condition {\bf (i)} of Rule {\bf A}.
Suppose now Step 3 fails. In particular, there is a  positive (resp. negative) occurrence of a subformula $\ada xG(x)$ (resp. $\ade xG(x)$) in $F$ such that $\predel\not\vdash H$, where $H$ is the result of replacing 
in $F$ the above occurrence by $G(y)$, with $y$ being the smallest variable not occurring in 
$F$. Let us write $H$ as $H(y)$. In view of Lemma \ref{newest}, for any variable $y'$ not occurring in $F$, we would also 
have $\predel\not\vdash H(y')$. This obviously means that no set $\vec{H}$ of formulas satisfies condition 
{\bf (ii)} of Rule {\bf A}.\vspace{5pt}

Each of the following two routines takes a finite amount of time for the same reasons as the routines of the above-described Steps 2 and 3 did.\vspace{4pt}

{\bf Testing Rule B1:} For each negative (resp. positive) surface occurrence of a subformula $G_1\adc\ldots\adc G_n$ (resp. 
$G_1\add\ldots\add G_n$) in $F$ and each $1\leq i\leq n$, see if $H$ is provable, where $H$ the result of replacing 
in $F$ the above occurrence by $G_i$. 
If one of such $H$s turns out to be provable, then the test has succeeded. Otherwise it has failed. Clearly $F$ is derivable by Rule {\bf B1} iff this test succeeds.\vspace{4pt}

{\bf Testing Rule B2:} For each negative (resp. positive) surface occurrence of a subformula $\ada xG(x)$ (resp. 
$\ade xG(x)$) in $F$, do the following:

{\em Step 1}: See if $H$ is provable, where $H$ the result of replacing 
in $F$ the above occurrence of $\ada xG(x)$ (resp. $\ade xG(x)$) by $G(y)$, where $y$ is the smallest variable not occurring in $F$.

{\em Step 2:} For each free term $t$ of $F$ such that (if $t$ is a variable) neither the above occurrence of $\ada xG(x)$ (resp. $\ade xG(x)$) 
in $F$ nor any of the free occurrences of $x$ in $G(x)$ are in the scope of $\cla t$ or $\cle t$, see if $H$ is provable,
where $H$ the result of replacing 
in $F$ the above occurrence of $\ada xG(x)$ (resp. $\ade xG(x)$) by $G(t)$.

If one of the above $H$s turns out to be provable, then the test has succeeded. Otherwise it has failed. 

With Lemma \ref{newest} in mind, a little thought can convince us that $F$ is derivable by Rule {\bf B2} iff this test succeeds.

\twocolumn

\begin{center}{\bf Index}\end{center}
 
arity of atom \pageref{x203}\\ 
arity of game \pageref{x1}\\ 
arity of predicate letter \pageref{x2}\\
atom \pageref{x3}\\
attached tuple \pageref{x4}\\
blind operations \pageref{bq}\\
branch: $e$-computation \pageref{x5}\\
branch: 
 $({{\mathcal E}},e,{{\mathcal H}})$ \pageref{x6}; $({{\mathcal H}},e,{{\mathcal E}})$  \pageref{x6}\\
classical model \pageref{jun25}\\
closed formula \pageref{clfrm}\\
choice operations \pageref{ad}\\
computable: see ``winnable"\\
compute: see ``win"\\
configuration \pageref{x7}\\
constant \pageref{x8}\\
constant game \pageref{x9}\\
constructive rule of induction \pageref{cri}\\
content (of game) \pageref{x10}\\
$\Delta_2$ complexity \pageref{x555}\\
$\Delta_2$-interpretation \pageref{x56}\\
depend (game on variable) \pageref{x11}\\
depth of game \pageref{x12}\\
distinctive: $F$-distinctive \pageref{dst}\\
elementary: game \pageref{x15}; formula \pageref{x16}\\ 
elementarization \pageref{x17}\\
EPM (easy-play machine) \pageref{x18}\\
fair: branch \pageref{x20}; EPM \pageref{x21}\\
{\bf FD} \pageref{x22}\\
$\predel$ \pageref{pred}\\
$\copredel$ \pageref{x23}\\
finitary game \pageref{x24}\\
finite-depth game \pageref{x25}\\
free game \pageref{x26}\\
game \pageref{game}\\
HPM (hard-play machine) \pageref{x27}\\
illegal: move \pageref{x28}; run \pageref{x29}; $\xx$-illegal \pageref{x30}\\
infinitary game \pageref{x31}\\
instable formula \pageref{x32}\\
instance (of formula) \pageref{x33}\\ 
instantiation \pageref{x34}\\
interactive algorithm \pageref{x35}\\
interpretation \pageref{x36}\\
interpretation: admissible \pageref{x37}\\ 
interpretation: perfect \pageref{pi}\\
knowledge \pageref{x38}\\
labmove (labeled move) \pageref{x39}\\ 
$\Last_i(c,c')$ \pageref{x800}\\
legal: move \pageref{x40}; run \pageref{x41}\\ 
$\legal{}{}$ \pageref{game}\\
linear reduction \pageref{linred}\\
logical atom \pageref{x201}\\
mapping reduction \pageref{june16}\\
move \pageref{x42}\\
parallel operations \pageref{mc}\\
negation \pageref{neg}\\
non-logical atom \pageref{x202};\\
occurrence: positive, negative  \pageref{x43}\\
occurrence: surface \pageref{x44}\\
$\xx$ \pageref{x45}\\
permission: granting \pageref{x46}; state \pageref{x47}\\
permutation: $(\vec{a},\vec{b})$-permutation \pageref{x400}\\
player \pageref{x45}\\
position \pageref{x48}\\
predicate \pageref{x49}\\
predicate letter \pageref{x50}\\ 
prefixation \pageref{prfx}\\
recurrence operations \pageref{x19}\\
run \pageref{x51}\\
run: spelled \pageref{x53}; cospelled \pageref{x52}\\
run:  ${\mathcal H}$ vs ${\mathcal E}$ \pageref{x54}\\
$\Sigma_2$ complexity \pageref{x55}\\
solvable: see ``winnable"\\
solve: see ``win"\\
specification  (specify) \pageref{x300}\\
stable formula \pageref{x57}\\
static game \pageref{static}\\
strict game \pageref{x58}\\
structure (of game) \pageref{x59}\\
substitution of terms  \pageref{x61}\\
substitution of variables \pageref{x60}\\
term \pageref{x62}; \pageref{jun24}\\
trivial game \pageref{x63}\\
unilegal run \pageref{x64}\\
unistructural game \pageref{x65}; $x$-, in $x$ \pageref{x66}\\
valid \pageref{val}; uniformly valid \pageref{uv}\\
valuation \pageref{x67}\\ 
variable \pageref{x68},\pageref{x69}\\
winnable (computable, solvable) \pageref{x71}\\
win: machine game \pageref{x72},\pageref{x73}\\
win:  machine against machine \pageref{x74}\\
$\win{}{}$ \pageref{game}\\
\ \\
\ \\
$\tlg$ as player \pageref{x45}\\
$\tlg$ as game \pageref{x63}\\
$\twg$ as player \pageref{x45}\\
$\twg$ as game \pageref{x63}\\
$\gneg$ when applied to games \pageref{neg}\\
$\gneg$ when applied to players \pageref{x45}\\
$\gneg$ when applied to runs \pageref{x75}\\
$\mlc$ \pageref{x76}\\
 $\mld$ \pageref{x77}\\
$\mli$  \pageref{x78}\\
$\cla$ \pageref{x79}\\
$\cle$ \pageref{x80}\\
$\adc$ \pageref{x81}\\
$\add$ \pageref{x82}\\
$\ada$  \pageref{x83}\\
$\ade$ \pageref{x84}\\
$\st$ \pageref{x19}\\
$A[x_1/t_1,\ldots,x_n/t_n]$ \pageref{x60}\\
$F[t_1/t'_1,\ldots,t_n/t'_n]$ \pageref{x61}\\
$A(x_1,\ldots,x_n)$ \pageref{x90}\\ 
$F(x_1,\ldots,x_n)$ \pageref{x91}\\
$\seq{\Phi}A$ \pageref{prfx}\\
$e[A]$ \pageref{x92}\\
$\emptyrun$ \pageref{x93}\\
$\elz{F}$ \pageref{x95}\\
$\Gamma^\gamma$ \pageref{jun23}\\
$\spadesuit$ \pageref{x13}\\
$\mapsto$ \pageref{mpst}\\
$\vdash$ \pageref{vdsh}\\
$\not\vdash$ \pageref{vdsh}\\
$\models$ \pageref{x94}

\newpage
\begin{thebibliography}{99}
\bibitem{Abr94}
S. Abramsky and R. Jagadeesan, {\em Games and full completeness for multiplicative linear logic}, {\bf Journal of Symbolic Logic} 59 (2) (1994), pp. 543-574.

\bibitem{Ben01} J. van Benthem, {\bf Logic in Games}, ILLC, University of Amsterdam, 2001 (preprint).
 
\bibitem{Bla92} A. Blass, {\em A game semantics for linear logic}, {\bf Annals of Pure and Applied Logic} 56 (1992), pp. 183-220.

\bibitem{Gir87} J.Y. Girard, {\em Linear logic}, {\bf Theoretical Computer Science} 50 (1) (1987), pp.  1-102.

\bibitem{gug} A. Guglielmi and L. Strassburger, {\em Non-commutativity and MELL in the calculus of structures}, 
{\bf Computer Science Logic} (Paris, 2001), pp. 54-68, Lecture Notes in Computer Science 2142, Springer, Berlin, 2001.

\bibitem{Hay93} J.M.E. Hyland and C.-H.L. Ong, {\em Fair games and full completeness for multiplicative linear logic without the MIX-rule}, Preprint, 1993.

\bibitem{Jap97} G. Japaridze, {\em A  constructive game semantics for the language of linear logic}, {\bf Annals of Pure and Applied Logic} 85 (2) (1997), pp. 87-156.

\bibitem{Jap98} G. Japaridze and D. de Jongh, {\em The logic of provability}, in: S. Buss (ed.), {\bf Handbook of Proof Theory}, Elsevier Science B.V., North Holland, 1998, pp. 475-546.

\bibitem{Jap00} G. Japaridze, {\em The propositional logic of elementary tasks}, 
{\bf Notre Dame Journal of Formal Logic} 41 (2000), N 2, pp.171-183.

\bibitem{Jap02a} G. Japaridze, {\em The logic of tasks}, {\bf Annals of Pure and Applied Logic} 117 (2002), pp. 263-295.

\bibitem{Jap03} G.Japaridze, {\em Introduction to computability logic}, {\bf Annals of Pure and Applied Logic} 123 (2003), 
pp.1-99.

\bibitem{Jap04} G. Japaridze, {\em Propositional computability logic I}, {\bf Transactions on Computational Logic} (to appear in 2006).

\bibitem{Japint} G. Japaridze, {\em Computability logic: a formal theory of interaction}, {\bf 
arXiv:cs.LO/0404024} (2004). 


\bibitem{Kle52} S.C. Kleene, {\bf Introduction to Metamathematics}, D. van Nostrand Company, New York / Toronto, 1952.

\bibitem{Lor59} P. Lorenzen, {\em Ein dialogisches Konstruktivit\"{a}tskriterium}, in: {\bf Infinitistic Methods}, in: PWN,  Proc. Symp. Foundations of Mathematics, Warsaw, 1961, pp. 193-200.

\end{thebibliography}
\end{document}